
\def\threebar{$\ol{{\bf 3}}$}

\def\nc{{\rm nc}}
\def\SM{{\sss SM}}

\def\Asl{\hbox{/\kern-.7500em\it A}} 

\def\rht{{\sss R}}
\def\lft{{\sss L}}
\def\sw{s_w}
\def\cw{c_w}

\def\MW{M_{\sss W}}
\def\MZ{M_{\sss Z}}

\def\mt{m_t}

\def\as{\alpha_s(M_{{\sss Z}})}

\def\leff{\Scl_{\rm eff}}

\def\gl{g_\lft}
\def\gr{g_\rht}
\def\gbl{{g^b_\lft}}
\def\gbr{{g^b_\rht}}

\def\gfl{{g^f_\lft}}
\def\gfr{{g^f_\rht}}

\def\bbar{\ol{b}}
\def\cbar{\ol{c}}

\def\zbb{Z\overline{b} b}
\def\Zbb{\zbb}

\def\ALR{A_{\sss LR}}
\def\AFB#1{A^0_{\sss FB}(#1)}

\def\Rb{$\hbox{\titlefont R}_{\hbox{\myfont b}}$}

\def\chisqminpdof{\chi_{\sss min}^2/\hbox{d.o.f.}}

\def\gLt{g_{\sss L}^t}
\def\gRt{g_{\sss R}^t}
\def\cL{c_\lft}
\def\sL{s_\lft}
\def\cR{c_\rht}
\def\sR{s_\rht}
\def\sLb{s_\lft^\ssb}
\def\cLb{c_\lft^\ssb}

\def\cRb{c_\rht^\ssb}
\def\gLtp{{\tilde g}_{\sss L}^{tt'}}
\def\gRtp{{\tilde g}_{\sss R}^{tt'}}

\def\Vc{{\cal V}}

\def\Tp{{T'}}

\def\lnrpr{\ln \left({r' \over r}\right)}

\def\yfp{y_{\lower2pt\hbox{$\sss f\phi$}}}
\def\yfpp{y_{\lower2pt\hbox{$\sss f'\phi'$}}}
\def\yfpnp{y_{\lower2pt\hbox{$\sss f'\phi$}}}
\def\yfnpp{y_{\lower2pt\hbox{$\sss f\phi'$}}}
\def\fp{{f'}}
\def\gr{g_{\sss R}}
\def\gl{g_{\sss L}}
\def\gs{g_{\sss S}}

\def\UR{{\cal U}_{\sss R}}
\def\UL{{\cal U}_{\sss L}}

\def\URT{{\cal U}_{\sss R}^\dagger}
\def\ULT{{\cal U}_{\sss L}^\dagger}

\def\cS{c_{\sss S}}\def\sS{s_{\sss S}}
\def\cR{c_{\sss R}}\def\sR{s_{\sss R}}
\def\cL{c_{\sss L}}\def\sL{s_{\sss L}}

\def\Pl{\gamma_{\sss L}}
\def\Pr{\gamma_{\sss R}}
\def\darr{\raise1.5ex\hbox{$\leftrightarrow$}\mkern-16.5mu \partial_\mu} 
\def\htr{{\tilde h}_{2,\rht}^{\raise2pt\hbox{$\sss -$}}}
\def\htwo{{h}_{2}^{\raise2pt\hbox{$\sss -$}}} 
\def\Wt{\widetilde{W}^{\raise2pt\hbox{$\sss -$}}}
\def\sgn{{\rm sign}} 


\font\titlefont = cmr10 scaled\magstep 4
\font\myfont    = cmr10 scaled\magstep 2
\font\sectionfont = cmr10
\font\littlefont = cmr5
\font\eightrm = cmr8 

\def\ss{\scriptstyle} 
\def\sss{\scriptscriptstyle} 

\newcount\tcflag
\tcflag = 0  

\ifnum\tcflag = 0 \magnification = 1200 \fi  

\global\baselineskip = 1.2\baselineskip
\global\parskip = 4pt plus 0.3pt
\global\abovedisplayskip = 18pt plus3pt minus9pt
\global\belowdisplayskip = 18pt plus3pt minus9pt
\global\abovedisplayshortskip = 6pt plus3pt
\global\belowdisplayshortskip = 6pt plus3pt


\def\endignore{}
\def\ignore #1\endignore{}

\newcount\dflag
\dflag = 0


\def\monthname{\ifcase\month
\or January \or February \or March \or April \or May \or June%
\or July \or August \or September \or October \or November %
\or December
\fi}

\newcount\dummy
\newcount\minute  
\newcount\hour
\newcount\localtime
\newcount\localday
\localtime = \time
\localday = \day

\def\advanceclock#1#2{ 
\dummy = #1
\multiply\dummy by 60
\advance\dummy by #2
\advance\localtime by \dummy
\ifnum\localtime > 1440 
\advance\localtime by -1440
\advance\localday by 1
\fi}

\def\settime{{\dummy = \localtime%
\divide\dummy by 60%
\hour = \dummy
\minute = \localtime%
\multiply\dummy by 60%
\advance\minute by -\dummy
\ifnum\minute < 10 
\xdef\spacer{0} 
\else \xdef\spacer{} 
\fi %
\ifnum\hour < 12 
\xdef\ampm{a.m.} 
\else 
\xdef\ampm{p.m.} 
\advance\hour by -12 %
\fi %
\ifnum\hour = 0 \hour = 12 \fi
\xdef\timestring{\number\hour : \spacer \number\minute%
\thinspace \ampm}}}



\def\endtitle{}
\def\title#1\endtitle{\vskip.5in\titlefont
\global\baselineskip = 2\baselineskip
#1\vskip.4in
\baselineskip = 0.5\baselineskip\rm}
 
\def\endauthors{}
\def\authors#1\endauthors{#1}

\def\endabstract{}
\def\abstract#1\endabstract{\vskip .3in%
\centerline{\sectionfont\bf Abstract}%
\vskip .1in
\noindent#1}

\def\nopageonenumber{\footline={\ifnum\pageno<2\hfil\else
\hss\tenrm\folio\hss\fi}}  

\newcount\nsection
\newcount\nsubsection

\def\section#1{\global\advance\nsection by 1
\nsubsection=0
\bigskip\noindent\centerline{\sectionfont \bf \number\nsection.\ #1}
\bigskip\rm\nobreak}

\def\subsection#1{\global\advance\nsubsection by 1
\bigskip\noindent\sectionfont \sl \number\nsection.\number\nsubsection)\
#1\bigskip\rm\nobreak}

\def\topic#1{{\medskip\noindent $\bullet$ \it #1:}} 
\def\endtopic{\medskip}

\def\appendix#1#2{\bigskip\noindent%
\centerline{\sectionfont \bf Appendix #1.\ #2}
\bigskip\rm\nobreak}


\newcount\nref
\global\nref = 1

\def\therefs{} 


\def\ref#1#2{\xdef #1{[\number\nref]}
\ifnum\nref = 1\global\xdef\therefs{\item{[\number\nref]} #2\ }
\else
\global\xdef\oldrefs{\therefs}
\global\xdef\therefs{\oldrefs\vskip.1in\item{[\number\nref]} #2\ }%
\fi%
\global\advance\nref by 1
}

\def\listrefs{\bigskip\section{References}\therefs}


\newcount\nfoot
\global\nfoot = 1

\def\foot#1#2{\xdef #1{(\number\nfoot)}
\footnote{${}^{\number\nfoot}$}{\eightrm #2}
\global\advance\nfoot by 1
}


\newcount\nfig
\global\nfig = 1
\def\thefigs{} 

\def\figure#1#2{\xdef #1{\number\nfig}
\ifnum\nfig = 1\global\xdef\thefigs{\item{\number\nfig} #2\ }
\else
\global\xdef\oldfigs{\thefigs}
\global\xdef\thefigs{\oldfigs\vskip.1in\item{\number\nfig} #2\ }%
\fi%
\global\advance\nfig by 1 } 

\def\fig#1{\xdef #1{\number\nfig}
\global\advance\nfig by 1 } 


\newcount\ntab
\global\ntab = 1

\def\table#1{\xdef #1{\number\ntab}
\global\advance\ntab by 1 } 


\newcount\cflag
\newcount\nequation
\global\nequation = 1
\def\eqlabel{(1)}

\def\nexteqno{\ifnum\cflag = 0
\global\advance\nequation by 1
\fi
\global\cflag = 0
\xdef\eqlabel{(\number\nequation)}}

\def\lasteqno{\global\advance\nequation by -1
\xdef\eqlabel{(\number\nequation)}}

\def\label#1{\xdef #1{(\number\nequation)}
\ifnum\dflag = 1
{\escapechar = -1
\xdef\draftname{\littlefont\string#1}}
\fi}

\def\clabel#1#2{\xdef\eqlabel{(\number\nequation #2)}
\global\cflag = 1
\xdef #1{\eqlabel}
\ifnum\dflag = 1
{\escapechar = -1
\xdef\draftname{\string#1}}
\fi}

\def\cclabel#1#2{\xdef\eqlabel{#2)}
\global\cflag = 1
\xdef #1{\eqlabel}
\ifnum\dflag = 1
{\escapechar = -1
\xdef\draftname{\string#1}}
\fi}


\def\eeq{}

\def\eqnn #1\eeq{$$ #1 $$}

\def\eq #1\eeq{
\ifnum\dflag = 0
{\xdef\draftname{\ }}
\fi 
$$ #1
\eqno{\eqlabel \rlap{\ \draftname}} $$
\nexteqno}



\def\eol{& \eqlabel \rlap{\ \draftname} \crcr
\nexteqno
\xdef\draftname{\ }}

\def\eeol{& \eqlabel \rlap{\ \draftname}
\nexteqno
\xdef\draftname{\ }}

\def\eolnn{\cr
\global\cflag = 0
\xdef\draftname{\ }}


\def\eqa #1\eeq{
\ifnum\dflag = 0
{\xdef\draftname{\ }}
\fi 
$$ \eqalignno{ #1 } $$
\global\cflag = 0}


\def\ie{{\it i.e.\/}}
\def\eg{{\it e.g.\/}}
\def\etc{{\it etc.\/}}
\def\etal{{\it et.al.\/}}
\def\apriori{{\it a priori\/}}

\def\vs{{\it vs.\/}}


\def\mpla#1#2#3{{\it Mod.\ Phys.\ Lett.} {\bf A#1}, (19#2) #3}

\def\npb#1#2#3{{\it Nucl.\ Phys.} {\bf B#1} (19#2) #3}
\def\plb#1#2#3{{\it Phys.\ Lett.} {\bf #1B} (19#2) #3}

\def\prd#1#2#3{{\it Phys.\ Rev.} {\bf D#1} (19#2) #3}

\def\prl#1#2#3{{\it Phys.\ Rev.\ Lett.} {\bf #1} (19#2) #3}

\def\zpc#1#2#3{{\it Zeit.\ Phys.} {\bf C#1} (19#2) #3}


\global\nulldelimiterspace = 0pt



\def\frac#1#2{{{#1} \over {#2}}\,}  
\def\hf{{1\over 2}}
\def\nth#1{{1\over #1}}


\def\Asl{\hbox{/\kern-.7500em\it A}} 
\def\Dsl{\hbox{/\kern-.6700em\it D}} 
\def\dsl{\hbox{/\kern-.5300em$\partial$}}
\def\pxpsl{\hbox{/\kern-.5600em$p$}}
\def\ssl{\hbox{/\kern-.5300em$s$}}
\def\epssl{\hbox{/\kern-.5100em$\epsilon$}}
\def\delsl{\hbox{/\kern-.6300em$\nabla$}}
\def\lxpsl{\hbox{/\kern-.4300em$l$}}
\def\elxpsl{\hbox{/\kern-.4500em$\ell$}}
\def\kxpsl{\hbox{/\kern-.5100em$k$}}
\def\qxpsl{\hbox{/\kern-.5000em$q$}}
\def\sla#1{\raise.15ex\hbox{$/$}\kern-.57em #1}
\def\Pl{\gamma_{\sss L}}
\def\Pr{\gamma_{\sss R}}



\def\twi{\widetilde}

\def\roughly#1{\mathrel{\raise.3ex\hbox{$#1$\kern-.75em\lower1ex\hbox{$\sim$}}}}
\def\lsim{\roughly<}
\def\gsim{\roughly>}

\def\tw#1{\tilde{#1}}
\def\ol#1{\overline{#1}}





\def\Scf{{\cal F}}

\def\Scl{{\cal L}}
\def\Scm{{\cal M}}

\def\Sct{{\cal T}}
\def\Scu{{\cal U}}


\def\ssb{{\sss B}}

\def\ssf{{\sss F}}

\def\ssl{{\sss L}}

\def\ssr{{\sss R}}
\def\ssS{{\sss S}}

\def\ssy{{\sss Y}}
\def\ssz{{\sss Z}}




\def\Avg#1{\left\langle #1 \right\rangle}



\def\hc{{\rm h.c.}}


\def\GeV{{\rm \ GeV}}


\input epsf

\font\myfont = cmr10 scaled\magstep 2

\overfullrule=0pt


\line{hep-ph/9602438 \hfill McGill-96/04}
\rightline{UdeM-GPP-TH-96-34}
\rightline{WIS-96/09/Feb-PH}

\vskip .01in

\title
\centerline{\Rb\ and New Physics:} 
\centerline{A Comprehensive Analysis}
\endtitle

\vskip .05in

\authors
\centerline{P.~Bamert${}^a$, C.P. Burgess${}^a$, J.M. Cline${}^a$,}
\vskip .05in
\centerline{D. London${}^{a,b}$ and E. Nardi${}^c$}
\vskip .1in
\centerline{\it ${}^a$ Physics Department, McGill University}
\centerline{\it 3600 University St., Montr\'eal, Qu\'ebec, Canada, H3A
2T8}
\vskip .1in
\centerline{\it ${}^b$ Laboratoire de Physique Nucl\'eaire, Universit\'e 
de Montr\'eal} 
\centerline{\it C.P. 6128, succ.~centre-ville, Montr\'eal, Qu\'ebec,
Canada, H3C 3J7} 
\vskip .1in
\centerline{\it ${}^c$ Department of Particle Physics }
\centerline{\it Weizmann Institute of Science, Rehovot 76100, Israel}
\endauthors

\vskip .15in

\abstract
We survey the implications for new physics of the discrepancy between the
LEP measurement of $R_b$ and its Standard Model prediction. Two broad
classes of models are considered: ($i$) those in which new $Z\bbar b$
couplings arise at tree level, through $Z$ or $b$-quark mixing with new
particles, and ($ii$) those in which new scalars and fermions alter the $Z
\bbar b$ vertex at one loop. We keep our analysis as general as possible in
order to systematically determine what kinds of features can produce
corrections to $R_b$ of the right sign and magnitude. We are able to
identify several successful mechanisms, which include most of those which
have been recently been proposed in the literature, as well as some earlier
proposals (\eg\ supersymmetric models). By seeing how such models appear as
special cases of our general treatment we are able to shed light on the
reason for, and the robustness of, their ability to explain $R_b$.
\endabstract


\vfill\eject

\section{Introduction}

\ref\lep{LEP electroweak working group and the LEP collaborations, ``A
Combination of Preliminary LEP Electroweak Results and Constraints on the
Standard Model'', prepared from summer 1995 conference talks.}

\ref\slc{SLC Collaboration, as presented at CERN by C. Baltay (June 1995).}

\ref\cdf{CDF Collaboration, F. Abe \etal , \prl{74}{95}{2626};
D0 Collaboration, S. Abachi \etal , \prl{74}{95}{2632}.}

The Standard Model (SM) of electroweak interactions has been tested and
confirmed with unprecedented precision over the past few years using
measurements of $e^+e^-$ scattering at the $Z$ resonance at LEP~\lep\ and
SLC~\slc. A particularly striking example of the impressive SM synthesis of
the data came with the discovery, at CDF and D0 \cdf, of the top quark with
a mass which is in excellent agreement with the value implied by the
measurements at LEP.

The biggest --- and only statistically important --- fly to be found so far
in the proverbial SM ointment is the experimental surplus of bottom quarks
produced in $Z$ decays, relative to the SM prediction. With the analysis of
the 1994 data as described at last summer's conferences \lep \slc , this
discrepancy has become almost a $4\sigma$ deviation between experiment and
SM theory. The numbers are:
\eq
\label\rbexth
R_b \equiv \Gamma_b/\Gamma_{\rm had} = 0.2219\pm0.0017 ,
\qquad \hbox{while } \qquad R_b(\hbox{SM}) = 0.2156 .
\eeq
The SM prediction assumes a top mass of $m_t=180\GeV$ and the strong
coupling constant $\alpha_s(\MZ)=0.123$, as is obtained by optimizing the
fit to the data.

There are other measurements which differ from their SM predictions at the
$\ge 2\sigma$ level: $R_c$ ($2.5 \sigma$), $\AFB{\tau}$ ($2.0\sigma$), and
the inconsistency ($2.4\sigma$) between $A^0_e$ as measured at LEP with
that obtained from $\ALR^0$ as determined at SLC \slc. In fact, since the
$R_c$ and $R_b$ measurements are correlated, and because they were
announced together, some authors refer to this as the ``$R_b$-$R_c$
crisis." One of the points we wish to make in this paper is that there is
no $R_c$ crisis. If the $R_b$ discrepancy can be resolved by the addition
of new physics, one then obtains an acceptable fit to the data. In other
words, $R_c$, as well as $\AFB{\tau}$ and $\ALR^0$, can reasonably be
viewed simply as statistical fluctuations.

On the other hand, it is difficult to treat the measured value of $R_b$ as
a statistical fluctuation. Indeed, largely because of $R_b$, the data at
face value now {\it exclude} the SM at the 98.8\% confidence level. If we
suppose that this disagreement is not an experimental artifact, then the
burning question is: What Does It Mean?

Our main intention in this paper is to survey a broad class of models to
determine what kinds of new physics can bring theory back into agreement
with experiment. Since $R_b$ is the main culprit we focus on explaining
both its sign and magnitude. This is nontrivial, but not impossible to do,
given that the discrepancy is roughly the same size as, though in the
opposite direction to, the large $\mt$-dependent SM radiative correction.
The result is therefore just within the reach of one-loop perturbation
theory.

Our purpose is to survey the theoretical possibilities within a reasonably
broad framework, and we therefore keep our analysis quite general, rather
than focusing on individual models. This approach has the virtue of
exhibiting features that are generic to sundry explanations of the $Z \to
b\bbar$ width, and many of the proposals of the literature emerge as
special cases of the alternatives which we consider.

In the end we find a number of possible explanations of the effect, each of
which would have its own potential signature in future experiments. These
divide roughly into two categories: those which introduce new physics into
$R_b$ at tree level, and those which do so starting at the one-loop level.

The possibilities are explored in detail in the remainder of the article,
which has the following organization. The next section discusses why $R_b$
is the only statistically significant discrepancy between theory and
experiment, and summarizes the kinds of interactions to which the data
points. This is followed by several sections, each of which examines a
different class of models. Section 3 studies the tree-level possibilities,
consisting of models in which the $Z$ boson or the $b$ quark mixes with a
hitherto undiscovered particle. We find several viable models, some of
which imply comparatively large modifications to the right-handed $b$-quark
neutral-current couplings. Sections 4 and 5 then consider loop
contributions to $R_b$. Section 4 concerns modifications to the $t$-quark
sector of the SM. Although we find that we can reduce the discrepancy in
$R_b$ to $\sim 2\sigma$, we do not regard this as sufficient to claim
success for models of this type. Section 5 then considers the general form
for loop-level modifications of the $Z\bbar b$ vertex which arise from
models with new scalars and fermions. The general results are then applied
to a number of illustrative examples. We are able to see why simple models,
like multi-Higgs doublet and Zee-type models fail to reproduce the data, as
well as to examine the robustness of the difficulties of a supersymmetric
explanation of $R_b$. Finally, our general expressions guide us to some
examples which {\it do} make experimentally successful predictions. 
Section 6 discusses some future experimental tests of various explanations
of the $R_b$ problem.
Our conclusions are summarized in section 7.

\section{The Data Speaks}

Taken at face value, the current LEP/SLC data excludes the SM at the 98.8\%
confidence level. It is natural to ask what new physics would be required
to reconcile theory and experiment in the event that this disagreement
survives further experimental scrutiny. Before digging through one's
theoretical repertoire for candidate models, it behooves the theorist first
to ask which features are preferred in a successful explanation of the
data.

\ref\bigfit{C.P. Burgess, S. Godfrey, H. K\"onig, D. London and I. Maksymyk,
 \prd{49}{94}{6115}.}

\ref\oldglobalfit{P. Bamert, C.P. Burgess and I. Maksymyk, \plb{356}{95}{282}.}

\ref\globalfit{P. Bamert, McGill University preprint McGill/95-64,
hep-ph/9512445, (1995).}

An efficient way to do so is to specialize to the case where all new
particles are heavy enough to influence $Z$-pole observables primarily
through their lowest-dimension interactions in an effective lagrangian.
Then the various effective couplings may be fit to the data, allowing a
quantitative statistical comparison of which ones give the best fit.
Although not all of the scenarios which we shall describe involve only
heavy particles, many of them do and the conclusions we draw using an
effective lagrangian often have a much wider applicability than one might
at first assume. Applications of this type of analysis to earlier data
\bigfit \oldglobalfit\ have been recently updated to include last summer's
data \globalfit, and the purpose of this section is to summarize the
results that were found.

\ref\pestak{M.E. Peskin and T. Takeuchi, \prl{65}{90}{964};
\prd{46}{92}{381}; W.J. Marciano and J.L. Rosner, \prl{65}{90}{2963};
D.C. Kennedy and P. Langacker, \prl{65}{90}{2967};
B. Holdom and J. Terning, \plb{247}{90}{88}.}

There are two main types of effective interactions which play an important
role in the analysis of $Z$-resonance physics, and we pause first to
enumerate briefly what these are. (For more details see Ref.~\bigfit.) The
first kind of interaction consists of the lowest-dimension deviations to
the electroweak boson self-energies, and can be parameterized using the
well-known Peskin-Takeuchi parameters $S$ and $T$ \pestak.\foot\utoo{The
third parameter, $\ss U$, also appears but doesn't play a role in the $\ss
Z$-pole observables.} The second class of interactions consists of
nonstandard dimension-four effective neutral-current fermion couplings,
which may be defined as follows:\foot\nohats{Here we introduce a
slight notation change relative to Ref.~\bigfit\ in that our couplings $\ss
\delta g_{\lft,\rht}^f$ correspond to $\ss \delta \hat{g}_{\lft,\rht}^f$ of
Ref.~\bigfit.}
\eq\label\efflagr
\leff^{\nc} = {e\over \sw \cw} \, Z_\mu \overline{f} \gamma^\mu
 \left[ \left( g_\lft^f + \delta g_\lft^f \right) \gamma_\lft +
\left(g_\rht^f + \delta g_\rht^f \right) \gamma_\rht \right] f .
\eeq
In this expression $g_\lft^f$ and $g_\rht^f$ denote the SM couplings, which
are normalized so that $g_\lft^f = I_3^f - Q^f\sw^2$ and $g_\rht^f =
-Q^f\sw^2$, where $I_3^f$ and $Q^f$ are the third component of weak isospin
and the electric charge of the corresponding fermion, $f$. $\sw =
\sin\theta_w$ denotes the sine of the weak mixing angle, and
$\gamma_{\lft(\rht)} = (1\mp \gamma_5)/2$. 

Fitting these effective couplings to the data leads to the following
conclusions. 

\topic{(1) What Must Be Explained} Although the measured values for several
observables depart from SM predictions at the $2\sigma$ level and more, at
the present level of experimental accuracy it is only the $R_b$ measurement
which really must be theoretically explained. After all, some $2\sigma$
fluctuations are not surprising in any sample of twenty or more independent
measurements. (Indeed, it would be disturbing, statistically speaking, if
all measurements agreed with theory to within $1\sigma$.) This observation
is reflected quantitatively in the fits of Ref.~\globalfit, for which the
minimal modification which is required to acommodate the $R_b$ measurement,
namely the addition of only new effective $Z\bbar b$ couplings, already
raises the confidence level of the fit to acceptable levels ($\chisqminpdof
= 15.5/11$ as compared to 27.2/13 for a SM fit). We therefore regard the
evidence for other discrepancies with the SM, such as the value of $R_c$,
as being inconclusive at present and focus instead on models which predict
large enough values for $R_b$.

\ref\shifman{M. Shifman, \mpla{10}{95}{605}; University of Minnesota
preprint TPI-MINN-95/32-T, hep-ph/9511469 (1995).}

\topic{(2) The Significance of $R_c$} Since the 1995 summer conferences
have highlighted the nonstandard measured values for the $Z$ branching
ratio into {\it both} $c$ and $b$ quarks, it is worth making the above
point more quantitatively for the particular case of the discrepancy in
$R_c$. This was addressed in Ref.~\globalfit\ by introducing effective
couplings of the $Z$ to both $b$ and $c$ quarks, and testing how much
better the resulting predictions fit the observations. Although the
goodness of fit to $Z$-pole observables {\it does} improve somewhat (with
$\chisqminpdof = 9.8/9$), it does so at the expense of driving the
preferred value for the strong coupling constant up to $\as = 0.180 \pm
0.035$, 
in disagreement at the level of $2\sigma$ with low-energy determinations,
which lie in the range $0.112 \pm 0.003$ \shifman.
This change in the fit value
for $\as$ is driven by the experimental constraint that the total $Z$ width
not change with the addition of the new $Z\cbar c$
couplings.\foot\asgoodforb{Introducing effective $\ss b$-quark couplings
have precisely the opposite effect --- since the SM prediction for $\ss
\Gamma_b$ is low and that for $\ss \Gamma_c$ is high relative to
observations --- lowering the strong coupling constant to $\ss
\alpha_s(M_Z)=0.103\pm 0.007$. } Once the low-energy determinations of
$\as$ are also included, $\chisqminpdof$ not only drops back to the levels
taken in the fit only to effective $Z\bbar b$ couplings, but the best-fit
prediction for $R_c$ again moves into a roughly $2\sigma$ discrepancy with
experiment.

\ref\Zprime{P. Chiappetta, J. Layssac, F.M. Renard and C. Verzegnassi,
Universit\'e de Montpellier preprint PM/96-05, hep-ph/9601306, 1996;
G. Altarelli, N. Di Bartolomeo, F. Feruglio, R. Gatto and M.L. Mangano,
CERN preprint CERN-TH/96-20, hep-ph/9601324, 1996.}

\ref\CCK{C.-H.V. Chang, D. Chang and W.-Y. Keung, National Tsing-Hua
University preprint NHCU-HEP-96-1, hep-ph/9601326, 1996.}

It is nevertheless theoretically possible to introduce new physics to
account for $R_c$ in a way which does not drive up the value of the strong
coupling constant. As argued on model-independent grounds in
Ref.~\globalfit, and more recently within the context of specific models
\Zprime \CCK, an alteration of the $c$-quark neutral-current couplings can
be compensated for in the total $Z$ width by also altering the
neutral-current couplings of light quarks, such as the $s$. We put these
types of models aside in the present paper, considering them to be
insufficiently motivated by the experimental data.

\ref\BPS{J Bernab\'eu, A. Pich and A. Santamar{\'\i}a, \plb{200}{88}{569}.
For other discussions of the radiative corrections to $\Zbb$, see
A.A. Akhundov, D. Yu. Bardin and T. Riemann, \npb{276}{86}{1};
W. Beenakker and W. Hollik, \zpc{40}{88}{141}; 
B.W. Lynn and R.G. Stuart, \plb{252}{90}{676};
J Bernab\'eu, A. Pich and A. Santamar{\'\i}a, \npb{363}{91}{326}.}

\topic{(3) LH vs.~RH Couplings} The data do not yet permit a determination
of whether it is preferable to modify the left-handed (LH) or right-handed
(RH) $Z\bbar b$ coupling. The minimum values for $\chi^2$ found in
Ref.~\globalfit\ for a fit involving either LH, RH or both couplings are,
respectively, $\chisqminpdof(\hbox{LH}) = 17.0/12$,
$\chisqminpdof(\hbox{RH}) = 16.1/12$ or $\chisqminpdof(\hbox{both}) =
15.5/11$. 

\figure\bfitfig
{A fit of the $\zbb$ couplings $\delta g^b_{\sss L,R}$
to $Z$-pole data from the 1995 Summer Conferences. The four solid lines 
respectively denote the $1\sigma$, $2\sigma$, $3\sigma$, and $4\sigma$
error ellipsoids. The SM prediction lies at the origin, $(0,0)$. This fit
yields $\alpha_s (M_Z) = 0.101\pm 0.007$ .}

\vskip 1truecm
\epsfbox{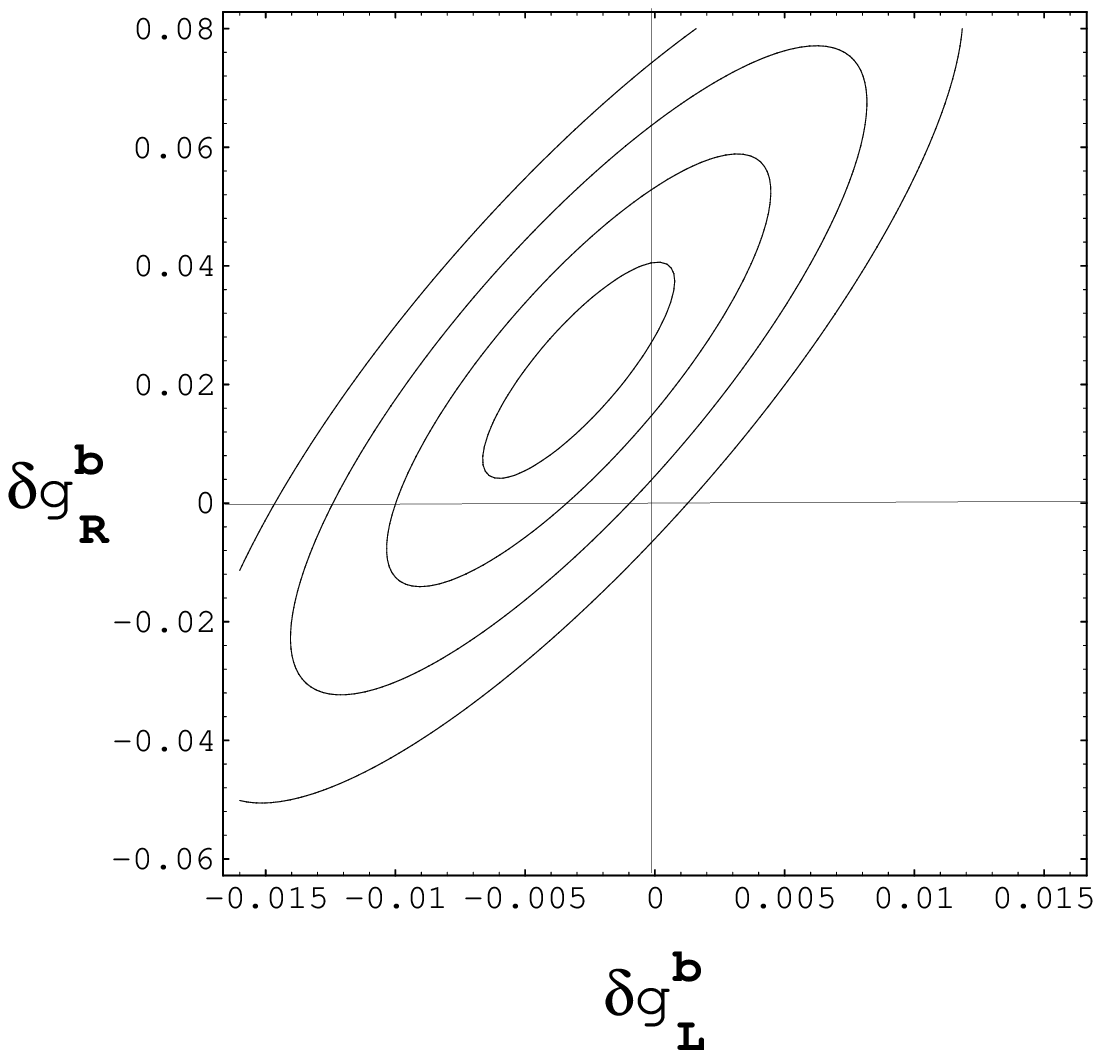}
\baselineskip = 0.8\baselineskip\noindent
{\eightrm Figure \bfitfig: A fit of the $\zbb$ couplings $\delta
g^b_{\sss L,R}$ to $Z$-pole data from the 1995 Summer Conferences. The
four solid lines respectively denote the $1\sigma$, $2\sigma$,
$3\sigma$, and $4\sigma$ error ellipsoids.  The SM prediction lies at
the origin, $(0,0)$. This fit yields $\alpha_s (M_Z) = 0.101\pm 0.007$.}
\vskip 0.5truecm
\baselineskip= 1.25\baselineskip

\table\bfittable

\topic{(4) The Size Required to Explain $R_b$} The analysis of
Ref.~\globalfit\ also indicates the size of the change in the
neutral-current $b$-quark couplings that is required if these are to
properly describe the data. The best fit values which are required are
displayed in Figure \bfitfig, and are listed in Table \bfittable.
Table \bfittable\ also includes for comparison the corresponding tree-level
SM couplings,
as well as the largest SM one-loop vertex corrections (those which depend
quadratically or logarithmically on the $t$-quark mass\foot\smradcor{More
precisely \BPS, we use $\ss \delta\gbl = \left( {\alpha_w \over 16\pi}
\right) [r + 2.88 \ln r]$, where $\ss r = \mt^2 / \MW^2$.} $\mt$),
evaluated at $\sw^2=0.23$. For making comparisons we take $\mt = 180$ GeV.

\midinsert
$$\vbox{\tabskip=0pt \offinterlineskip
\halign to \hsize{\strut \tabskip=1em plus 2em minus .5em \hfil#\hfil &
\hfil#\hfil &\hfil#\hfil &\hfil#\hfil &\hfil#\hfil \cr
\noalign{\hrule}\noalign{\smallskip}\noalign{\hrule}\noalign{\medskip}
Coupling & $g$(SM tree) & $\delta g$(SM top loop) & $\delta g$(Individual Fit) 
& $\delta g$(Fit to Both) \cr
\noalign{\medskip}\noalign{\hrule}\noalign{\smallskip}
$\gbl$ & $-0.4230$ & $0.0065$ & $-0.0067 \pm 0.0021$ & $-0.0029 \pm
0.0037$ \cr
\noalign{\smallskip}
$\gbr$ & $0.0770$ & 0 & $0.034 \pm 0.010$ & $0.022 \pm 0.018$ \cr 
\noalign{\medskip}\noalign{\hrule}\noalign{\smallskip}\noalign{\hrule} }}
$$
\centerline{{\bf Table \bfittable}}
\smallskip\noindent
{\eightrm Required Neutral-Current $\ss b$-Quark Couplings: The last two
columns display the size of the effective correction to the left- and
right-handed SM $\ss Z\bbar b$ couplings which best fit the data. The
``individual fit'' is obtained using only one effective chiral coupling in
addition to the SM parameters $\ss \mt$ and $\ss \as$. The ``fit to both''
includes both couplings. Also shown for comparison are the SM predictions
for these couplings, both the tree-level contribution (``SM tree''), and the
dominant $\ss\mt$-dependent one-loop vertex correction, evaluated at
$\ss\sw^2=0.23$ (``SM top loop'').}

\endinsert

As we now describe, the implications of the numbers appearing in Table
\bfittable\ depend on the handedness (LH \vs\ RH) of effective new-physics
$Z\bbar b$ couplings.

\topic{(4a) LH Couplings} Table \bfittable\ shows that the required change
in the LH $Z\bbar b$ couplings must be negative and comparable in magnitude
to the $\mt$-dependent loop corrections within the SM. The sign must be
negative since the prediction for the $Z \to b\bbar$ width must be
increased relative to the SM result in order to agree with experiment. This
requires $\delta \gbl$ to have the same sign as the tree-level value for
$\gbl$, which is negative. As we shall see, this sign limits the kinds of
models which can produce the desired effect. Comparison with the SM loop
contribution shows that the magnitude required for $\delta \gbl$ is
reasonable for a one-loop calculation. Since the size of the
$\mt$-dependent part of the SM loop is enhanced by a factor of $\mt^2
/\MW^2$, the required new-physics effect must be {\it larger} than a
generic electroweak loop correction.

\topic{(4b) RH Couplings} Since the SM tree-level RH coupling is opposite
in sign to its LH counterpart and is some five times smaller, the
new-physics contribution required by the data, $\delta\gbr$, is positive
and comparable in size to the tree-level coupling. This makes it likely
that any new-physics explanation of the data which relies on changing
$\gbr$ must arise at tree level, rather than through loops.

\topic{(5) Absence of Oblique Corrections} A final proviso is that any
contribution to $\gbl$ or $\gbr$ should not be accompanied by large
contributions to other physical quantities. For example, Ref.~\globalfit\
finds that the best-fit values for the oblique parameters $S$ and $T$ are
\eq\label\oblique
\eqalign{S&= -0.25\pm 0.19\cr
T&= -0.12\pm 0.21\cr}
\eeq
(with a relative correlation of $0.86$) even when $\delta g_{\sss L,R}^b$
are free to float in the fit. Since $T$ often gets contributions similar in
size to $\delta\gbl$, these bounds can be quite restrictive.

Notice that we need not worry about the possibility of having large
cancellations between the new-physics contributions to the oblique
parameters and $\delta\gbl$ in $R_b$. It is true that such a partial
cancellation actually happens for $\Gamma_b$ in the SM, where the loop
contributions proportional to $\mt^2$ in $T$ and $\delta\gbl$ exactly
cancel in the limit that $\sw^2 = \nth{4}$, and so end up being suppressed
by a factor $\sw^2 - \nth{4}$. We nevertheless need not consider such a
cancellation in $R_b$ since the oblique parameters (especially $T$) almost
completely cancel between $\Gamma_b$ and $\Gamma_{\rm had}$.
Quantitatively, we have \bigfit:
\eq\label\nocancel 
\eqalign{
\Gamma_b &= \Gamma_b^\SM \Bigl(1 - 4.57 \, \delta \gbl + 0.828 \, \delta \gbr
- 0.00452 \, S + 0.0110 \, T \Bigr) \cr
\Gamma_{\rm had} &= \Gamma_{\rm had}^\SM \Bigl(1 - 1.01 \, \delta \gbl +
0.183 \, \delta \gbr - 0.00518 \, S + 0.0114 \, T \Bigr) \cr
\hbox{so} \quad R_b &= R_b^\SM \Bigl(1 - 3.56 \, \delta \gbl +
0.645 \, \delta \gbr + 0.00066 \, S - 0.0004 \, T \Bigr) .\cr}
\eeq

\endtopic

We now turn to a discussion of the circumstances under which the above
conditions may be achieved in a broad class of models. 

\section{Tree-Level Effects: Mixing}

\ref\vectri{E. Ma, University of California at Riverside preprint UCRHEP-T153,
hep-ph/9510289, 1995}

\ref\vecdou{T. Yoshikawa, Hiroshima University preprint HUPD-9528,
hep-ph/9512251, 1995.}

\ref\vecsin{G. Bhattacharyya, G.C. Branco and W.-S. Hou, CERN preprint
CERN-TH/95-326, hep-ph/9512239, 1995.}

At tree level the $Z\bbar b$ couplings can be modified if there is mixing
amongst the charge $-\nth{3}$ quarks, or the neutral, colourless vector
bosons. Being a tree-level effect it is relatively easy and straightforward
to analyze and compare different scenarios. Also, since mixing effects can
be large, mixing can provide comparatively large corrections to the
$\zbb$-coupling, such as is needed to modify $R_b$ through changes to
$\gbr$. Not surprisingly, a number of recent models \Zprime \CCK, \vectri
-\vecsin\ use mixing to try to resolve the $R_b$ (and $R_c$)
discrepancy. 
Our aim here is to survey the possibilities in a reasonably general
way. We therefore postpone for the moment a more detailed
phenomenological analysis of the various options.

In general we imagine that all particles having the same spin, colour and
electric charge can be related to one another through mass matrices (some
of whose entries might be constrained to be zero in particular models due
to gauge symmetries or restrictions on the Higgs-field representations). 
We denote the colour-triplet, charge $Q=-{1\over 3}$, quarks in the flavour
basis by $B^\alpha$, and label the corresponding mass
eigenstates\foot\smbasis{We imagine having already diagonalized the SM mass
matrices so that in the absence of this nonstandard mixing one of the $\ss
B^\alpha$ reduces to the usual $\ss b$ quark, with a diagonal mass matrix
with the $\ss d$ and $\ss s$ quarks.} by $b^i$. The mass-eigenstate quarks,
$b^i$, are obtained from the $B^\alpha$ by performing independent unitary
rotations, $\left({\cal U}^\dagger_{\ssl,\ssr}\right)^{\alpha i}$, amongst
the left- and right-handed fields. The $b$ quark that has been observed in
experiments is the lightest of the mass eigenstates, $b = b^{1}$, and all
others are necessarily much heavier than this state.

Similar considerations also apply for colourless, electrically-neutral
spin-one particles. In this case we imagine the weak eigenstates,
$Z^w_\mu$, to be related to the mass eigenstates, $Z^m_\mu$, by an
orthogonal matrix, $\Scm^{wm}$. We take the physical $Z$, whose properties
are measured in such exquisite detail at LEP and SLC, to be the lightest of
the mass eigenstates: $Z_\mu \equiv Z_\mu^{1}$.

\ref\nrtprd{E. Nardi, E. Roulet and D. Tommasini, \prd{46}{92}{3040}.}

Assuming that all of the $b^i$ and $Z^m$ (except for the lightest ones, the
familiar $b$ and $Z$ particles) are too heavy to be directly produced at
$Z$-resonance energies, we find that the flavour-diagonal effective 
neutral-current couplings relevant for $R_b$ are
\eq\label\genrotation
\eqalign{
g^b_{\ssl,\ssr} \equiv (g_{m=1})_{\ssl,\ssr}^{11} &= \sum_{\alpha\beta w}
(g_w)_{\ssl,\ssr}^{\alpha\beta} \, \Scu_{\ssl,\ssr}^{\alpha 1*} \,
\Scu_{\ssl,\ssr}^{\beta 1} \,
\Scm^{w1} \cr
 &= \sum_{\alpha w} (g_w^\alpha)_{\ssl,\ssr} \, \Bigl|
\Scu_{\ssl,\ssr}^{\alpha 1} \Bigr|^2 
\, \Scm^{w1} ~, \cr}
\eeq
where the neutral-current couplings are taken to be diagonal in the flavour
$B^\alpha$ basis.\foot\mzshift{Eq.~\genrotation\ describes the most relevant
effects for the $\ss R_b$ problem, namely the mixing of $\ss Z$ and $\ss b$
with new states. However, in general other indirect effects are also
present, such as, for example, a shift in $\ss \MZ$ due to the mixing with
the $\ss Z'$. For a detailed analysis of the simultaneous effects of mixing
with a $\ss Z'$ and new fermions, see Ref.~\nrtprd.}

This expression becomes reasonably simple in the common situation for which
only two particles are involved in the mixing. In this case we may write
$B^\alpha = \pmatrix{B \cr B' \cr}$, $b^i = \pmatrix{b \cr b' \cr}$ and $Z^w =
\pmatrix{Z \cr Z' \cr}$, and take $\Scu_\ssl$, $\Scu_\ssr $ and $\Scm$ to
be two-by-two rotation matrices parameterized by the mixing angles
$\theta_\ssl$, $\theta_\ssr$ and $\theta_\ssz$. In this case
eq.~\genrotation\ reduces to
\eq\label\genrtwobytwo 
g^b_{\ssl,\ssr} = \Bigl[ (g^\ssb_\ssz)_{\ssl,\ssr} c_{\ssl,\ssr}^2 + 
(g^{\ssb'}_\ssz)_{\ssl,\ssr} s_{\ssl,\ssr}^2 \Bigr] \, c_\ssz +
\Bigl[ (g^\ssb_{\ssz'})_{\ssl,\ssr} c_{\ssl,\ssr}^2 + 
(g^{\ssb'}_{\ssz'})_{\ssl,\ssr} s_{\ssl,\ssr}^2 \Bigr] \, s_\ssz ,
\eeq
where $s_\ssl$ denotes $\sin\theta_\ssl$, \etc\ Increasing $R_b$ requires
increasing the combination $(\gbl)^2 + (\gbr)^2$. To see how this works we
now specialize to more specific alternatives.

\eject

\subsection{$Z$ Mixing}

\ref\Holdom{B. Holdom, \plb{351}{95}{279}.}

First consider the case where two gauge bosons mix. Then eq.~\genrtwobytwo\
reduces to
\eq\label\zmixingonly
g^b_{\ssl,\ssr} = (g^\ssb_\ssz)_{\ssl,\ssr} \, c_\ssz +
(g^\ssb_{\ssz'})_{\ssl,\ssr} \, s_\ssz ,
\eeq
where $(g^\ssb_\ssz)_{\ssl,\ssr}$ is the SM coupling in the absence of $Z$
mixing, and $(g^\ssb_{\ssz'})_{\ssl,\ssr}$ is the $b$-quark coupling to the
new field $Z'_\mu$ (which might itself be generated through $b$-quark
mixing). It is clear that so long as the $Z'\bbar b$ coupling is nonzero,
then it is always possible to choose the angle $\theta_\ssz$ to ensure that
the total effective coupling is greater than the SM one,
$(g^\ssb_\ssz)_{\ssl,\ssr}$.
This is because the magnitude of any function of the form $f(\theta_\ssz)
\equiv A \, c_\ssz + B \, s_\ssz$ is maximized by the angle
$\tan\theta_\ssz = B/A$, for which $\left| f \right|_{\rm max} =\left|
A/c_\ssz\right| \ge |A| $.

The model-building challenge is to ensure that the same type of
modifications do not appear in an unacceptable way in the effective $Z$
couplings to other fermions, or in too large an $\MZ$ shift due to the
mixing. This can be ensured using appropriate choices for the
transformation properties of the fields under the new gauge symmetry, and
sufficiently small $Z$-$Z'$ mixing angles. Models along these lines have
been recently discussed in Refs.~\Zprime,\Holdom.

\ref\vtb{T.J. LeCompte (CDF Collaboration), Fermilab preprint 
FERMILAB-CONF-96/021-E.}

\subsection{$b$-Quark Mixing}

The second natural choice to consider is pure $b$-quark mixing, with no new
neutral gauge bosons.
We consider only the simple case of $2\times 2$  mixings, since  
with only one new $B'$ quark mixing with the SM
bottom quark, eq.~\genrtwobytwo\ simplifies considerably. 
As we will discuss below, we believe  this to be  sufficient  
to elucidate most of the features of the possible $b$-mixing solutions
to the $R_b$ problem.  

Let us first establish some notation. We denote the weak $SU(2)$
representations of the SM $B_{\lft,\rht}$ and  of the $B'_{\lft,\rht}$
as  $R_{\lft,\rht}$ and $R'_{\lft,\rht}$, respectively, where
$R=(I,I_3)$. The SM $B$-quark assignments are $R_\lft= \left( {1\over 2},
-{1\over 2} \right)$ and $R_\rht= (0,0)$. By definition, a
$B'$ quark must have electric charge $Q =-1/3$, but may in principle have
arbitrary weak isospin $R'_{\lft,\rht} = (I'_{\lft,\rht},I'_{3\lft,\rht})$.

In terms of the
eigenvalues $I'_{3\ssl}$ and $I'_{3\ssr}$ of the weak-isospin generator
$I_3$ acting on $B'_\ssl$ and $B'_\ssr$, the combination of couplings 
which controls $\Gamma_b$ becomes
\eq\label\gammab 
\Gamma_b \propto (g_\ssl^b)^2+(g_\ssr^b)^2 = \left( -{c_\ssl^2\over 2} +
{\sw^2\over3} + s_\ssl^2 I'_{3\ssl} \right)^2 + \left( {\sw^2\over 3} + 
s_\ssr^2 I'_{3\ssr} \right)^2 . 
\eeq 
In order to increase $\Gamma_b$ using this expression, $\theta_\lft$ and
$\theta_\rht$ must be such as to make $g_\ssl^b$ more negative, 
$g_\ssr^b$ more positive, or both. Two ways to ensure this are to choose
\eq\label\isospins 
I'_{3\ssl} <-{1\over 2} \qquad \hbox{or} \qquad I'_{3\ssr} > 0 .
\eeq
There are also two other alternatives, involving large mixing angles or
large $B'$ representations: $I'_{3\ssl} > 0$, with
$\sL^2(I'_{3\lft}+{1\over 2}) >1- 2\sw^2/3 \simeq 0.85$, and
$I'_{3\ssr} < 0$, with $s^2_\ssr \, |I'_{3\ssr}| > 2\sw^2/3 \simeq
0.15$.  
Note that, in the presence of LH mixing, the CKM elements $V_{qb}$
($q=u,c,t$) get rescaled as $V_{qb}\to \cL V_{qb}$, thus leading to a
decrease in rates for processes in which the $b$ quark couples to a
$W$. Therefore charged-current data can in principle put constraints
on large LH mixing. For example, future measurements of the various
$t$-quark decays at the Tevatron will allow the extraction of $V_{tb}$
in a model-independent way, thus providing a lower limit on $c_\lft$.
At present, however, when the assumption of three-generation unitarity
is relaxed (as is implicit in our cases) the current measurement of
$BR(t\to Wb)/BR(t\to Wq)$ implies only the very weak limit
$|V_{tb}|>0.022$ (at 95\% c.l.) \vtb. Hence, to date there are still no
strong constraints on large LH mixing solutions. Regarding the RH
mixings, as discussed below there is no corresponding way to derive
constraints on $c_\rht$, and so large $s_\rht$ solutions are always
possible.

We proceed now to classify the models in which the SM bottom quark
mixes with other new $Q=-1/3$  fermions.  Although there are endless
possibilities for the kind of exotic quark one could consider,  the
number of possibilities can be drastically reduced, and a complete
classification  becomes possible, after the following two assumptions
are made:

\topic{(i)}  There are no new Higgs-boson
representations  beyond doublets and singlets. 

\topic{(ii)} The usual $B$-quark mixes with a single $B'$, producing
the mass eigenstates $b$ and $b'$. This constrains the mass matrix to
be $2\times 2$: 
\eq
\label\bmass
\left( \matrix{ {\bar B} & {\bar B'} \cr} \right)_\lft
\left( \matrix{ M_{11} & M_{12} \cr M_{21} & M_{22} \cr}\right)
\left( \matrix{ B \cr B' \cr} \right)_\rht ~. 
\eeq

\endtopic

\ref\LL{See, for example, P. Langacker and D. London, \prd{38}{88}{886}.}

We will examine all of the alternatives consistent with these
assumptions, both of which we believe to be well-motivated, and indeed
not very restrictive. The resulting models include the ``standard''
exotic fermion scenarios \LL\ (vector singlets, vector doublets, mirror
fermions), as well as a number of others.

Let us first discuss assumption {\it (i)}.  From Table \bfittable\ and
eq.~\gammab\ one sees that the mixing angles must be at least as large
as 10\% to explain $R_b$, implying that the off-diagonal entries in
the mass matrix eq. \bmass\ which give rise to the mixing are of order
$s_{\lft,\rht} M_{22}\gsim O(10)$ GeV. If these entries are generated
by Higgs fields in higher than doublet representations, such large
VEVs would badly undermine the agreement between theory and experiment
for the $\MW/\MZ$ mass ratio.\foot\higgs{The contribution of these
relatively large non-standard VEVs cannot be effectively compensated
by new loop-effects.  On the other hand, beyond Higgs doublets, the
next case of a Higgs multiplet preserving the tree-level ratio is that
of $\ss I_{3{\sss H}}=3$, $\ss Y_{\sss H} = 2$.  We do not consider
such possibilities, which would also require the mixed $\ss B'$ to
belong to similarly high-dimensional representations.  We also neglect
alternative scenarios invoking, for example, more Higgs triplets and
cancellations between different VEVs, since these suffer from severe
fine-tuning problems. }

According to assumption {\it (i)}, the permitted Higgs representations
are $R_{\sss H} = \left({1\over 2}, \pm {1\over 2} \right)$ and
$(0,0)$. It is then possible to specify which representations
$R'_{\lft,\rht}$ allow the $B'$ to mix with the $B$ quark of the SM:

\topic {(1)} 
Since the $B'$ should be relatively heavy, we require that $M_{22}\neq 0$. 
Then the restriction {\it (i)} on the possible Higgs representations   
implies that
\eq
\label\firstbisorel
|I'_\lft - I'_\rht| = 0\,,\,{1\over 2}\,;
\eeq
and
\eq
\label\secondbisorel
|I'_{3\lft} - I'_{3\rht}| = 0\,,\,{1\over 2}\,.
\eeq

\topic {(2)} To have $b$--$b'$ mixing, at least one of the off-diagonal
entries, $M_{12}$ or $M_{21}$, must be nonzero. These terms arise
respectively from the gauge-invariant products $R_{\sss H} \otimes
R_{\lft} \otimes \bar R'_{\rht}$ and $R_{\sss H} \otimes R_{\rht} \otimes
\bar R'_{\lft}$ 
so that $\bar R'_{\lft(\rht)}$ must transform as the conjugate
of the tensor product $R_{\sss H}\otimes R_{\rht (\lft)}$: 
\eq
\label\thirdbisorel
R'_{\lft} = R_{\sss H} \otimes R_{\rht}=(0,0) \, ,\, \left({1\over
2},\pm {1\over 2} \right)\,, 
\eeq
or
\eq
\label\lastbisorel
R'_{\rht} = R_{\sss H} \otimes R_{\lft}=(0,0) \,,\, \left( {1\over 2}, 
-{1\over 2}\right) \,,\,(1,-1)\,,\,(1,0).
\eeq
Thus the only possible representations for the $B'$ are those with
$I'_{\rht}=0\,,\,{1 \over 2}\,,\,1$ and $I'_{\lft}=0\,,\,{1\over
2}\,,\,1\,,\,{3\over 2}$, subject to the restrictions 
\firstbisorel\--\lastbisorel.

As for assumption {\it (ii)}, it is of course possible that several
species of $B'$ quarks mix with the $B$, giving rise to an $N\times N$
mass matrix, but it seems reasonable to study the allowed types of
mixing one at a time.  After doing so it is easy to extend the
analysis to the combined effects of simultaneous mixing with multiple
$B'$ quarks.  Thus {\it (ii)} appears to be a rather mild assumption.


There is one sense in which {\it (ii)} might appear to restrict the
class of phenomena we look at in a qualitative way: it is possible to
obtain mixing between the $B$ and a $B'$ in one of the higher
representations we have excluded by ``bootstrapping'', that is, by
intermediate mixing with a $B'_1$ in one of the allowed
representations. The idea is that, if the SM $B$ mixes with such a
$B'_1$, but in turn the latter mixes with a $B'_2$ of larger isospin,
this would effectively induce a $B$--$B'_2$ mixing, which is not
considered here. However, since mass entries directly coupling $B$ to
$B_2'$ are forbidden by assumption {\it (i)}, the resulting $B$--$B_2'$
mixing will in general be proportional to the $B$--$B_1'$ mixing,
implying that these additional effects are subleading, \ie\ of higher
order in the mixing angles. This means that if the dominant $B$--$B'_1$
mixing effects are insufficient to account for the measured value of
$R_b$, adding more $B'$ quarks with larger isospin will not
qualitatively change this situation.

There is, however, a loophole to this argument. If the mass matrix has
some symmetry which gives rise to a special ``texture,''  then it is
possible to have large mixing angles and thus evade the suppression due
to products of small mixing angles alluded to above.  Indeed, we have
constructed several examples of $3\times 3$  quasi-degenerate matrices
with three and four texture zeros, for which the $B$--$B_2'$ mixing is
not suppressed and, due to the degeneracy, can be maximally large.  For
example, let us choose $B'_1$ in  a vector doublet with
$I_{3\lft,\rht}=+1/2$ and  $B'_2$  in  a vector triplet with
$I_{3\lft,\rht}=+1$. Because of our assumption of no Higgs triplets,
direct $B$ mixing with such a $B'_2$ is forbidden, and
$M_{13}=M_{31}=M_{12}=0$.  It is easy to check that for a generic
values of the nonvanishing mass matrix elements, the induced $s^{\sss
13}_{\lft,\rht}$ mixings are indeed subleading with respect to $s^{\sss
12}_{\lft,\rht}$.  However, if we instead suppose that all the nonzero
elements are equal to some large mass $\mu$, then there are two nonzero
eigenvalues $m_{b'}\sim \mu$ and $m_{b''}\sim 2\mu$ while the
$B-B'_2$  mixing angles $s^{\sss 13}_\lft\sim \sqrt{1/3}$ and $s^{\sss
13}_\rht\sim \sqrt{3/8}$  are unsuppressed relative to
 $s^{\sss 12}_{\lft,\rht}$.\foot\mb{A small perturbation of the order
of a few GeV can be added to some of the nonzero mass entries to lift
the degeneracy and give a nonzero value for $m_b$.}  Although it may be
unnatural to have near-equality of the mass entries generated by
singlet and doublet Higgs VEVs, as is needed in this case and in most
of the other examples we found, it is still possible that some
interesting solutions could be constructed along these lines.

Apart from some special cases analogous to the one outlined above, we
can therefore conclude that neither does assumption {\it (ii)}
seriously limit the generality of our results.

\table\enritableone

\def\asas{\phantom{{$^(**)$}}}

\midinsert
$$\vbox{\tabskip=1em plus 4em minus .5em \offinterlineskip
\halign to \hsize{\strut \tabskip=1em plus 2em minus .5em 
\hfil#\hfil & \hfil#\hfil & \hfil#\hfil &\hfil#\hfil &\qquad#\hfil &\hfil#\hfil   \cr
\noalign{\hrule}\noalign{\smallskip}\noalign{\hrule}\noalign{\medskip}
 $I'_\lft$ & $I'_{3\lft}$ & $I'_\rht$ & $I'_{3\rht}$  & Model                               & Mixing     \cr
\noalign{\medskip}\noalign{\hrule}
\noalign{\smallskip}
 $  0   $  &     0      &     0     & $ 0 \> $\ \     & ${\it 1}$\asas 
\qquad Vector Singlet          & $L$          \cr
\noalign{\medskip}
 \omit     & \omit      &    $1/2$  & $-{1 / 2} $     & ${\it 2}^{(**)}$ 
\qquad Mirror Family         &  $L,\,R$      \cr
\noalign{\smallskip}
 \omit     & \omit      &    \omit  & $+{1 / 2} $     & ${\it 3}^{(*)}$                     &  $(L),\,R$    \cr
\noalign{\medskip}
   $1/2$   & $-{1 / 2}$ &   0       & $ 0 \> $  \ \   & ${\it 4}$\asas 
\qquad $4^{th}$ Family         &    --         \cr
\noalign{\medskip}
 \omit     & \omit      &  ${1/2}$  &   $-{1/2}$ \  \ & ${\it 5}^{(**)}$ 
\qquad Vector Doublet (I)      &      $R$       \cr
\noalign{\medskip}
 \omit     & \omit      &    $1$    &   $-1$  \   \   & ${\it 6}^{(**)}$                           & $R$            \cr
\noalign{\smallskip}
 \omit     & \omit      &  \omit    &   $0$  \   \    & ${\it 4}'$                          & --            \cr
\noalign{\medskip}
 \omit     & $+{1 / 2}$ &     0     & $ 0 \> $ \ \    & ${\it 7}$                           & $L$          \cr
\noalign{\medskip}
 \omit     & \omit      &  ${1/2}$  &   $+{1/2}$ \  \ & ${\it 8}^{(*)}$ \  
\qquad Vector Doublet (II) & $(L),\,R$      \cr
\noalign{\medskip}
 \omit     & \omit      &    $1$    &   $0$  \   \    & ${\it 7}'$                           & $L$         \cr
\noalign{\smallskip}
 \omit     & \omit      &  \omit    &   $+1$  \   \   & ${\it 9}^{(*)}$                      & $(L),\,R$     \cr
\noalign{\medskip}
 $  1  $   &    $-1$    &    1      & $ -1   $ \ \    & ${\it 10}^{(*)}$  
\qquad  Vector Triplet (I) & $L,\,(R)$   \cr
\noalign{\medskip}
 \omit     & \omit      &  ${1/2}$  & $-{1 / 2}$    \ & ${\it 11}^{(*)}$                     & $L,\,(R)$      \cr
\noalign{\medskip}
 \omit     &     0      &      1    & $ 0 $  \ \      & ${\it 1}'$\asas   
\qquad    Vector Triplet (II)   &   $L$        \cr
\noalign{\medskip}
 \omit     & \omit      &   ${1/2}$ &   $-{1 / 2}$  \ & ${\it 2}'$                           & $L,\,(R)$      \cr
\noalign{\medskip}
 $3 / 2$   & $ -{3/2} $ & $  1    $ & $ -1 \> $\ \    & ${\it 12}^{(*)}$                     & $L,\,(R)$        \cr
\noalign{\medskip}
 \omit     & $ -{1/2} $ &    1      & $ -1 \> $\ \    & ${\it 6}'$                           &     $(R)$       \cr
\noalign{\smallskip}
 \omit     & \omit      & \omit     & $ 0 \> $\ \     & ${\it 4}''$                          &   --           \cr
\noalign{\medskip}
 \omit     & $ +{1/2} $ &    1      & $ 0 \> $\ \     & ${\it 7}''$                          &  $L$         \cr
\noalign{\smallskip}
\noalign{\medskip}\noalign{\hrule}\noalign{\smallskip}\noalign{\hrule} }}$$
\centerline{{\bf Table \enritableone}}
\medskip\noindent
\centerline{Models and Charge Assignments}
{\eightrm All the possible models for $\ss B$--$\ss B'$ mixing allowed
by the assumptions that {\it (i)} here are no new Higgs
representations beyond singlets and doublets, and {\it (ii)} only
mixing with a single $\ss B'$ is considered.  The presence of LH or RH
mixings which can affect the $\ss b$ neutral current couplings is
indicated under `Mixing'. Subleading mixings, quadratically
suppressed, are given in parenthesis. Equivalent models, for the
purposes of $\ss R_b$, are indicated by a prime $\ss (^\prime)$ in the
`Model' column, while models satisfying eq.~\isospins\ and which can
account for the deviations in $\ss R_b$ with small mixing angles, are
labeled by an asterisk $\ss ^{(*)}$.  Large RH mixing solutions are
labeled by a double asterisk $\ss ^{(**)}$, while models {\it 7, 7}$\,'$
and {\it 7}$\,''$ allow for a solution with large LH mixing.}

\endinsert

\ref\nrtnpb{E. Nardi, E. Roulet and D. Tommasini, \npb{386}{92}{239}.} 
\ref\nrtplb{E. Nardi, E. Roulet and D. Tommasini, \plb{344}{95}{225}.}
\ref\jade{JADE Collaboration, E. Elsen et al., \zpc{46}{90}{349}.}
\ref\cello{CELLO Collaboration, H.J. Behrend et al., \zpc{47}{90}{333}.}

We can now enumerate all the possibilities allowed by assumptions {\it
(i)} and {\it (ii)}.  

With the permitted values of $I'_{\rht}$ and
$I'_{\lft}$ listed above, and the requirement that at least one of the
two conditions \thirdbisorel--\lastbisorel\ is satisfied, there are 19
possibilities, listed in Table \enritableone.  Although not all of them
are anomaly-free, the anomalies can always be canceled by adding other
exotic fermions which have no effect on $R_b$. Since only the values of
$I'_{3\lft}$ and $I'_{3\rht}$ are important for the $b$ neutral current
couplings, for our purpose models with the same $I'_{3\lft,\rht}$
assignments are equivalent, regardless of $I'_{\lft,\rht}$ or
differences in the mass matrix or mixing pattern.  Altogether there
are 12 inequivalent possibilities. Equivalent models are indicated by a
prime $(^\prime)$ in the `Model' column in Table \enritableone.

Due to gauge invariance and to  the restriction {\it (i)}\ on the Higgs
sector, in several cases one of the off-diagonal entries $M_{12}$ or
$M_{21}$ in eq.~\bmass\ vanishes, leading to a hierarchy between the LH
and the RH mixing angles.  If the $b'$ is much heavier than the $b$,
$M_{12}=0$ yields $s_\rht \sim M_{21}/M_{22}$, while the LH mixing is
suppressed by $M_{22}^{-2}$. If on the other hand $M_{21}=0$, then the
suppression for $s_\rht$ is quadratic, leaving $s_\ssl$ as the dominant
mixing angle. For these cases, the subdominant mixings are shown
in parentheses in the `Mixing' column in Table \enritableone. Notice that
while models {\it 2} and {\it 6} allow for a large right-handed mixing 
angle solution of the $R_b$ anomaly, the ``equivalent'' models {\it 2}$\,'$ 
and {\it 6}$\,'$ do not, precisely because of such a suppression.

Six choices satisfy one of the two conditions in eq.~\isospins, and
hence can solve the $R_b$ problem using small mixing angles.  They are
labeled by an asterisk $^{(*)}$ in Table \enritableone. Three of these
models ({\it 10,11,12}) satisfy the first condition for solutions using
small LH mixings.  Since for all these  cases  $I'_{3\rht}<0$, a large
RH mixing could alternatively yield a solution but because $s_\rht$ is
always suppressed with respect to $s_\lft$, this latter possibility is
theoretically disfavored.  The other three choices (models {\it 3,8,9})
satisfy the second condition for solutions using small RH mixing.  It
is noteworthy that in all six models the relevant mixing needed to
explain $R_b$ is automatically the dominant one, while the other, which
would exacerbate the problem, is quadratically suppressed and hence
negligible in the large $m_{b'}$ limit. 
There are two choices (models {\it 5,6}) for which $I'_{3\rht}<0$ and
there is only RH mixing, and one (model {\it 2}) for which
$I'_{3\rht}<0$ and $s_\rht$ is unsuppressed with respect to $s_\lft$.
These three cases allow for solutions with large RH mixings, and are
labeled by a double asterisk $^{(**)}$. Finally, a solution with large
LH mixing is possible (models  {\it 7, 7}$\,'$ and {\it 7}$\,''$) in which
$I'_{3\lft}=+1/2$, and  $I'_{3\rht}=0$ implies no RH mixing effects.


\figure\mirrorfig 
{The experimentally allowed mixing angles for a
mirror family. The thick line covers the entire area of values for
$s_\ssl$ and $s_\ssr$ which are needed to agree with the experimental
value for $R_b$ to the $2\sigma$ level or better. The thin line
represents the one-parameter family of mixing angles which reproduce
the SM prediction. Notice that the small-mixing solution, which passes
through $s_\ssl = s_\ssr = 0$, is ruled out since $I'_\ssl = 0$ 
implies that any LH mixing will {\it reduce} $\gbl$ and thus increases the
discrepancy with experiment.}

In the light of Table \enritableone\ we now discuss in more detail  
the most popular models, as well as some other more exotic possibilities.

\vskip 1truecm
\epsfbox{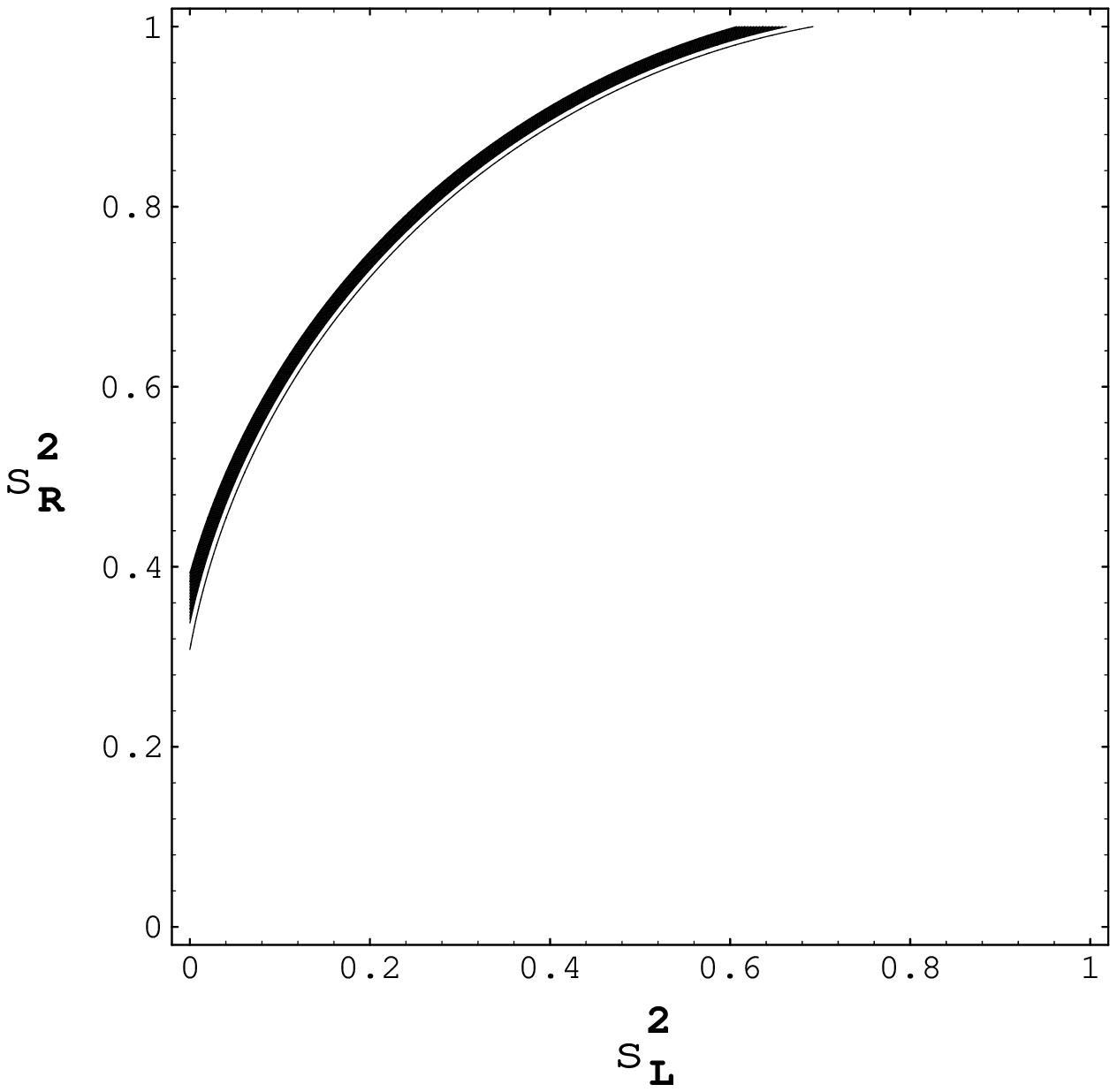}
\baselineskip = 0.8\baselineskip\noindent
Figure \mirrorfig:
{\eightrm The experimentally allowed mixing angles for a
mirror family. The thick line covers the entire area of values for
$s_\ssl$ and $s_\ssr$ which are needed to agree with the experimental
value for $R_b$ to the $2\sigma$ level or better. The thin line
represents the one-parameter family of mixing angles which reproduce
the SM prediction. Notice that the small-mixing solution, which passes
through $s_\ssl = s_\ssr = 0$, is ruled out since $I'_\ssl = 0$ implies
that any LH mixing will {\it reduce} $\gbl$ and thus increases the
discrepancy with experiment.}
\vskip 0.5truecm
\baselineskip= 1.25\baselineskip

\topic{Vector singlet} Vector fermions by definition have identical left-
and right-handed gauge quantum numbers.  A vector singlet (model {\it 1})
is one for which $I'_\ssl = I'_\ssr =0$.  Inspection of eq.~\gammab\ shows
that mixing with such a vector-singlet quark always acts to reduce
$R_b$.\foot\branco{A $\ss Q=+2/3$ vector singlet can however be used to
reduce $\ss R_c$ \CCK \vectri \vecsin, provided that steps are taken, as
suggested in Section 2 above, to avoid the resulting preference for an
unacceptably large value for $\ss\as$).}

\topic{Mirror family} A mirror family (model {\it 2}) is a fourth family
but with the chiralities of the representations interchanged.  Because
$I'_{3\ssl}$ vanishes, LH mixing acts to reduce the magnitude of $\gbl$,
and so tends to make the prediction for $R_b$ worse than in the SM. For
sufficiently large RH mixing angles, however, this tendency may be
reversed. As was discussed immediately below eq.~\isospins, since
$I'_{3\ssr}$ is negative a comparatively large mixing angle of $s_\ssr^2
\gsim 1/3$ is needed to sufficiently increase $R_b$. Such a large RH
mixing angle is phenomenologically permitted by all off-resonance
determinations of $g_\ssr^b$ \nrtnpb. In fact, the $b$-quark production
cross section and asymmetry, as measured in the $\gamma$--$Z$ interference
region \jade \cello, cannot distinguish between the two values
$s^2_\ssr=0$ and $4\sw/3$, which yield exactly the same
rates.\foot\sbrlimit{The current 90\% c.l. upper bound $\ss
s^2_\ssr<0.010$ \nrtplb\ holds in the small mixing angle region $\ss
s^2_\ssr\ll 1/3$.} Hence this kind of model can solve the $R_b$ problem,
though perhaps not in the most aesthetically pleasing way. 
As is shown in Fig.~\mirrorfig, the allowed range of mixing angles is
limited to a narrow strip in the $s_\ssl^2 - s_\ssr^2$ plane.

\ref\negs{P. Bamert and C.P. Burgess, \zpc{66}{95}{495} (hep-ph/9407203).}

\topic{Fourth family} A fourth family (model {\it 4}) cannot resolve
$R_b$ via tree-level effects because the new $B'$ quark has the same
isospin assignments as the SM $b$ quark, and so they do not mix in the
neutral current.\foot\fourth{These models have the further difficulty
that, except in certain corners of parameter space \negs, they produce
too large a contribution to the oblique parameters, $\ss S$ and $\ss T$, 
to be
consistent with the data.} Two other possibilities (models ${\it 4}'$
and ${\it 4}''$) yield the same $I'_{3\lft,\rht}$ assignments as the
fourth family model, and are similarly unsuccessful in explaining $R_b$
since they do not modify the $b$ quark neutral current couplings.

\topic{Vector doublets} There are two  possibilities which permit a
$Q=- \nth{3}$ quark to transform as a weak isodoublet, and 
in  both cases mixing with the SM $b$ is allowed. 
They can be labeled by the different hypercharge value
using the usual convention $Q = I_3 + Y$.

With the straightforward  choice $I'_{3\ssl} = I'_{3\ssr} = -1/2$
(model {\it 5}), we have  $Y'_\ssl = Y'_\ssr= 1/6$.  This type of model
is discussed in Ref.~\vecdou, where the isopartner of the $B'$ is a top-like
quark $T'$ having charge $+ {2\over 3}$. Since these are the same
charge assignments as for the standard LH $b$-quark, this leads to no
mixing in the neutral current amongst the LH fields, and therefore only
the right-handed mixing angle $s_\ssr$ is relevant for $R_b$. Since
$I'_{3\ssr}$ is negative a comparatively large mixing angle of
$s_\ssr^2 \gsim 1/3$ is needed to sufficiently increase $R_b$,   in
much the same way as we found for the mirror-family scenario discussed
above. The required mixing
angle that gives the experimental value, $R_b = 0.2219 \pm 0.0017$, is
\eq\label\evdmixone s_\ssr^2 = 0.367^{+0.013}_{-0.014} . 
\eeq

The other way to fit a $Q=-1/3$ quark into a vector doublet corresponds
to $I'_{3\ssl} = I'_{3\ssr} = +1/2$ (model {\it 8})  and so $Y' = -5/6$
\CCK. The partner of the $B'$ in the doublet is then an exotic quark,
$R$, having $Q= -4/3$.  Here $I'_{3\ssl}$ has the wrong sign for
satisfying eq.~\isospins\ and so mixing {\it decreases} the magnitude
of $\gbl$. On the other hand, $I'_{3\ssr}$ has the right sign to
increase $\gbr$. Whether this type of model can work therefore depends
on which of the two competing effects in $R_b$ wins.  It is easy to see
that in  this model the  $M_{21}$ entry in the $B$--$B'$ mass matrix
eq.~\bmass\  vanishes, which  as discussed above results in a
suppression of $s_\ssl$ quadratic in the large mass, but only a linear
suppression for $s_\ssr$.  Hence, $s_\ssl$ becomes negligible in the
large $m_{b'}$ limit, leaving $s_\ssr$ as the dominant mixing angle in
$R_b$.  The mixing angle which reproduces the experimental value for
$R_b$ then is
\eq\label\evdmix 
s_\ssr^2 = 0.059^{+0.013}_{-0.015} .
\eeq 
However, in order to account for such a large value of the
mixing angle in a natural way, the $b'$ cannot be much heavier than $\sim
100$ GeV. 

Similarly to the $Y' = -5/6$ vector doublet case, models {\it 3} and
{\it 9} also provide a solution through RH mixings.  In model {\it 3},
the subdominant competing effect of  $s_\lft$ is further suppressed by
a smaller $I'_{3\lft}$, while in model {\it 9} the effect of $s_\rht$
is enhanced by  $I'_{3\rht}=+1$, and hence a mixing angle a factor of 4
smaller that in \evdmix\ is sufficient to explain $R_b$.

\topic{Vector triplets} There are three possibilities for placing a
vector $B'$ quark in an isotriplet representation:
$I'_{3\ssl}=I'_{3\ssr}= -1,0,+1$. The last does not allow for $b$
mixing, if only Higgs doublets and singlets are present, and  for our
purposes, $I'_{3\ssl}=I'_{3\ssr}=0$ (model ${\it 1}'$) is equivalent to
the vector singlet case already discussed.  Only the assignment
$I'_{3\ssl}=I'_{3\ssr}=-1$ (model {\it 10}) allows for a 
resolution of the $R_b$
problem, and it was proposed in Ref.~\vectri. If $B'$ is the
lowest-isospin member of the triplet there is an exotic quark of charge
$Q=+5/3$ in the model. Again in the limit of large $b'$ mass one
combination of mixing angles  (in this case $s_\ssr$) is negligible,
due to the vanishing of $M_{12}$ in eq.~\bmass.  As a result,  $s_\ssl$
plays the main role in $R_b$.  Agreement with experiment requires
\eq\label\evtripsoln 
s_\ssl^2 = 0.0127 \pm 0.0034 .
\eeq 
Since the resulting change to $\gbl$ is so small, such a slight mixing
angle would have escaped detection in all other experiments to date. 

Similarly to  this case,  models {\it 11} and {\it  12} also provide a
solution through LH mixings.  In model {\it   11} the unwanted  effects
of  $s_\rht$ are further suppressed, while for model {\it   12} a LH
mixing somewhat smaller than in \evtripsoln\ is sufficient to explain
the data.

\endtopic

Our analysis of tree level effects shows that both $Z$-mixing and
$b$-mixing can resolve the $R_b$ discrepancy.  $b$-quark mixing
solutions satisfying the two assumptions that {\it (i)} there are no
new Higgs representations beyond singlets and doublets, and {\it (ii)}
only mixing with a single $ B'$ is relevant, have been completely
classified. The list of the exotic new $B'$ quarks with the right
electroweak quantum numbers is  given in Table \enritableone.
Solutions with small $s_\rht$ and $s_\lft$ mixing angles are possible
when the $B'_\rht$ is the member with highest $I'_{3\rht}$ in an
isodoublet or isotriplet, or when $B'_\lft$ is the member with lowest
$I'_{3\lft}$ in an isotriplet or isoquartet.  In all these cases, new
quarks with exotic electric charges are also present.  Some other
possible solutions correspond to $I'_{3\rht} < 0 $ and are due to
mixing amongst the RH $b$-quarks involving rather large mixing angles,
while for $I'_{3\lft}=+1/2$ we find another solution requiring even
larger LH mixing. It is intriguing that such large mixing angles are
consistent with all other $b$-quark phenomenology.  We have not
attempted to classify  models in which mixing with new states with very
large values of $I'_{3\lft\rht}$ can arise as a result of
bootstrapping through some intermediate $B'$ mixing.  Under special
circumstances,  they could allow for additional solutions.

For some of the models considered, the contributions to the oblique
parameters could be problematic, yielding additional constraints.
However, for the particular class of vectorlike models (which includes
two of the  small mixing angle solutions) loop effects are sufficiently
small to remain acceptable.\foot\anomalies{Vectorlike models have the
additional advantage of being automatically anomaly free.} This is
because, unlike the top quark which belongs to a chiral multiplet,
vectorlike heavy $b'$ quarks tend to decouple in the limit that their
masses get large.  Introducing mixing with other fermions does produce
nonzero oblique corrections, but these remain small enough to have
evaded detection.  Exceptions to this statement are models involving a
large number of new fields, like entire new generations, since these
tend to accumulate large contributions to $S$ and $T$.


\section{One-Loop Effects: $t$-Quark Mixing}

We now turn to the modifications to the $Z\bbar b$ couplings which can
arise at one loop. Recall that this option can only explain $R_b$ if the LH
$b$-quark coupling, $\gbl$, receives a negative correction comparable in
size to the SM $\mt$-dependent contributions. As was argued in section 2,
it is the LH coupling we are interested in because a loop-level change in
$\gbr$ is too small to fix the discrepancy between the SM and experiment.

The fact that the $R_b$ problem could be explained if the $\mt$-dependent
one-loop contributions of the SM were absent naturally leads to the idea
that perhaps the $t$-quark couples differently to the $b$-quark than is
supposed in the SM. If the $t$ quark mixes significantly with a new $t'$
quark one might be able to significantly reduce the relevant contributions
below their SM values. In this section we show that it is at best possible
to reduce the discrepancy to $\sim 2\sigma$ in models of this type, and so
they cannot claim to completely explain the $R_b$ data.

Our survey of $t$-quark mixing is organized as follows. We first describe
the framework of models within which we systematically search,    
and we identify all of the possible exotic $t$-quark quantum numbers 
which can potentially work. This study is carried out 
much in the spirit of the analysis of $b$ mixing 
presented in  section 3. 
We then describe the possible $t'$ loop contributions to the
neutral-current $b$ couplings. Since this calculation is very similar to
computing the $\mt$-dependent effects within the SM, we briefly review the
latter. Besides providing a useful check on our final expressions, we find
that the SM calculation also has several lessons for the more general
$t$-quark mixing models.

\subsection{Enumerating the Models} 

In this section we identify a broad class of models in which the SM top
quark mixes with other exotic top-like fermions.  As in the previous
section concerning $b$-quark mixing, we denote the electroweak
eigenstates by capitals, $T^\alpha$, and the mass eigenstates by lower-case
letters, $t^i$.  To avoid confusion, quantities which specifically
refer to the $b$ sector will be labeled with the superscript $^{\ss B}$.
By definition, a $T'$ quark must have electric charge $Q =2/3$, but may
in principle have arbitrary weak isospin $R'_{\lft,\rht} =
(I'_{\lft,\rht},I'_{3\lft,\rht})$.  Following closely the discussion in
the previous section, we make three assumptions which allow for a
drastic simplification in the analysis, without much loss of
generality:

\topic{(i)} First, the usual $T$-quark is only allowed to mix with a single
$T'$ quark at a time, producing the mass eigenstates $t$ and $t'$.

\topic{(ii)} Second, for the Higgs-boson representations, we assume
only one doublet and singlets. Additional doublets would complicate the
analysis of the radiative corrections in a model-dependent way due to
the extra diagrams involving charged Higgs bosons.

\topic{(iii)} Finally, certain $T'$-quark representations also contain new
$B'$ quarks. We denote the $B'_\lft $ and $B'_\rht $ as `exotic'
whenever they have non-standard weak isospin assignments, that is,
$I'^\ssb_{3\lft}\neq -{1\over 2}$ or $I'^\ssb_{3\rht}\neq 0$. As we
have already discussed, for exotic $B'$ quarks $b$--$b'$ mixing will
modify the $b$ neutral-current couplings at tree level, overwhelming
the loop-suppressed $t$--$t'$ mixing effects in $R_b$. We therefore
carry out our analysis under the requirement that any $b$--$b'$ mixing
affecting the $b$ neutral-current couplings be absent.

\endtopic

Our purpose is now to examine all of the alternatives which can arise
subject to these three assumptions.
According to $(i)$, the $T$--$T'$ mass matrices we consider are $2\times
2$, and can be written in the general form 
\eq
\label\tmass
\left( \matrix{ {\bar T} & {\bar\Tp} \cr} \right)_\lft
\left( \matrix{ M_{11} & M_{12} \cr M_{21} & M_{22} \cr}\right)
\left( \matrix{ T \cr T' \cr} \right)_\rht ~. 
\eeq
Due to our restriction $(ii)$ on the Higgs sector, certain elements of this
mass matrix are nonzero only for particular values of the $T'$ weak
isospin. Moreover, whenever $T'_\rht$ belongs to a multiplet which also
contains a $Q=-1/3$ $B'_\rht$ quark, the $M_{12}^\ssb$ and $M_{12}$ entries
of the $B$--$B'$ and $T$--$T'$ mass matrices are the same. In those cases
in which the $B'$ quark is exotic, assumption $(iii)$ then forces us to set
$M_{12}=0$. In contrast, the $M_{21}$ entries are unrelated -- for example,
the choice $M_{21}^\ssb =0$ is always possible even if $M_{21}\neq 0$ for
the $T$ and $T'$ quarks.

In order to select those representations, $R'_{\lft,\rht}$, which can mix
with the SM $T$ quark, we require the following conditions to be satisfied:

\topic {(1)} 
In order to ensure a large mass for the $t'$, we require $M_{22}\neq 0$. 
Analogously to \firstbisorel\ and \secondbisorel,  this implies
\eq
\label\firstisorel
|I'_\lft - I'_\rht| = 0\,,\,{1\over 2}\,;
\eeq
and
\eq
\label\secondisorel
|I'_{3\lft} - I'_{3\rht}| = 0\,,\,{1\over 2}\,.
\eeq

\topic {(2)} To ensure a non-vanishing $t$--$t'$ mixing we require at least
one of the two off-diagonal entries, $M_{12}$ or $M_{21}$, to be
non-vanishing. This translates into the following conditions on 
$R'_{\lft}$ and $R'_{\rht}$: 
\eq
\label\thirdisorel
R'_{\lft} = R_{\sss H} \otimes R_{\rht}=(0,0) \, ,\, \left({1\over
2},\pm {1\over 2} \right)\,, 
\eeq
or
\eq
\label\lastisorel
R'_{\rht} = R_{\sss H} \otimes R_{\lft}=(0,0) \,,\, \left( {1\over 2}, 
+{1\over 2}\right) \,,\,(1,0)\,,\,(1,+1).
\eeq

\topic {(3)} Whenever $R'_\rht$ contains a $Q=-1/3$ quark, and either 
$B'_\lft$ or $B'_\rht$ have non-standard isospin assignments, we require
$M_{12}=0$. This ensures that at tree level the neutral current $b$
couplings are identical to those of the SM. Clearly, in the cases in which
the particular $R'_\lft$ representation implies a vanishing $M_{21}$
element, imposing the condition $M_{12}=0$ completely removes all $t$--$t'$
mixing.

We now may enumerate all the possibilities. From
eqs.~\firstisorel--\lastisorel, it is apparent that as in the  
$B'$ case the only allowed 
representations must have $I'_{\rht}=0\,,\,{1 \over 2}\,,\,1$ and
$I'_{\lft}=0\,,\,{1\over 2}\,,\,1\,,\,{3\over 2}$.

Consider first $I'_\lft=1$ or ${3\over 2}$. In this case, from
eq.~\thirdisorel, $M_{21}=0$. Thus, we need $M_{12}\ne 0$ if there is to be
any $t$--$t'$ mixing. The four possibilities for $R'_\rht$ are shown in
eq.~\lastisorel. Of these, $R'_\rht=(0,0)$ is not allowed since
eq.~\firstisorel\ is not satisfied. In addition,
$R'_\rht=({1\over2},{1\over2})$ and $(1,0)$ both contain exotic $B'$ quarks
($I'^{\ssb}_{3\rht}=-{1\over2}$ or $-1$) and so $M_{12}$ is forced to
vanish, leading to no $t$--$t'$ mixing. This leaves $R'_\rht=(1,1)$ as a
possibility, since the $B'_\rht$ is not exotic ($I'^\ssb_{3\rht} =0$). If
we choose $R'_\lft$ such that $I'^\ssb_{3\lft}=-{1\over 2}$, then both 
$B'_{\lft}$ and $B'_{\rht}$ are SM-like, and $b$--$b'$ mixing is not
prohibited since it does not affect the $b$ neutral current couplings. 
Thus, the combination $R'_\lft = ({3\over2},{1\over2})$, $R'_\rht=(1,1)$ is
allowed.

Next consider $I'_\lft = 0$ or ${1\over 2}$. Here, regardless of the value
of $I'_{3\lft}$, $M_{21}$ can be nonzero. Thus any $R'_\rht$ representation
which satisfies eqs.~\firstisorel\ and \secondisorel\ is permitted. It is
straightforward to show that there are 11 possibilities.

\table\davidtableone

The list of the allowed values of $I'_{3\lft}$ and $I'_{3\rht}$ which under
our assumptions lead to $t$--$t'$ mixing is shown in Table \davidtableone.
There are twelve possible combinations, including 
fourth-generation fermions, vector singlets, vector doublets, and mirror
fermions. Not all of these possibilities are anomaly-free, but as already noted  
one could always cancel anomalies by adding other exotic fermions which give 
no additional effects in $R_b$.

\midinsert
$$\vbox{\tabskip=1em plus 4em minus .5em \offinterlineskip
\halign to \hsize{\strut \tabskip=1em plus 2em minus .5em 
\hfil#\hfil & \hfil#\hfil & \hfil#\hfil &\hfil#\hfil &\hfil#\hfil &\hfil#\hfil
 \cr
\noalign{\hrule}\noalign{\smallskip}\noalign{\hrule}\noalign{\medskip}
$I'_\lft$ & $I'_{3\lft}$ & $I'_\rht$ & $I'_{3\rht}$ & Model & Group \cr
\noalign{\medskip}\noalign{\hrule}
\noalign{\smallskip}
 $3 / 2$ & $+{1 / 2}$ & 1 & $ +1 \> $\ \ & \omit & $A_1$ \cr
\noalign{\medskip}
 $1 / 2$ & $+{1 / 2}$ & 1 & $ +1 \> $\ \ & \omit & $A_2$ \cr
\noalign{\smallskip}
 \omit & \omit & \omit & $ 0 $ & \omit & $B_1$ \cr
\noalign{\smallskip}
 \omit & \omit & $1 / 2$ & $+{1 / 2} $ & Vector Doublet (I) & $B_2$ \cr
\noalign{\smallskip}
 \omit & \omit & $ 0 $ & $ 0 $ & $4^{th}$ Family & $C_1$ \cr
\noalign{\medskip}
 $1 / 2$ & $-{1 / 2}$ & 1 & $ 0 $ & \omit & $B_3$ \cr
\noalign{\smallskip}
 \omit & \omit & \omit & $ -1 \> $\ \ & \omit & $B_4$ \cr
\noalign{\smallskip}
 \omit & \omit & $1 / 2$ & $-{1 / 2} $ & Vector Doublet (III) & $B_5$ \cr
\noalign{\smallskip}
 \omit & \omit & $ 0 $ & $ 0 $ & \omit & $C_2$ \cr
\noalign{\medskip}
 0 & 0 & $1 / 2$ & $+{1 / 2} $ & Mirror Fermions & $B_6$ \cr
\noalign{\smallskip}
 \omit & \omit & \omit & $-{1 / 2} $ & \omit & $B_7$ \cr
\noalign{\smallskip}
 \omit & \omit & $ 0 $ & $ 0 $ & Vector Singlet & $C_3$ \cr
\noalign{\medskip}\noalign{\hrule}\noalign{\smallskip}\noalign{\hrule} }}$$

\centerline{{\bf Table \davidtableone}}
\medskip\noindent
\centerline{Models and Charge Assignments}
{\eightrm Values of the weak isospin of $\ss T'_\lft$ and $\ss T'_\rht$
which, under the only restrictions of singlet and doublet Higgs
representations, lead to nonzero $\ss t$--$\ss t'$ neutral current 
mixing. The `Model' column, labels the more familiar
possibilities for the $\ss T'$ quarks: Vector Singlets, Mirror
Fermions, Fourth Family and Vector Doublets. The other models are more
exotic.}

\endinsert

It is useful to group the twelve possibilities into three different
classes, according to the particular constraints on the form of the
$T$--$T'$ mass matrix in eq.~\tmass.

The first two entries in Table \davidtableone, which we have assigned to
group $A$, correspond to the special case in which the $B_{\lft,\rht}$ and
$B'_{\lft,\rht}$ have the same third component of weak isospin, hence
leaving the $b$ neutral current unaffected by mixing. Because both
$B'_\lft$ and $B'_\rht$ appear in the same multiplets with $T'_{\lft}$ and
$T'_{\rht}$, two elements of the $B$-quark and $T$-quark mass matrices are
equal:
\eq\label\firstmassrel
M_{12} = M^\ssb_{12} ~,~~~~~ M_{22} = M^\ssb_{22} ~.
\eeq
As we will see, this condition is important since it implies a relation
between the mixings and the $m_t$, $m_{t'}$ mass eigenvalues.
Although outside the subject of this paper, it is noteworthy that for these
models the simultaneous presence of both  $b$--$b'$ and $t$--$t'$ RH mixing
generates new effects in the charged currents: 
right-handed $W{\bar t}b$ charged currents  get induced, proportional to the
product of the $T$ and $B$ quark mixings $s_\rht\, s^\ssb_\rht$. 
Compared to the modifications in the neutral currents and in the 
LH charged currents, these effects are of higher order in the mixing 
angles  \LL \nrtnpb\ and, most importantly, they can only change  
the RH $b$ coupling. But as noted above,  $g_\rht^b$
is far too small to account for the measured $R_b$ value using loop  effects 
of this kind. Therefore the mixing-induced RH currents allowed in 
models $A_1$ and $A_2$ are ineffective for fixing the $R_b$ discrepancy, 
and will not be considered in the remainder of this paper.  

For the models in group $B$, the condition 
\eq\label\lastmassrel
M_{12} = 0 
\eeq
holds. In the four cases corresponding to $R'_\rht=(1,0)$ (models
$B_1,\,B_3$) and $R'_\rht=(1/2,1/2)$ (models $B_2,\,B_6$), an exotic
$B'_\rht$ quark is present in the same $T'_{\rht}$ multiplet. Hence
$M_{12}$ has to be set to zero in order to forbid the unwanted tree-level
$b$ mixing effects.
In the other three cases belonging to group $B$, $T'_\rht$ corresponds to
the lowest component of non-trivial multiplets: $R'_\rht=(1,-1)$ (model
$B_4$) and $R'_\rht= (1/2,-1/2)$ (models $B_5$, $B_7$). For these values
of $I'_{3\rht}$, $M_{12}=0$ is automatically ensured, due to our
restriction to Higgs singlets or doublets. Furthermore, these
representations do not contain a $B'_\rht$ quark, and no $B'_\lft$ quark
appears in the corresponding $R'_\lft$. There is therefore no $b$--$b'$
mixing.

We should also remark that in model $B_3$ no $B'_\lft$-quark appears in
$R'_\lft$. However, a $B'_\lft$ is needed as the helicity partner of the
$B'_\rht$ present in $R'_\rht=(1,0)$. Because of our restriction on the
allowed Higgs representations, $B'_\lft$ must belong to $R'_\lft=(1,0)$ or
$R'_\lft=(1/2,1/2)\,$, which in turn contain a new $T''_\lft\neq T'_\lft$.
While the first choice corresponds to a type of $T''_\lft$ mixing which we
have already excluded from our analysis, the second choice is allowed and
corresponds to model $B_1$. Following assumption $(i)$, even in this case
we neglect possible $T''_\lft$ mixings of type $B_1\,$, when analysing $B_3$. 

Finally, the remaining three models constitute group $C$, corresponding to
$R'_\rht=(0,0)$. In this group, $T'_\rht$ is an isosinglet, as is the SM
$T_\rht$, implying that only LH $t$--$t'$ mixing is relevant. For $C_2$ and
$C_3$, $R'_\lft$ does not contain a $B'_\lft$, while for $C_1$ the
$B'_\lft$ is not exotic. Hence in all the three cases the $b$
neutral-current couplings are unchanged relative to the SM, and we need not
worry about tree-level $b$-mixing effects.


\subsection{$t$-Quark Loops Within the Standard Model} 

\figure\davidfigone{The Feynman diagrams through which the top
quark contributes to the $Z\bbar b$ vertex within the Standard
Model.}

Before examining the effect of $t$--$t'$ mixing on the radiative
correction to $\Zbb$, we first review the SM computation. We follow the
notation and calculation of Bernab\'eu, Pich and Santamar{\'\i}a \BPS
(BPS). The corrections are due to the 10 diagrams of Fig.~\davidfigone.
All diagrams are calculated in 't Hooft-Feynman gauge, and we neglect
the $b$-quark mass as well as the difference $|V_{tb}|^2 - 1$.

Due to the neglect of the $b$-quark mass, and due to the LH character
of the charged-current couplings, the $t$-quark contribution to the
$Z\bbar b$ vertex correction preserves helicity. Following BPS we write
the helicity-preserving part of the $Z\to b\bbar$ scattering amplitude
as
\eq
\Sct = - \, \left( {e \over \sw\cw} \right) \, {\bar b}(p_1,\lambda_1) \,
\Gamma^\mu b(p_2,\lambda_2) \, \epsilon_\mu(q,\lambda),
\eeq
with
\eq\label\Zbs 
\Gamma^\mu = \Gamma^\mu_0 + \delta\, \Gamma^\mu ~,~~~~~~
\delta\, \Gamma^\mu = 
{\alpha\over 2\pi} \, \gamma^\mu \gamma_\lft I(s,r)~.
\eeq
\noindent
where $\delta\, \Gamma^\mu$ represents the loop-induced correction to the
$Z\bbar b$ vertex. $I(s,r)$ is a dimensionless and Lorentz-invariant form
factor which depends, \apriori, on the three independent ratios: $r\equiv
m_t^2/\MW^2$, $s \equiv \MZ^2/ \MW^2$ and $q^2/\MW^2$. For applications at
the $Z$ resonance only two of these are independent due to the mass-shell
condition $q^2 = \MZ^2$. Moreover, for an on-shell $Z$, non-resonant
box-diagram contributions to $e^+e^- \to b\bbar$ are unimportant, and
$I(s,r)$ can be treated as an effectively gauge-invariant quantity.

\vskip 1truecm
\epsfysize=5in\epsfbox{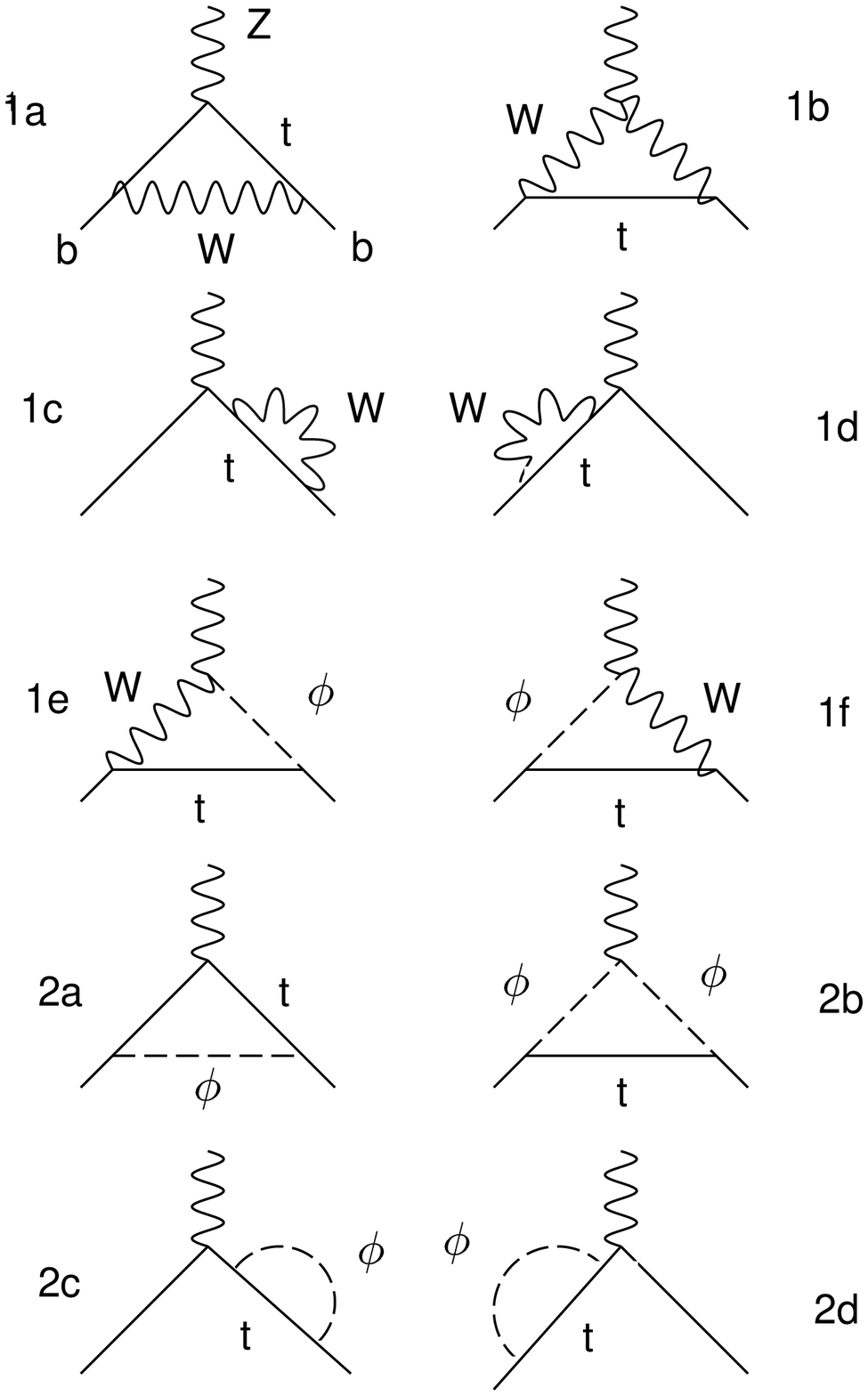}
\baselineskip = 0.8\baselineskip\noindent
{\eightrm Figure \davidfigone: The Feynman diagrams through which the top
quark contributes to the $Z\bbar b$ vertex within the Standard
Model}.
\vskip 0.5truecm
\baselineskip= 1.25\baselineskip

The contributions due to the $t$-quark may be isolated from other
radiative corrections by keeping only the $r$-dependent part of $I(s,r)$. 
BPS therefore define the difference
\eq 
\label\formf
F(s,r) \equiv I(s,r) - I(s,0)~. 
\eeq
Given this function, the $\mt$-dependence of the width $Z\to b\bbar$ is 
obtained using 
\eq \label\gbshift
\Gamma_b^\SM(r) = \Gamma_b^\SM(r=0) \left[ 1 + {\alpha\over \pi} \; 
\left(
{\gbl \over (\gbl)^2 + (\gbr)^2} \right) \; F^\SM(s,r) + \hbox{V.P.}(s,r) 
\right] .
\eeq
In this last equation V.P.($s,r$) denotes the $\mt$-dependent contributions
which enter $\Gamma_b$ through the loop corrections to the gauge-boson
vacuum polarizations. 

The function $F^\SM(s,r)$ 
is straightforward
to compute. Although the resulting expressions are somewhat obscure, 
the special case $s=0$ reveals some interesting features which are also
present in our new-physics calculations, and so we show
the $s=0$ limit explicitly here. For $s = 0$, an evaluation of the graphs
of Fig.~\davidfigone\ gives the following expressions:
\eqa
\label\SMfirst
F^{1(a)} & = - \, {1 \over 2\sw^2}\left\{ {\gLt \over 2} \, 
\left[ {r(r-2)\over(r-1)^2} \, \ln r + {r\over r-1} \right] + \gRt
\left[{r\over (r-1)^2} \, \ln r - 
{r \over r-1} \right]\right\}~~~~~~~~~~~~ \eol
F^{1(b)} & = {3\cw^2\over 4\sw^2} \, \left[ {r^2 \over (r-1)^2}\, \ln r -
{r\over r-1} \right] \eol
F^{1(c)+1(d)} & = {1\over 12}\left[1 - {3\over 2\sw^2}\right] \left\{
{r^2\over (r-1)^2} \, \ln r - {r\over r-1} \right\} \eol
F^{1(e)+1(f)} & = {r\over 2} \, \left[ {r \over (r-1)^2}\, \ln r 
- {1\over r-1} \right] \eol
F^{2(a)} & = - {r\over 4\sw^2} \, \left\{ {\gRt\over 2} 
\left[ \Delta + {r(r-2)\over(r-1)^2} \, \ln r + {2r-1\over r-1} \right] 
+ \gLt\, \left[ {r\over (r-1)^2} \, \ln r - {r\over r-1} \right] \right\} \eol 
F^{2(b)} & = -{1\over 8}\left[1 - {1\over 2\sw^2}\right] r \left[
\Delta + {r^2\over (r-1)^2} \, \ln r - {1\over r-1} \right] \eol
\label\SMlast
F^{2(c)+2(d)} & = {1\over 24}\left[1 - {3\over 2\sw^2}\right] r
\left[ \Delta + {r^2\over (r-1)^2} \, \ln r - {1\over r-1} \right], \eeol
\eeq
with
\eq
\Delta \equiv {2\over n-4} + \gamma + \ln(\MW^2/4\pi\mu^2) - {3\over 2}~,
\eeq
where $n$ is the spacetime dimension arising in dimensional regularization,
and 
\eq\label\SMtcouplings 
\gLt\ = {1\over 2} - {2\over 3}\,\sw^2 ~,~~~~~
\gRt\ = -\, {2\over 3}\, \sw^2\ ~.
\eeq
The picture becomes much simpler after summing the diagrams to 
obtain the total SM contribution:
\eq
\label\SMresult
F^\SM (s=0,r) = \sum_{i=1(a)}^{2(d)} F^i = {1\over 8\sw^2} 
\left[ {r^2 \over r-1} - 6 \, {r \over r-1} + {r(3r+2) \over (r-1)^2} \, 
\ln r \right] .
\eeq
There are two points of interest in this sum. First, it is ultraviolet
finite since all of the divergences $\propto 1/(n-4)$ have cancelled.
This is required on general grounds since there can be no $r$-dependent
divergences in $I^\SM(s,r)$, and so these must cancel in $F^\SM(s,r)$.
A similar cancellation also occurs when new physics is included,
provided that it respects the $SU_\ssl(2) \times U_\ssy(1)$
gauge symmetry and that the complete set of new contributions is 
carefully included. 

The second interesting feature of eq.~\SMresult\ lies in its dependence
on the weak mixing angle, $\sw$. Each of the contributions listed in
eqs.~\SMfirst\ through \SMlast\ has the form $F^i = (x^i + y^i
\sw^2)/\sw^2$; however all of the terms involving $y^i$ have cancelled
in the sum, eq.~\SMresult. This very general result also applies to
all of the new-physics models we consider in subsequent sections. As
will be proved in Section 5, the cancellation is guaranteed by
electromagnetic gauge invariance, because the terms subleading in
$\sw^2$ are proportional to the {\it electromagnetic} $b$-quark vertex
at $q^2 = 0$, which must vanish. This gives a powerful check on all of
our calculations.

Rather than using complete expressions for $F(s,r)$, we find it more
instructive to quote our results in the limit $r \gg 1$, where powers
of $1/r$ and $s/r$ may be neglected. We do the same for the ratio of
masses of other new particles to $\MW^2$ when these arise in later
sections. Besides permitting compact formulae, this approximation also
gives numerically accurate expressions for most of the models'
parameter range, as is already true for the SM, even though $r$ in
this case is only $\sim 4$. In the large-$r$ limit $F^\SM(s,r)$
becomes
\eq
\label\largemt
F^\SM (r) \to {1\over 8\sw^2} \left[ r + \left( 3 - {s \over 6} \, (
1 - 2 \sw^2) \right) \ln r \right] + \cdots, 
\eeq
where the ellipsis denotes terms which are finite as $r \to \infty$. 
Several points are noteworthy in this expression.

\topic{1} The $s$-dependent term appearing in eq.~\largemt\ is
numerically very small, changing the coefficient of $\ln r$ from 3 to
2.88. This type of $s$-dependence is of even less interest when we
consider new physics, since our goal is then to examine whether the new
physics can explain the discrepancy between theory and experiment in
$R_b$. That is, we want to see if the radiative corrections can have
the right sign and magnitude to change $\Gamma_b$ by the correct
amount. For these purposes, so long as the inclusion of $q^2$-dependent
terms only changes the numerical analysis by factors $\lsim 25\%$ (as
opposed to changing its overall sign) they may be neglected.

\topic{2} The above-mentioned cancellation of the terms proportional to
$\sw^2$ when $s=0$ no longer occurs once the $s$-dependence is
included. This is as expected since the electromagnetic Ward identity
only enforces the cancellation at $q^2 = 0$, corresponding to $s=0$ in
the present case. Notice that the leading term, proportional to $r$, is
$s$-independent, and because of the cancellation it is completely
attributable to graph (2a) of Fig.~\davidfigone. All of the other
graphs cancel in the leading term.  Due to its intrinsic relation with
the cancellation of the $\sw^2$-dependent terms, the fact that only one
graph is responsible for the leading contribution to $\delta \gbl$
still holds once new physics is included.  This will prove useful for
identifying which features of a given model control the overall sign of
the new contribution to $\delta \gbl$.

\topic{3} Since the large-$r$ limit corresponds to particle masses (in
this case $m_t$) that are large compared to $\MW$ and $\MZ$, this is
the limit where the effective-lagrangian analysis described in Section
2 directly applies. Then the function $F$ can be interpreted as the
effective $Z\bbar b$ coupling generated when the heavy particle is
integrated out. Quantitatively, $\delta \gbl$ is related to $F$ by
\eq \label\normofF
\delta\gbl = \left( { \alpha \over 2 \pi } \right) \; F.
\eeq

\topic{4} The vacuum polarization contributions to $\Gamma_b$ of
eq.~\gbshift\ have a similar interpretation in the heavy-particle
limit. In this case the removal of the heavy particles can generate
oblique parameters, which also contribute to $\Gamma_b$. In the
heavy-particle limit eq.~\gbshift\ therefore reduces to the first of
eqs.~\nocancel.

\endtopic

\subsection{$\delta\gbl$ in the $t$-Quark Mixing Models}

We may now compute how mixing in the top-quark sector can affect the loop
contributions to the process $Z \to b \bbar$. 
As in the SM analysis, we set $m_b=0$. In addition, following the
discussion in the previous subsection, we neglect the $s$-dependence in all
our expressions. We also ignore all vacuum-polarization effects, knowing
that they essentially cancel in $R_b$. Finally, in the CKM matrix, we set
$|V_{id}|= |V_{is}|=0$ where $i=t,t'$. Thus, the charged-current couplings
of interest to us are described by a $2 \times 2$ mixing matrix, just as in
the neutral-current sector. In the absence of $t$--$t'$ mixing this
condition implies $|V_{tb}|=1$.

For $t$--$t'$ mixing, independent of the weak isospin of the $T'$, we write 
\eq
\label\tmix
\pmatrix{T \cr T' \cr}_{\lft,\rht} ={\cal U}_{\lft,\rht}
\pmatrix{t \cr t' \cr}_{\lft,\rht} ~,~~~~~
{\cal U}_\lft = \pmatrix{\cL & \sL &\cr -\sL & \cL \cr} ~,~~~~~
{\cal U}_\rht = \pmatrix{\cR & -\sR &\cr \sR & \cR \cr} ~.
\eeq
where $\cL \equiv \cos\theta_{\sss L}$, \etc. The matrices ${\cal
U}_{\lft,\rht}$ are analogous to the $b$--$b'$ mixing matrices defined in
eq.~\genrotation\ in our tree-level analysis of $b$ mixing.

In the presence of $t$--$t'$ mixing, the diagonal neutral-current couplings
are modified:
\eq
\label\trotation
g^i_{\ssl,\ssr} = \sum_{a=T,T'} g_{\ssl,\ssr}^{a} \, 
\left( {\cal U}_{\ssl,\ssr}^{a i} \right)^2
\equiv g^{t,\SM}_{\ssl,\ssr} + {\tilde g}^i_{\lft,\rht}~, 
\eeq
where $i=t,t'$, and $g^{t,\SM}_{\ssl,\ssr}$ are the SM couplings defined in
eq.~\SMtcouplings. The new terms $ {\tilde g}^i_{\lft,\rht}$ explicitly
read
\eqa
\label\firsttildegdef
{\tilde g}^t_\lft = & \left( I'_{3\lft} - {1\over 2} \right) \sL^2 ~,
~~~~~~~~~~ {\tilde g}^t_\rht = I'_{3\rht} \sR^2 ~, \eol
\label\secondtildegdef
{\tilde g}^{t'}_\lft = & \left( I'_{3\lft} - {1\over 2} \right) \cL^2 ~,
~~~~~~~~~~ {\tilde g}^{t'}_\rht = I'_{3\rht} \cR^2 ~. \eeol
\eeq
In addition,  whenever the $T'_{\lft,\rht}$ has nonstandard isospin
assignments, $I'_{3\lft} \ne 1/2$ or $I'_{3\rht} \ne 0$,
flavour-changing neutral-current (FCNC) couplings are also  induced :
\eq
\label\fcncrotation
g^{i j}_{\ssl,\ssr} = \sum_{a=T,T'} g_{\ssl,\ssr}^{a} \, 
{\cal U}_{\ssl,\ssr}^{a i}
{\cal U}_{\ssl,\ssr}^{a j}
\equiv {\tilde g}^{ij}_{\lft,\rht}~, 
\eeq
where $i,j=t,t'$, and $i\ne j$. Here,
\eq
\label\lasttildegdef
\gLtp = \left({1\over 2} - I'_{3\lft} \right) \sL\cL ~,
~~~~~~~ \gRtp = I'_{3\rht} \sR\cR ~.
\eeq

Eq.~\tmix\ determines the effective $t$ and $t'$ neutral-current couplings
(eqs.~\trotation--\lasttildegdef). However, the charged-current couplings
depend on the matrix $\Vc = {{\cal U}_\lft}^\dagger\, {\cal U}_\lft^\ssb$. 
Hence we need to  consider also $b$ mixing, since, as discussed in Sec.~4.1,
in those cases in which the $B'$ quark is not exotic
($I'^\ssb_{3\lft}=-1/2$, $I'^\ssb_{3\rht}=0$), we have no reason to require
${\cal U}_\lft^\ssb = I$ (\ie\ no $b$-$b'$ mixing). We then define the
$2\times 2$ charged current mixing matrix 
\eq
\label\Vmatrix
\Vc = {{\cal U}_\lft}^\dagger \, {\cal U}_\lft^\ssb;~~~~~~~~
\Vc_{tb} \equiv \cL \cLb + \sL \sLb ~,~~~~~
\Vc_{t'b} \equiv \sL \cLb - \cL \sLb ~, 
\eeq
which trivially satisfies the orthogonality conditions
$\Vc\,\Vc^\dagger=\Vc^\dagger\,\Vc=I$. In the absence of $b$--$b'$ mixing,
clearly $\Vc_{tb}\, \to \cL$, $ \Vc_{t'b}\,\to \sL$. We also note that, by
assumption, whenever $\Vc\neq {\cal U}_\lft$ we necessarily have $
I'_{3\lft} =+1/2$ (so that $ I'^\ssb_{3\lft} =-1/2$) in order to guarantee
that the $B'_\lft$ is not exotic. From eqs.~\firsttildegdef,
\secondtildegdef\ and \lasttildegdef, this implies that ${\tilde
g}^t_{\lft} = {\tilde g}^{t'}_{\lft} = {\tilde g}^{t t'}_{\lft} = 0$, that
is, the mixing effects on the LH $t$ and $t'$ neutral-current couplings vanish. 

The Feynman rules of relevance for computing the $\Zbb$ vertex loop
corrections in the presence of a mixing in the top-quark sector can now be
easily written down:
\eq
\label\frules
\eqalign{
W{\bar t_i}b : & ~ {ig \over \sqrt{2}} \, \Vc_{t_ib}\,\gamma_\mu
\gamma_\lft\cr 
\phi{\bar t_i}b : & ~ {ig \over \sqrt{2} \MW} \, \Vc_{t_ib}\, m_i\,
\gamma_\lft\cr 
Z{\bar t}_it_i : & ~ {ig\over\cw} \, \gamma_\mu \, \left[
\left( g^{t,\SM}_\lft\, \gamma_\lft + g^{t,\SM}_\rht\, \gamma_\rht \right) + 
\left( {\tilde g}^{t_i}_{\lft}\, \gamma_\lft + {\tilde g}^{t_i}_{\rht}
\, \gamma_\rht \right) \right] ,\cr 
Z{\bar t}t' : & ~ {ig\over\cw} \, \gamma_\mu \, \left[
{\tilde g}^{t t'}_{\lft}\, \gamma_\lft + {\tilde g}^{t t'}_{\rht}
\, \gamma_\rht \right] ,\cr }
\eeq
where $\phi$ are the unphysical charged scalars, and $t_i = t,\,t'$. The
vertices listed in eq.~\frules\ reduce to the SM Feynman rules in the limit
of no mixing.

As pointed out at the end of subsection 4.1, in some groups of models
equalities can be found between some elements of the $T$--$T'$ and
$B$--$B'$ mass matrices. These have important consequences. 
In particular, once expressed in terms of the physical masses and mixing
angles, the equalities of eq.~\firstmassrel\ (which hold in the models of
group $A$) can be written 
\eq
\label\newmassconditions
\big[ {\cal U}_\lft M_{\rm diag}\,{\cal U}_\rht^\dagger\big]_{a2}=
\big[ {\cal U}^\ssb_\lft M^\ssb_{\rm diag}\,{\cal
U}^{\ssb\dagger}_\rht\big]_{a2}= 
\big({\cal U}^\ssb_\lft\big)_{a2} \, m_{b'} \, \cRb~;~~~~~~~~~~~~ (a=1,2)~, 
\eeq
where $M_{\rm diag} = {\rm diag}\> [m_t,\, m_{t'}]$, and we have used 
$M^\ssb_{\rm diag}=\delta_{i2}\, m_{b'}$ (recall that we take $m_b=0$).
Multiplying now on the left by $ \big({\cal
U}^{\ssb\dagger}_\lft\big)_{1a}$ and summing over $a$ we obtain 
\eq
\label\firstrrelation
\big[ \Vc^\dagger\, M_{\rm diag}\,{\cal U}_\rht^\dagger\big]_{12}=
 m_t \, \Vc_{tb}\, \sR + m_{t'} \, \Vc_{t'b}\, \cR = 0~.
\eeq
For the models in group $B$, the vanishing of $M_{12}$ implies no $b$
mixing. Then $\Vc= {\cal U}_\lft$, and eq.~\newmassconditions\ still holds
in the limit $\Vc_{tb}\, \to \cL$, $ \Vc_{t'b}\,\to \sL$. For the models in
group $C$ no particular relation between masses and mixing angles can be
derived. For example, it is clear that in the $4^{th}$ family model $C_1$,
eq.~\firstrrelation\ does not hold. However, for all these models
$I'_{3\rht}=0$. Hence, noting that all the ${\tilde g}_{\rht}$ couplings in
eqs.~\firsttildegdef, \secondtildegdef\ and \lasttildegdef\ are
proportional to $I'_{3\rht}$, and defining $r'=m_{t'}^2/\MW^2$, squaring
eq.~\firstrrelation\ yields a relation which holds for all models in Table
\davidtableone:
\eq
\label\rrprelation
 \Vc_{tb}^2\, {\tilde g}^{t}_{\rht}\,r \, = 
 \, \Vc_{t'b}^2\, {\tilde g}^{t'}_{\rht}\, r' = 
- \,\Vc_{tb}\,\Vc_{t'b}\, {\tilde g}^{tt'}_{\rht}\,\sqrt{rr'} ~.
\eeq
This relation is used extensively in the calculation which follows.

\figure\davidfigtwo{The additional Feynman diagrams which are
required for models in which the $t$ quark mixes with an exotic,
heavy $t'$ quark.}

\vskip 1truecm
\epsfysize=3in\epsfbox{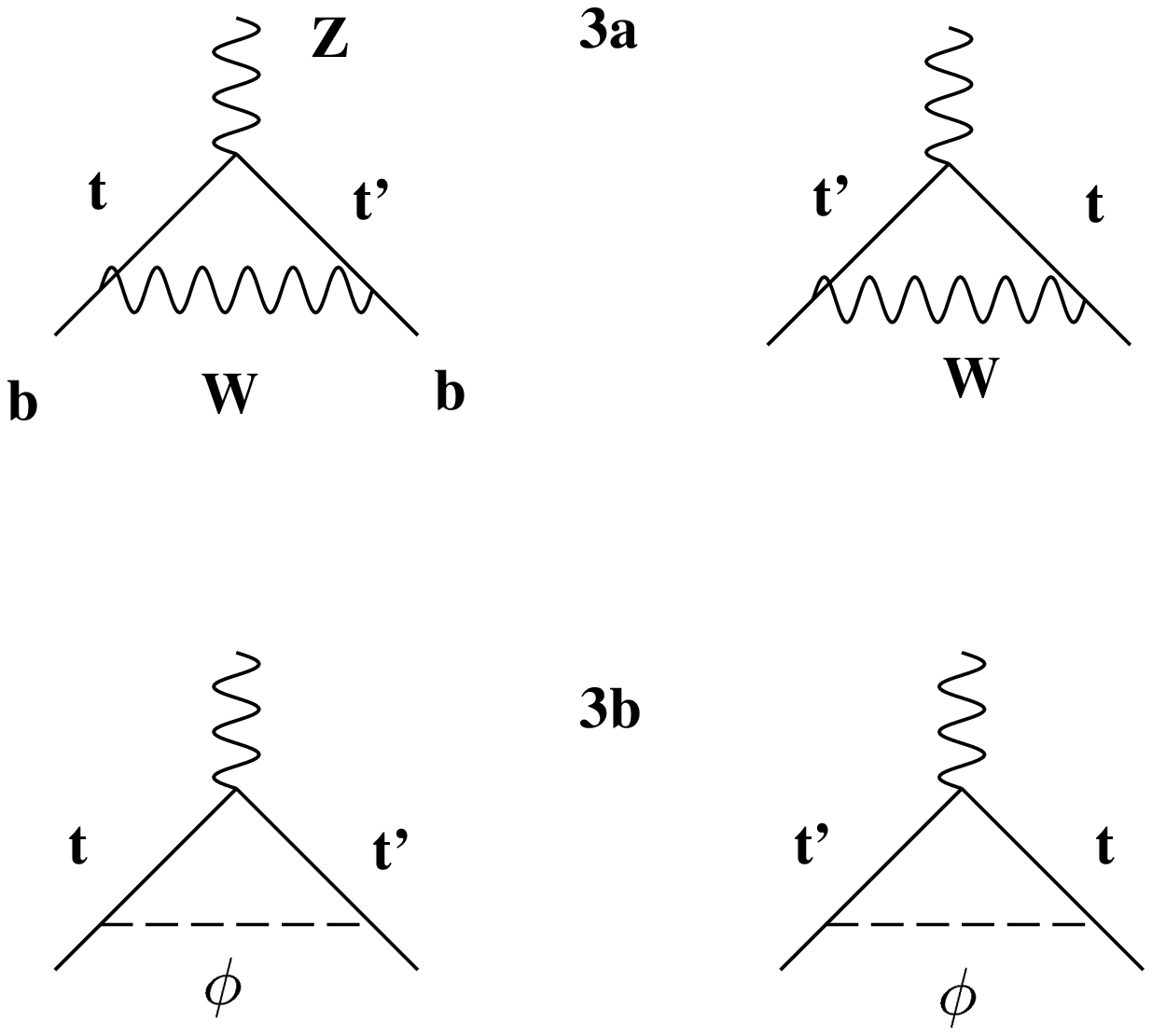}
\baselineskip = 0.8\baselineskip\noindent
{\eightrm Figure \davidfigtwo: The additional Feynman diagrams which are
required for models in which the $t$ quark mixes with an exotic,
heavy $t'$ quark.}
\vskip 0.5truecm
\baselineskip= 1.25\baselineskip

How do we generalize the SM radiative correction to include $t$--$t'$
mixing? First note that for each of the diagrams in Fig.~\davidfigone,
there is also a diagram in which all the $t$-quarks are replaced by
$t'$-quarks. Second, there are 2 new diagrams (Fig.~\davidfigtwo) due to
the FCNC coupling of the $Z$ to the $t$ and $t'$. So to generalize the SM
result to the case of mixing, three things have to be done:

\topic {(i)} multiply eqs.~\SMfirst-\SMlast\ by $\Vc_{tb}^2$ for the $t$
contribution and $\Vc_{t'b}^2$ for $t'$ (with $r\to r'$ ), 

\topic {(ii)} replace $g^t_{\lft,\rht}$ by the modified couplings in
eq.~\trotation, adding eqs.~\firsttildegdef\ and \secondtildegdef\
respectively for $t$ and $t'$,

\topic {(iii)} include diagrams 3(a) and 3(b) (Fig.~\davidfigtwo)
corresponding to the FCNC couplings (eqs.~\fcncrotation-\lasttildegdef).

\endtopic

A glance at the Feynman rules in eq.~\frules\ shows that in the first step
$(i)$, a correction proportional to $g^{t,\SM}_{\lft,\rht}$, and
independent of the ${\tilde g}_{\lft,\rht}$ couplings, is generated. This
correction is common to all models in Table \davidtableone\ -- it appears
even in the case in which the $t$ NC couplings are not affected ($4^{th}$
family). In contrast, steps $(ii)$ and $(iii)$ generate corrections which
differ for different models. It is useful to recast them into two types,
one proportional to the LH neutral current couplings 
($\propto\,\Vc_{ib}\,\Vc_{jb}\, {\tilde g}_{\lft}$), and the other
proportional to the RH neutral current couplings $ (\propto
\Vc_{ib}\,\Vc_{jb}\,{\tilde g}_{\rht}$). The LH and RH corrections vanish
respectively for $I'_{3\lft}=+1/2$ and $I'_{3\rht}=0$, when the
corresponding neutral-current couplings are not affected by the mixing.

In the presence of mixing, the correction due to the diagrams of
Fig.~\davidfigone\ involving internal $t$-quarks becomes
\eq\label\equationi
\sum_{i=1(a)}^{2(d)} F^i = \Vc_{tb}^2 \left[ F^\SM (r) + {\tilde F}
({\tilde g}^t_{\lft,\rht},r) \right] ~,
\eeq
where $F^\SM (r)$ is given by eq.~\SMresult\ and
%
\eq\label\ii
{\tilde F}({\tilde g}^t_{\lft,\rht},r) = {1\over 8\sw^2} 
\left[ {\tilde g}_\lft^t\, r\, \left(2 - {4 \over r-1} \, \ln r
\right) 
- \, {\tilde g}_\rht^t \, r\, \left( \Delta + { 2r - 5 \over r -
1 } + {r^2-2r+4 \over (r-1)^2} \, \ln r\right) \right]. 
\eeq
The third step {\it (iii)} gives rise to a new contribution
\eq 
F^{3(a)}+F^{3(b)} = \Vc_{tb}\,\Vc_{t'b}\>{\tilde F} ({\tilde
g}^{tt'}_{\lft,\rht},r,r')~. 
\eeq
Evaluating diagrams 3(a) and 3(b) (Fig.~\davidfigtwo) we find
\eqa
\label\threea
F^{3(a)} & = -\, {1\over\sw^2} \Vc_{tb}\,\Vc_{t'b}\, \left\{ {1\over 2} \,
\gLtp \, {1 \over r' - r} \left[ {{r'}^2 \over r' - 1}\ln r' - 
{r^2 \over r - 1}\ln r \right] \right. \eolnn
& ~~~~~~~~~~~~~~~~~~~~~~~~~~~ \left. - \gRtp \, \sqrt{r r'}\, {1\over r'-r}
\left[ {r' \over r' - 1}\ln r' - {r \over r - 1}\ln r \right] \right\} \eol
\label\threeb
F^{3(b)} & = {1\over 4 \sw^2} \Vc_{tb}\,\Vc_{t'b}\, \left\{ 2 \, \gLtp
\, {r r' \over r' - r} \left[ {r' \over r' - 1}\ln r' - 
{r \over r - 1}\ln r \right] \right. \eolnn
& ~~~~~~~~~~~~~~~~~~~~~ \left. - \gRtp \sqrt{r r'}
\left( \Delta + 1 + {1\over r'-r} \left[
{{r'}^2 \over r' - 1}\ln r' - 
{r^2 \over r - 1}\ln r \right] \right) \right\}~.\eeol
\eeq

Putting all the contributions together, for the general case we find
\eq\label\totalresult
F =\sum_{i=1(a)}^{3(b)} F^i =
\sum_{j=1,2}\,\Vc_{t_jb}^2\, \left[ F^\SM (r_j) +
 {\tilde F} ({\tilde g}^{t_j}_{\lft,\rht},r_j) \right] +
\Vc_{tb}\,\Vc_{t'b}\,{\tilde F} ({\tilde g}^{tt'}_{\lft,\rht},r,r')~. 
\eeq
where $t_j=t,\,t'$ and $r_j=r,\,r'$. We note that due to eq.~\rrprelation\
all the divergent terms proportional to $ {\tilde g}_\rht\,\Delta $ cancel
in the sum. Now, the {\it correction} $\delta \gbl = {\alpha \over 2\pi} \,
X_{corr}$ to the SM result can be explicitly extracted from
eq.~\totalresult\ by means of the relation $\Vc_{tb}^2 = 1 - \Vc_{t'b}^2$. 
Moreover, as anticipated it is possible to divide the various contributions
to $X_{corr}$ into three different pieces: a universal correction, a
correction due to LH mixing only, and a correction due to the RH mixing.
Hence we write 
\eq
\label\Xcorr
X_{corr} \equiv F - F^\SM = X_{corr}^{Univ} + X_{corr}^{\sss LH} +
X_{corr}^{\sss RH}~, 
\eeq
where 
\eqa
\label\univcorr
X_{corr}^{Univ} & = \, \Vc_{t'b}^2\, \left[ F^\SM (r') - F^\SM (r) \right]~,
 \eol
\label\lhcorr
X_{corr}^{\sss LH} & = \, \Vc_{tb}^2 \, {\tilde F} ({\tilde g}^t_{\lft},r)
+ \Vc_{t'b}^2\, {\tilde F} ({\tilde g}^{t'}_{\lft},r') 
+ \Vc_{tb}\,\Vc_{t'b}\,{\tilde F} ({\tilde g}^{tt'}_{\lft},r,r')~, \eol
\label\rhcorr
X_{corr}^{\sss RH} & = \,\Vc_{tb}^2 \, {\tilde F} ({\tilde g}^t_{\rht},r)
+ \Vc_{t'b}^2\, {\tilde F} ({\tilde g}^{t'}_{\rht},r') 
+ \Vc_{tb}\,\Vc_{t'b}\,{\tilde F} ({\tilde g}^{tt'}_{\rht},r,r')~. \eol 
\eeq

Using the explicit expressions for ${\tilde g}^t_{\lft,\rht}$, 
${\tilde g}^{t'}_{\lft,\rht}$ and ${\tilde g}^{tt'}_{\lft,\rht}$ as given in 
eqs.~\firsttildegdef, \secondtildegdef\ and \lasttildegdef\ above, 
together with relation \rrprelation\ for the RH piece, these read
\eqa
\label\univcorrf
X_{corr}^{Univ} = & \, \Vc_{t'b}^2 \, f_1^{corr}(r,r') \eol
\label\lhcorrf
X_{corr}^{\sss LH} = & \, \left( 1 - 2 I'_{3\lft} \right) \, \Vc_{tb}\,\Vc_{t'b} \,\sL \cL \,
f_2^{corr}(r,r') \eol
\label\rhcorrf
X_{corr}^{\sss RH} = & \, \left( 2 I'_{3\rht} \right) \,\Vc_{tb}^2 \, \sR^2 \, 
f_3^{corr}(r,r') \eeol 
\eeq
with
\eqa
\label\univcorrfn
f_1^{corr}(r,r') = & \, {1\over 8\sw^2} \left\{ 
{r'\,(r'-6) \over r'-1} + {r'(3r'+2) \over (r'-1)^2} \ln r' -
{r\,(r-6) \over r-1} - {r(3r+2) \over (r-1)^2} \ln r \right\}~, \eolnn
&\eol
\label\lhcorrfn
f_2^{corr}(r,r') = & \, {1\over 8\sw^2} \left\{
{\cL \Vc_{t'b}\over \sL\Vc_{tb}} \left( - r' + {2 \,r' \over r'-1} \, \ln r' 
\right) +
{\sL \Vc_{tb}\over \cL\Vc_{t'b}} \left( - r + {2 \,r \over r-1} \, \ln r \right) \right. \eolnn
& ~~~~~~~~~~~~~~~~~~~ \left. 
+ \, {2 \, {r'}^2\, (r-1)\over (r'-1)\,(r'-r) } \, \ln r' 
- \, {2 \, {r}^2\, (r'-1)\over (r-1)\,(r'-r) } \, \ln r \> \right\}~, \eolnn
&\eol
\label\rhcorrfn
f_3^{corr}(r,r') = & \, {1\over 8\sw^2} \, 
r \left\{ - {1\over 2} \left[ {2 r -5\over r-1} + {r^2 - 2r + 4
\over (r-1)^2} \, \ln r \right] - {1\over 2} \left[ {2 r' -5\over r'-1} +
{{r'}^2 - 2r' + 4 \over (r'-1)^2} \, \ln r' \right] \right. \eolnn
- 4  & \left. \, {1 \over r' - r} \, 
\left[ {r' \over r' - 1}\ln r' - {r \over r - 1}\ln r \right] 
+ \left[ 1 + {1 \over r' - r} \, 
\left( {{r'}^2 \over r' - 1} \ln r' 
- {r^2 \over r - 1}\ln r \right) \right] \right\}. \eolnn
& \eeol
\eeq
Note that a value of $V_{tb}$ different from unity can be easily accounted 
for by using the unitary condition $ \left| \Vc_{tb} \right|^2 +
\left|\Vc_{t'b} \right|^2 = \left| V_{tb} \right|^2 \equiv 1-\left|
V_{ts}\right|^2 + \left| V_{td} \right|^2$ in eqs.~\univcorrf--\rhcorrfn.
 
As we have already pointed out, because of our requirement of no $B$--$B'$
mixing when the $B'$ is exotic, only when $I'_{3\lft}=+1/2$ can we have 
$\cL\neq \Vc_{tb}$, $\sL\neq \Vc_{t'b}$. However, in this case 
$X_{corr}^{\sss LH}$ vanishes. Hence, without loss of generality, we can
set the LH neutral current mixing equal to the charged current mixing in
$X_{corr}^{\sss LH}$, obtaining 
\eqa
\label\newlhcorrf
& X_{corr}^{\sss LH}  = \, \left( 1 - 2 I'_{3\lft} \right) \, 
\Vc_{tb}^2\,\Vc_{t'b}^2 \,
f_2^{corr}(r,r'), \eol
\label\newlhcorrfn
& f_2^{corr}(r,r') =  \, {1\over 8\sw^2}  \left\{- (r+r') +
 {2\,r\,r'\over r'-r}\, \ln{r'\over r}\> \right\}~.  \eeol 
\eeq
{}From eqs.~\univcorrf, \rhcorrf\ and \newlhcorrf\ we see that there are only
two independent mixing parameters relevant for the complete analysis of our
problem: the LH matrix element $\Vc_{tb}$ and the RH mixing $\sR$. 
Furthermore, note that as $r' \to r$, all the corrections in
eqs.~\univcorrfn, \rhcorrfn\ and \newlhcorrfn\ vanish, independent of the
mixing angles. This comes about because of a GIM-like mechanism for all the
pieces which do not depend on $I'_{3\rht}$. The $I'_{3\rht}$-dependent
contribution from the RH fermions coupling to the $Z$ vanishes in the limit
$r' \to r$ as a consequence of eq.~\firstrrelation. 

In the limit $r,r' \gg 1$, for the functions $f_i^{corr}(r,r')$ we obtain 
\eqa
f_1^{corr}(r,r') & \to  {1\over 8\sw^2} \left\{r' - r + 3 \ln \left({r'
 \over r}\right)\right\} ,
\eol
f_2^{corr}(r,r') & \to {1\over 8\sw^2} \left\{-\left( r + r' \right) + 
 {2\, r r' \over r' - r}
 \lnrpr\right\} \, ,
 \eol
f_3^{corr}(r,r') & \to {1\over 8\sw^2} \left\{
- r + {1\over 2} \left( 1 + {r \over r'} \right) 
{ r r' \over r' - r} \lnrpr -  {3 \, r
\over r' - r} \, \lnrpr + {3\over 2} \, \left( 1 + {r \over r'} \right)\right\}.
 \eeol 
\eeq

\table\davidtabletwo
\noindent

Let us now consider the numerical values of these corrections in more
detail. Using $m_t = 180$ GeV, $\MW=80$ GeV, and $\sw^2 = 0.23$,
eq.~\SMresult\ gives a SM radiative correction of
\eq
F^\SM = 4.01~.
\eeq
The question is whether it is possible to cancel this correction, thus 
eliminating the $R_b$ problem, by choosing particular values of $m_{t'}$
and the mixing angles. For various values of $m_{t'}$, the value of
$X_{corr}$ (eq.~\Xcorr) is shown in Table \davidtabletwo.

\midinsert\vskip -0.5cm
$$\vbox{\tabskip=1em plus 2em minus .5em\offinterlineskip
\halign to \hsize{\strut \tabskip=1em plus 2em minus .5em \hfil#\hfil &
\hfil#\hfil \cr
\noalign{\hrule}\noalign{\smallskip}\noalign{\hrule}\noalign{\medskip}
$m_{t'}$ & $X_{corr}$ \cr
\noalign{\medskip}\noalign{\hrule}
\noalign{\smallskip}
75 GeV & $ -3.31 \, \Vc_{t'b}^2 \> - \> 1.21 \left(1 - 2 I'_{3\lft} \right)
\Vc_{t'b}^2 \Vc_{tb}^2 \> + \> 1.39 \left( 2 I'_{3\rht} \right) \Vc_{tb}^2
\sR^2 $ \cr 
\noalign{\smallskip}
100 GeV & $ -2.70 \, \Vc_{t'b}^2 \> - \> 0.71 \left(1 - 2 I'_{3\lft}
\right) \Vc_{t'b}^2 \Vc_{tb}^2 \> + \> 0.59 \left( 2 I'_{3\rht} \right)
\Vc_{tb}^2 \sR^2 $ \cr 
\noalign{\smallskip}
125 GeV & $ -1.97 \, \Vc_{t'b}^2 \> - \> 0.34 \left(1 - 2 I'_{3\lft}
\right) \Vc_{t'b}^2 \Vc_{tb}^2 \> + \> 0.22 \left( 2 I'_{3\rht} \right)
\Vc_{tb}^2 \sR^2 $ \cr
\noalign{\smallskip}
150 GeV & $ -1.14 \, \Vc_{t'b}^2 \> - \> 0.10 \left(1 - 2 I'_{3\lft}
\right) \Vc_{t'b}^2 \Vc_{tb}^2 \> + \> 0.05 \left( 2 I'_{3\rht} \right)
\Vc_{tb}^2 \sR^2 $ \cr
\noalign{\smallskip}
175 GeV & $ -0.20 \, \Vc_{t'b}^2 \> - \> 0.003 \left(1 - 2 I'_{3\lft}
\right) \Vc_{t'b}^2 \Vc_{tb}^2 \> + \> 0.001 \left( 2 I'_{3\rht} \right)
\Vc_{tb}^2 \sR^2 $ \cr
\noalign{\smallskip}
200 GeV & \phantom{\hbox{$-$}}$ 0.84 \, \Vc_{t'b}^2 \> - \> 0.04 \left(1 -
2 I'_{3\lft} \right) \Vc_{t'b}^2 \Vc_{tb}^2 \> + \> 0.02 \left( 2
I'_{3\rht} \right) \Vc_{tb}^2 \sR^2 $ \cr
\noalign{\smallskip}
225 GeV & \phantom{\hbox{$-$}}$ 1.97 \, \Vc_{t'b}^2 \> - \> 0.23 \left(1 -
2 I'_{3\lft} \right) \Vc_{t'b}^2 \Vc_{tb}^2 \> + \> 0.07 \left( 2
I'_{3\rht} \right) \Vc_{tb}^2 \sR^2 $ \cr
\noalign{\smallskip}
250 GeV & \phantom{\hbox{$-$}}$ 3.20 \, \Vc_{t'b}^2 \> - \> 0.55 \left(1 -
2 I'_{3\lft} \right) \Vc_{t'b}^2 \Vc_{tb}^2 \> + \> 0.15 \left( 2
I'_{3\rht} \right) \Vc_{tb}^2 \sR^2 $ \cr
\noalign{\smallskip}
275 GeV & $ \phantom{\hbox{$-$}}4.52 \, \Vc_{t'b}^2 \> - \> 1.01 \left(1 -
2 I'_{3\lft} \right) \Vc_{t'b}^2 \Vc_{tb}^2 \> + \> 0.24 \left( 2
I'_{3\rht} \right) \Vc_{tb}^2 \sR^2 $ \cr
\noalign{\smallskip}
300 GeV & $ \phantom{\hbox{$-$}}5.93 \, \Vc_{t'b}^2 \> - \> 1.61 \left(1 -
2 I'_{3\lft} \right) \Vc_{t'b}^2 \Vc_{tb}^2 \> + \> 0.34 \left( 2
I'_{3\rht} \right) \Vc_{tb}^2 \sR^2 $ \cr
\noalign{\medskip}\noalign{\hrule}\noalign{\smallskip}\noalign{\hrule} }}
$$
\centerline{{\bf Table \davidtabletwo}}
\smallskip\noindent
{\eightrm Dependence of the $\ss t$--$\ss t'$ Mixing Results on $\ss m_{t'}$:
This table indicates the dependence on the mass of the $\ss t'$ quark of
the corrections to $\ss \gbl$ due to $\ss t$--$\ss t'$ mixing, with the
$\ss t$ mass fixed at 180 GeV.} 

\endinsert

We see that even for $m_{t'} > m_t$, it {\it is} possible to choose
$I'_{3\lft}$, $I'_{3\rht}$, and the LH and RH mixing angles such that the
correction is negative. So the discrepancy in $R_b$ between theory and
experiment can indeed be reduced via $t$--$t'$ mixing. 

\ref\CDFone{F. Abe et al.\ (CDF Collaboration), \prd{45}{92}{3921}.}
Referring to the models listed in Table \davidtabletwo, the optimal
choice for the weak isospin of the $T'$ is $I'_{3\lft} = -1/2$ and
$I'_{3\rht} = -1$, regardless of the value of $m_{t'}$. Furthermore,
maximal RH mixing, $\sR^2 \sim 1$, is also preferred. However, even
with these choices, it is evidently impossible to completely remove the
$R_b$ problem. From the above table, the best we can do is to take
$m_{t'} = 75$ GeV and $\Vc_{t'b}^2 = \sL^2 = 0.6$, in which case the
total correction is $X_{corr} = -3.68$. This leaves a $1.5\sigma$
discrepancy in $R_b$, which would put it in the category of the other
marginal disagreements between experiment and the SM. However, such a
light $t'$ quark has other phenomenological problems. In particular,
CDF has put a lower limit of 91 GeV on charge $2/3$ quarks which decay
primarily to $Wb$ \CDFone. Unless one adds other new physics to evade
this constraint, the lightest $t'$ allowed is about $m_{t'} \sim 100$
GeV. In this case, maximal LH mixing ($\Vc_{t'b}^2 = \sL^2 \sim 1$)
gives the largest effect:  $X_{corr} = - 2.7$. The predicted value of
$R_b$ is then still some $2\sigma$ below the measured number.

Another possibility is that the charge $2/3$ quark observed by CDF is in
fact the $t'$, while the real $t$-quark is much lighter, say $m_t \sim 100$
GeV. Assuming small $t$--$t'$ mixing,
and that the $t'$ is the lightest member of the new multiplet, 
the $t'$ will then decay to $Wb$, as
observed by CDF, but the SM radiative correction will be reduced. 
This situation is essentially identical to that discussed above, in which
the LH $t$--$t'$ mixing is maximal, and $m_{t'} \sim 100$ GeV: the SM value
of $R_b$ will still differ from the experimental measurement by about
$2\sigma$. The only way for such a scenario to work is if $m_t < \MW$.
However, new physics is then once again required to evade the constraint
from Ref.~\CDFone.

For all the possibilities of this section our conclusion is therefore
the same:
it is not possible to completely explain $R_b$ through $t$--$t'$ mixing.
The best we can do is reduce the discrepancy between theory and experiment
to about $2\sigma$, which might turn out to be sufficient, depending on
future measurements.

\section{One-Loop Effects: Other Models}

Another way to change $\gbl$ at the one-loop level is to introduce exotic
new particles that couple to both the $Z$ and the $b$ quark. One-loop
graphs involving such particles can then modify the $Z\bbar b$ vertex as
measured at LEP and SLC. Recall once more the conclusion from Section 2:
agreement with experiment requires the LH $b$-quark coupling, $\gbl$, to
get a negative correction comparable in size to the SM $\mt$-dependent
contributions since loop-level changes to $\gbr$ are too small to be
detectable.

In this section we first exhibit the general one-loop correction due to
exotic new scalar and spin-half particles, with the goal of identifying the
features responsible for the overall sign and magnitude of the result. We
then use this general result to investigate a number of more specific
cases.

The answer is qualitatively different depending on whether or not the new
scalars and fermions can mix, and thus have off-diagonal couplings to the
$Z$ boson. We therefore treat these two alternatives separately. The
simplest case is when all $Z$ couplings are diagonal, so that the one-loop
results depend only upon two masses, those of the fermion and the scalar in
the loop. Then the correction to the $Zb\bar b$ vertex is given by a very
simple analytic formula, which enables us to easily explain why a number of
models in this category give the `wrong' sign, reducing $\Gamma_b$ rather
than increasing it.

More generally however, the new particles in the loops have couplings to
the $Z$ which are diagonal only in the flavour basis but not the mass
eigenstate basis, so the expressions become significantly more
complicated. This occurs in supersymmetric extensions of the standard
model, for example. After proposing several sample models which can
resolve the $R_b$ problem, we use our results to identify which features
of supersymmetric models are instrumental in so doing. 

\subsection{Diagonal Couplings to the $Z$: General Results}

We now present formulae for the correction to the $Z\bbar b$ vertex due to
a loop involving generic scalar and spin-half particles. In this section we
make the simplifying assumption that all of the $Z$-boson couplings are
flavour diagonal. This condition is relaxed in later sections where the
completely general expression is derived. The resulting formulae make it
possible to see at a glance whether a given model gives the right sign for
alleviating the discrepancy between experiment and the SM prediction for
$R_b$.

\figure\figone{The one-loop vertex correction and self-energy contributions to
the $Zb\bar b$ vertex due to fermion-scalar loops.}

\figure\figtwo{The one-loop contributions to the $Z\bbar b$ vertex due to the
gauge-boson vacuum polarizations.}

The one-loop diagrams contributing to the decay $Z\to b\bar b$ can be
grouped according to whether the loop attaches to the $b$ quark (\ie\ the
vertex correction and self-energy graphs of Fig.~\figone) or whether the
loop appears as part of the gauge boson vacuum polarization (Fig.~\figtwo).
For the types of models we consider these two classes of graphs are
separately gauge invariant and finite, and so they can be understood
separately. This is particularly clear in the limit that the particles
within the loop are heavy compared to $\MZ$, since then the vacuum
polarization graphs represent the contribution of the oblique parameters,
$S$ and $T$, while the self-energy and vertex-correction graphs describe
loop-induced shifts to the $b$-quark neutral current couplings, $\delta
g^b_{\lft,\rht}$.

Furthermore, although we must ensure that the oblique parameters do not
become larger than the bound of eq.~\oblique, eq.~\nocancel\ shows that
they largely cancel in the ratio $R_b$. We therefore restrict our attention
in this section to the diagrams of Fig.~\figone\ by themselves. The sum of
the contributions of Fig.~\figone\ is also finite as a result of the Ward
identity which was alluded to in Section 3. This Ward identity relates the
vertex-part graphs of Fig.~\figone a,b to the self-energy graphs of
Fig.~\figone c,d.  Since this cancellation is an important check of our
results, let us explain how it comes about.

We first consider an unbroken $U(1)$ gauge boson with a tree-level
coupling of $g_b$ to the the $b$-quark. This gives rise to the familiar
Ward identity from quantum electrodynamics: for external fermions with
four-momenta $p$ and $p'$,
\label\scai
\eq 
(p-p')^\mu \, \Gamma_\mu = g_{\rm eff}(S_\ssf^{-1}(p)-S_\ssf^{-1}(p')) ,
\eeq
where $\Gamma^\mu$ is the one-particle-irreducible vertex part and
$S_\ssf(p)$ is the fermion propagator. If we denote the vertex-part
contributions (Fig.~\figone a,b) to the effective vertex at zero momentum
transfer by $\delta g_b$, and the self-energy-induced wave function
renormalization of the $b$ quark by $Z_b$, then at one loop the Ward
identity \scai\ reduces to $g_b (1+Z_b) (\pxpsl-\pxpsl') = (g_b+\delta g_b)
(\pxpsl-\pxpsl')$, or
\eq \label\wardidzm
\delta g_b - g_b Z_b = 0. 
\eeq
This last equation is the more general context for the cancellation which
we found in Section 3; it states that the self-energy graphs (Fig.~\figone
c,d) must precisely cancel the vertex part (Fig.~\figone a,b) in the limit
of zero momentum transfer. Another way of understanding eq.~\wardidzm\ is
to imagine computing the effective $b$-photon vertex due to integrating out
a heavy particle. Eq.~\wardidzm\ is the condition that the two effective
operators $\bbar \, \dsl b$ and $\bbar \, \Asl b$ have the right relative
normalization to be grouped into the gauge-covariant derivative: $\bbar \,
\Dsl b$.

But for the external $Z$ boson, the Ward identity only applies to those
parts of the diagrams which are insensitive to the fact that the $U(1)$
symmetry is now broken. These include the $1/(n-4)$ poles from dimensional
regularization, and also the contributions to the $b$ neutral-current
coupling proportional to $\sw^2$, since the latter arise only through
mixing from the couplings of the photon.

We now return to the diagrams of Fig.~\figone. The first step is to
establish the Feynman rules for the various vertices which appear. Since we
care only about the LH neutral-current couplings, it suffices to consider
couplings of the new particles to $b_\lft$:
\label\scaiii
\eq
{\cal L}_{\rm scalar} = \yfp \,\phi \,\bar f \Pl b + {\rm h.c.}
\eeq
and we write the $Z$ coupling to $f$ and $\phi$ as
\label\scaii
\eq
{\cal L}_{\rm nc} = \left( {e \over \sw\cw} \right) \; Z^\mu \, \Bigl[ \bar f
\gamma_\mu ( \gfl \Pl + \gfr \Pr ) f + i \gs \; \phi^\dagger
\darr\phi \Bigr]. 
\eeq
The couplings, $g = \{ \gfl, \gfr, \gs \}$, are normalized so that $g
= I_3 - Q \sw^2$ for all fields, $f^a_{\lft,\rht}$ and $\phi^m$.

In the examples which follow, the field $f$ can represent either an
ordinary spinor (\eg, $t$) or a {\it conjugate} spinor (\eg, $t^c$). This
difference must be kept in mind when inferring the corresponding charge
assignments for the neutral-current couplings of the $f$. For example, the
left-handed top quark has $I_{3\ssl} = +\frac12$, so $\gfl = \hf - \frac23
\sw^2$ and $I_{3\ssr} = 0$, so $\gfr = - \frac23 \sw^2$. If the internal
fermion were a top {\it anti}quark, however, we would instead have $\gfr =
- \hf + \frac23 \sw^2$ and $\gfl = + \frac23 \sw^2$. The latter couplings
follow from the former using the transformation of the neutral current
under charge conjugation: $\gamma_\mu \Pl \leftrightarrow -\gamma_\mu\Pr$.

We quote the results for evaluating the graphs of Fig.~\figone\ in the
limit where $\MZ$ (and of course $m_b$) are negligible compared to $m_f$
and $M_\phi$, since they are quite simple and illuminating in this
approximation. It will be shown that the additional corrections due to the
nonzero mass of the $Z$ boson are typically less than 10\% of this leading
contribution.

\vskip 1truecm
\epsfxsize=5.5in\epsfbox{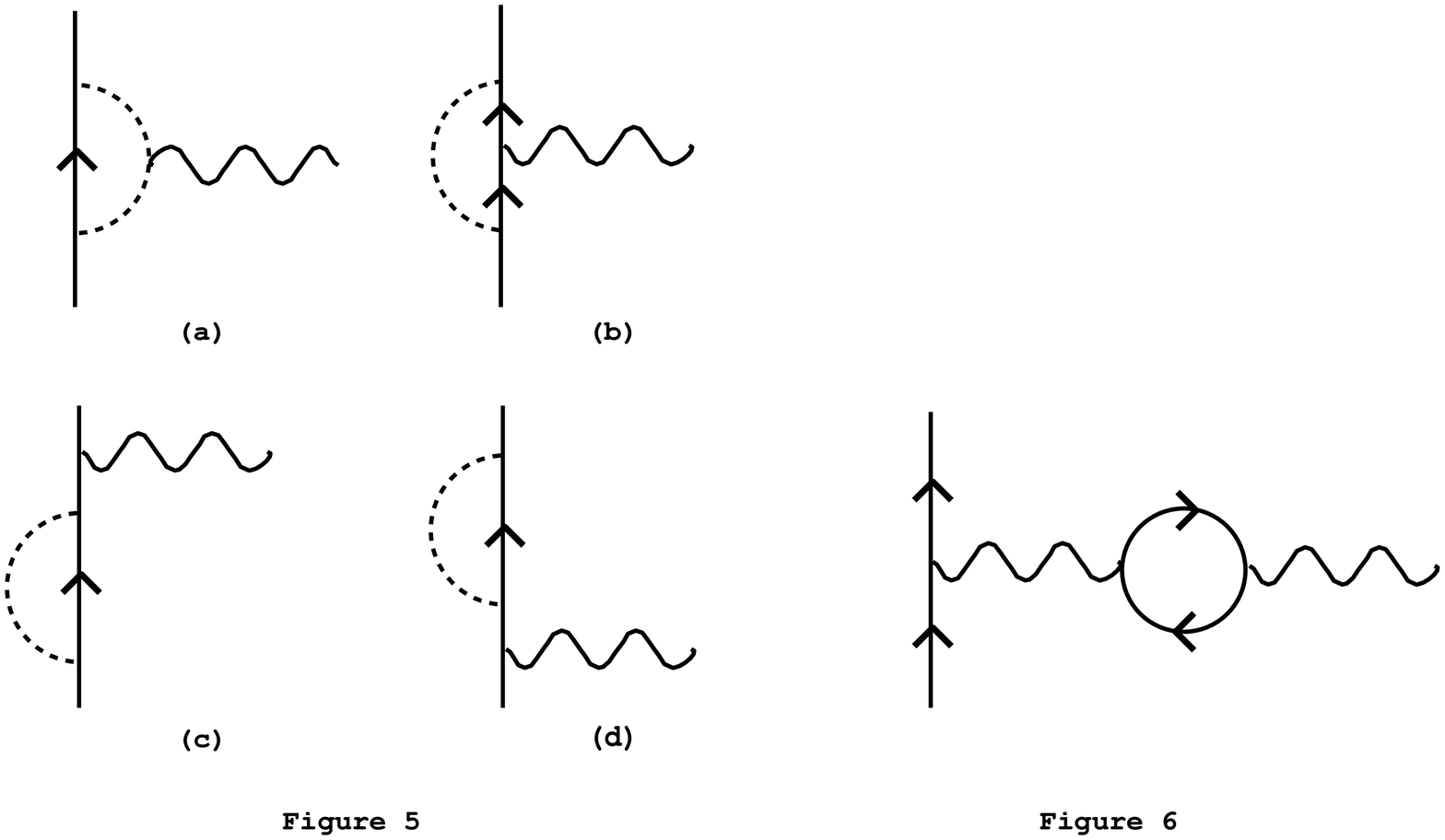}
\baselineskip = 0.8\baselineskip\noindent
{\eightrm Figure \figone: The one-loop vertex correction and
self-energy contributions to the $Zb\bar b$ vertex due to
fermion-scalar loops.} \hfill\break
\noindent{\eightrm Figure \figtwo: The one-loop
contributions to the $Z\bbar b$ vertex due to the gauge-boson vacuum
polarizations.}
\vskip 0.5truecm
\baselineskip= 1.25\baselineskip

 We find that
\eq\label\nomixresult
\delta \gbl = {1 \over 32 \pi^2} \sum_{f \phi} n_c \, |\yfp|^2 \Bigl[ 2(\gfl -
\gfr) \Scf(r) + (-\gfr + \gbl + \gs) \Bigl( \Delta_\phi - \twi\Scf(r) \Bigr)
\Bigr], \eeq
where $\Scf(r)$ and $\twi\Scf(r)$ are functions of the mass ratio $r =
m_f^2 /M_\phi^2$, 
\eqa
\Scf(r) & \equiv { r \over (r - 1)^2} \; \Bigl[ r - 1 - \ln r \Bigr], \eol
\twi\Scf(r) & \equiv { r \over (r - 1)^2} \; \Bigl[ r - 1 - r \ln r
\Bigr] . \eeol
\eeq
$\Delta_\phi$ denotes the divergent combination $\Delta_\phi \equiv
{2\over n-4} + \gamma + \ln(M^2_\phi/4\pi\mu^2) + {1\over 2}$, and $n_c$ is
a colour factor that depends on the $SU_c(3)$ quantum numbers of the fields
$\phi$ and $f$. For example, $n_c = 1$ if $\phi\sim{\bf 1}$ or $f\sim{\bf
1}$ (colour singlets); $n_c = 2$ if $f\sim$ \threebar\ and
$\phi\sim$ \threebar\ or ${\bf 6}$; $n_c = {16 \over 3}$ if $f\sim{\bf 3}$
and $\phi\sim{\bf 8}$.

The cancellation of divergences we expected on general grounds is now
evident in the present example, because electroweak gauge invariance of the
scalar interaction \scaiii\ implies that the neutral-current couplings are
related by
\label\scaiiia
\eq
 \gs + \gbl - \gfr = 0 .
\eeq 
This forces the term proportional to $\twi\Scf$ to vanish in
eq.~\nomixresult. As advertised the remaining term is both ultraviolet
finite and independent of $\sw^2$, which cancels in the combination $\gfl -
\gfr$.

We are left with the compact expression
\label\scaiv 
\eq
\delta \gbl = {1 \over 16 \pi^2} \sum_{f \phi} n_c \, |\yfp|^2 \, (\gfl -
\gfr) \; \Scf(m_f^2 /M_\phi^2).
\eeq
\figure\figthree{From top to bottom, the functions 
$\Scf(m^2_f/M^2_\phi)$, $F_\ssr(r) = F_\ssS(r)$ and $F_\ssl(r)$
which appear in the
loop contribution to the left-handed $Zb\bar b$ vertex, sections
5.1 and 5.3.}
%
Interestingly, it depends only on the axial-vector coupling of the internal
fermion to the gauge boson $W_3$ and not on the vector coupling. The
function of the masses $\Scf(r)$ is positive and monotonically increasing,
with $\Scf(r) \sim r$ as $r \to 0$ and $\Scf(\infty) = 1$, as can be seen
in Fig.~\figthree.

\vskip 1truecm
\epsfxsize=4in\epsfbox{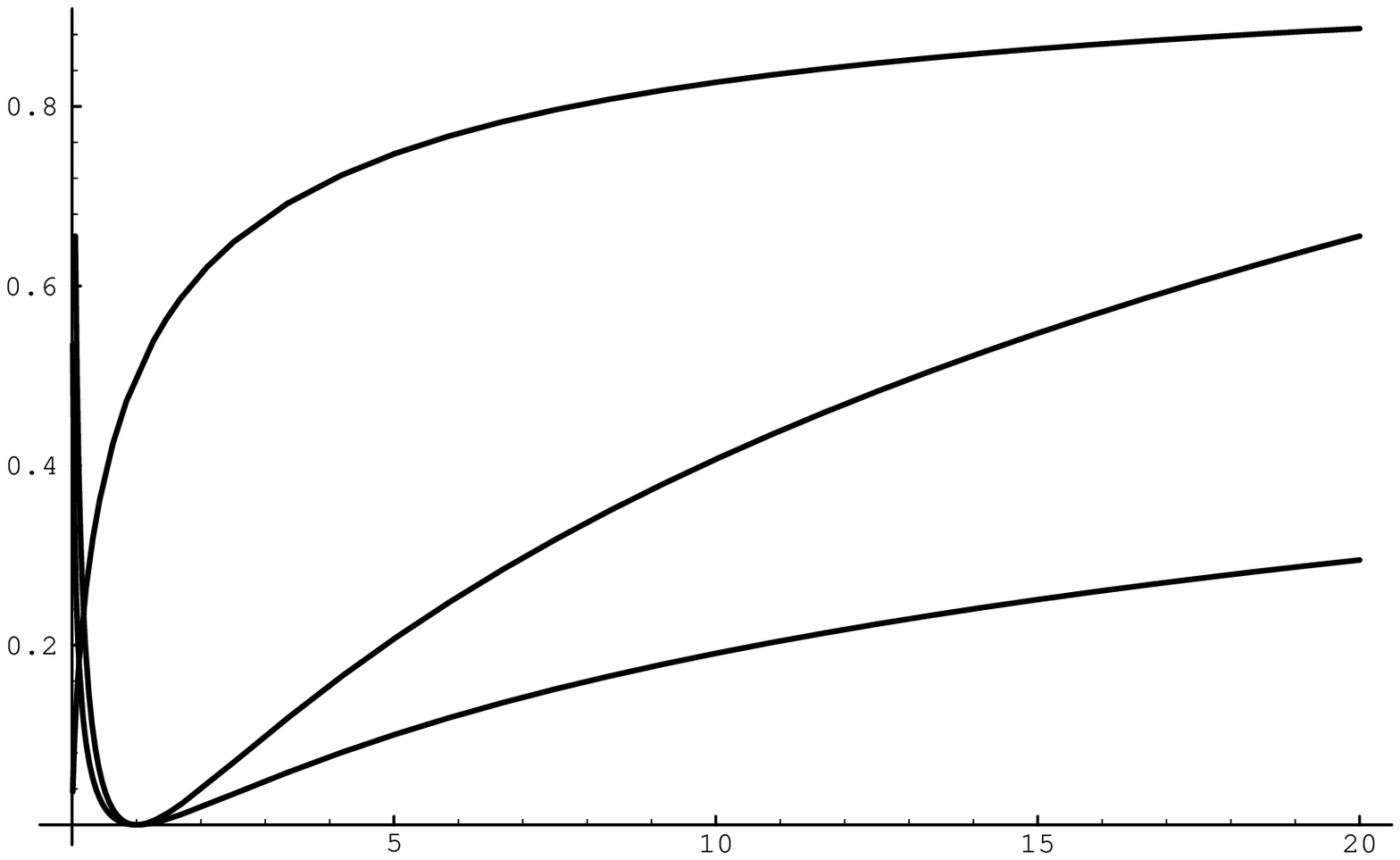}
\baselineskip = 0.8\baselineskip\noindent
{\eightrm Figure \figthree: From top to bottom, the functions
$\Scf(m^2_f/M^2_\phi)$, $F_\ssr(r) = F_\ssS(r)$ and $F_\ssl(r)$ which
appear in the loop contribution to the left-handed $Zb\bar b$ vertex,
sections 5.1 and 5.3.}
\vskip 0.5truecm
\baselineskip= 1.25\baselineskip

It is straightforward to generalize eq.~\scaiv\ to include the effect of
the nonzero $Z$ boson mass. Expanding to first order in $\MZ^2$, one
obtains an additional correction to the effective vertex,
\label\scavii
\eq
\delta_{\sss Z} \gbl = \sum_{f\phi} {|\yfp|^2 n_c \over 96\pi^2}
\left({\MZ^2\over m^2_f}\right)
\int_0^1\! dx\,\left\{ {x^3 (\gbl-\gfr) + 2(1-x)^3 \gfr \over x 
(M^2_\phi/m^2_f-1) + 1} 
+ {(1-x)^3 \gfl \over (x(M^2_\phi/m^2_f -1) + 1)^2} \right\}.
\eeq
To see that this is typically an unimportant correction, consider the limit
in which the scalar and fermion masses are equal, $r=1$. Then the total
correction \scaiv$+$\scavii\ is
\label\MZcorr
\eq
 \delta \gbl +\delta_{\sss Z} \gbl = \sum_{f\phi} {|\yfp|^2 n_c \over
32\pi^2} \left( \gfl - \gfr + {\MZ^2 \over 12\, m^2_f}\,
\left(\gbl + \gfl + \gfr \right) \right).
\eeq
Although the $\MZ^2$ correction can be significant if $\gfl = \gfr$, the
total correction would then be too small to explain the $R_b$ discrepancy,
and would thus be irrelevant.

\subsection{Why Many Models Don't Work}

What is important for applications is the relative sign between the tree
and one-loop contributions of eq.~\scaiv. In order to increase $R_b$ so as
to agree with the experimental observation, one needs for them both to have
the same sign, and so $\delta\gbl \propto (\gfl - \gfr) < 0$ in eq.~\scaiv.
Thus an internal fermion with the quantum numbers of the $b$-quark has
$\gfl - \gfr = - \hf$ and would increase $R_b$. Conversely, a fermion like
the $t$-quark has $\gfl - \gfr = +\hf$ and so causes a decrease. Moreover,
because the combination $(\gl - \gr)$ is invariant under charge
conjugation, the same statements hold true for the antiparticles: a $\bar
b$ running in the loop would increase $R_b$ whereas a $\bar t$ would
decrease it.

\ref\MaNg{E.~Ma and D.~Ng, \prd{53}{96}{255}.}

It thus becomes quite easy to understand which models with diagonal
couplings to the $Z$ boson can improve the prediction for $R_b$.
Multi-Higgs-doublet models have a hard time explaining an $R_b$ excess
because typically it is the top quark that makes the dominant contribution
to the loop diagram, since it has the largest Yukawa coupling, $\yfp\sim
1$, and the largest mass, to which the function $\Scf$ is very sensitive.
However for very large $\tan\beta$ (the ratio of the two Higgs VEV's), the
Yukawa coupling of the $t$ quark to the charged Higgs can be made small and
that of the $b$ quark can be made large, as in Ref.~\MaNg. Fig.~\figthree\
shows that, in fact, one must go to extreme values of these parameters,
because in addition to needing to invert the natural hierarchy between
$y_t$ and $y_b$, one must overcome the big suppression for small fermion
masses coming from the function $\Scf$.

Precisely the same argument applies to a broad class of Zee-type models,
where the SM is supplemented by scalar multiplets whose weak isospin and
hypercharge permit a Yukawa coupling to the $b$ quark and one of the other
SM fermions. So long as the scalars do not mix and there are no new
fermions to circulate in the loop, all such models have the same difficulty
in explaining the $R_b$ discrepancy. Below we will give some examples of
models which, in contrast, {\it are} able to explain $R_b$.

\subsection{Generalization to Nondiagonal $Z$ Couplings }

We now turn to the more complicated case where mixing introduces
off-diagonal couplings among the new particles. Because of mixing the
couplings of the fermions to the $Z$ will be matrices in the mass basis.
Similar to eqs.~\trotation\ and \fcncrotation\ we write
\eq 
(g_{\ssl,\ssr})^{ff'} = \sum_a \left[ ({\cal U}^{af}_{\lft,\rht})^* \,
{\cal U}_{\lft,\rht}^{af'} \, I^a_{3\ssl,\ssr} -\delta^{ff'} \, 
Q^a \sw^2 \right],
\eeq
where ${\cal U}^{af}_{\lft,\rht}$ are the mixing matrices. An analogous
expression gives the off-diagonal scalar-$Z$ coupling in terms of the
scalar mixing matrix, ${\cal U}_\ssS^{a\phi}$. Of course if all of the
mixing particles share the same value for $I_3$, then unitarity of the
mixing matrices guarantees that the couplings retain this form in any
basis.

This modification of the neutral-current couplings has two important
effects on the calculation of $\delta \gbl$. One is that the
off-diagonal $Z$ couplings introduce the additional graphs of the type
shown in Fig.~\figone a,b, where the fermions or scalars on either side
of the $Z$ vertex have different masses. The other is that the mixing
matrices spoil the relationship, eq.~\scaiiia, whereby the term
proportional to $\twi\Scf$ canceled in eq.~\nomixresult. But this is
only because of the mass-dependence of $\Scf$ and $\Delta_\phi$.
Therefore the cancellation still occurs if all of the particles that
mix with each other are degenerate, as one would expect. Moreover the
ultraviolet divergences still cancel since they are mass-independent.

Evaluation of the graphs gives the following result at $q^2 = \MZ^2 = 0$: 
\eq\label\defofGterms
\delta \gbl = {1 \over 32 \pi^2} \Bigl[ G_{\rm diag} + G_{f f'} +
G_{\phi\phi'} \Bigr] , 
\eeq
where $G_{\rm diag}/32\pi^2$ represents the contribution involving only the
diagonal $Z$ couplings, and so is identical to the previously derived
eq.~\nomixresult. It is convenient to write it as
\eq \label\diagpart
G_{\rm diag} = \sum_{f \phi} n_c \, |\yfp|^2 \Big\{ 2(g_\lft -
g_\rht)^{ff} \Scf(r) + \Bigl[-(g_\rht)^{ff} + \gbl + (\gs)^{\phi\phi} \Bigr]
\Bigl( \Delta_\phi - \twi\Scf(r) \Bigr) \Big\} .
\eeq
Here and in the following expressions we use the notation $r = m_f^2 /
M^2_\phi$ and $r' = m_\fp^2/M_\phi^2$. As before $\Delta_\phi$ denotes the
UV-divergent quantity $\Delta_\phi \equiv {2\over n-4} + \gamma +
\ln(M^2_\phi/4\pi\mu^2) + {1\over 2}$. 

The remaining terms in eq.~\defofGterms\ come from the new graphs of
Fig.~\figone a,b, where the scalars or fermions on either side of the
$Z$ vertex have different masses, due to mixing:
%
\eqa \label\fermoffdiag
G_{ff'} &= \sum_{\phi,f \ne f' } n_c \, \yfp \,  
\yfpnp^* 
\Bigl[
2(g_\lft)^{ff'} \Scf_\lft(r,r') - (g_\rht)^{ff'} \Bigl(\Delta_\phi -
\Scf_\rht(r,r') \Bigr) \Bigr], \eol 
\label\scalaroffdiag
G_{\phi\phi'} &= \sum_{f,\phi \ne \phi'} n_c \, \yfp \,
\yfnpp^*
(\gs)^{\phi\phi'} \Bigl[ \Delta_\phi - \Scf_\ssS(x,x') \Bigr], \eeol
\eeq
where $\Scf_\lft(r,r')$, $\Scf_\rht(r,r')$ and $\Scf_\ssS(x,x')$ are given
by: 
\eqa \label\LHfn
\Scf_\lft(r,r') &= {\sqrt{rr'} \over r - r'} \; \left[ {r \over r - 1} \; 
\ln r - {r'\over r' - 1} \; \ln r' \right], \eol
\label\RHfn
\Scf_\rht(r,r') &= {1 \over r - r'} \; \left[ {r^2 \over r - 1} \; 
\ln r - {r'^2\over r' - 1} \; \ln r' \right], \eol
\label\Sfn
\Scf_\ssS(x,x') &={1 \over (x-1)(x'-1)} \; \Bigl( 1 + \ln x \Bigr) + 
{x'^2 \over(x' -1)(x'-x)} \left( 1 + \ln {x\over x'} \right) + 
{x^2 \over (x-1)(x-x')}, \eolnn &\eeol 
\eeq
and $x$, $x'$ are the mass ratios $x = M^2_\phi/m^2_f$ and $x' =
M^2_{\phi'} / m^2_f$. These expressions have several salient features
which we now discuss. First, eqs.~\diagpart, \fermoffdiag\ and
\scalaroffdiag\ are obviously much more complicated than eq.~\scaiv. In
particular, it is no longer straightforward to simply read off the sign of
the result.

Second, the sum of the UV divergences in eqs.~\diagpart, \fermoffdiag\ and
\scalaroffdiag,
\eq\label\divcancel
G_{\Delta} \propto \sum_{ff'\phi\phi'} \yfp \, \yfpp^* \; \Bigl[
-(g_\rht)^{ff'} \delta^{\phi\phi'} + \gbl \delta^{ff'} \delta^{\phi\phi'} +
(\gs)^{\phi\phi'} \delta^{ff'} \Bigr], 
\eeq
is {\it basis independent} since a unitary transformation of the fields
cancels between the Yukawa and neutral-current couplings. Thus it can be
evaluated in the electroweak basis where the neutral-current couplings are
diagonal and proportional to $-\gfr + \gbl + \gs$, which vanishes due to
conservation of weak isospin and hypercharge at the scalar-fermion vertex.
We are therefore free to choose the renormalization scale $\mu^2$ in
$\ln(M^2_\phi/\mu^2)$ to take any convenient value. The $M_\phi$-dependence
of $\Delta_\phi$ makes $G_{\phi\phi'}$ look unsymmetric under the
interchange of $\phi$ and $\phi'$, but this is only an artifact of the way
it is expressed. For example when there are only two scalars,
$G_{\phi\phi'}$ is indeed symmetric under the interchange of their masses.

Third, all the contributions {\it except} those of $G_{\phi\phi'}$ are
suppressed by powers of $m_f/M_\phi$ in the limit that the scalars are much
heavier than the fermions. Thus to get a large enough correction to $\gbl$
requires that: ($i$) not all of the scalars be much heavier than the
fermions which circulate in the loop, or ($ii$) the scalars mix
significantly and have the right charges so that $G_{\phi\phi'}$ is
nonnegligible and negative. We use option ($ii$) in what follows to
construct another mechanism for increasing $R_b$.

Finally, even if the two fermions are degenerate, one does not generally
recover the previous expression \nomixresult\ that applied in the absence
of mixing. This is because Dirac mass matrices are diagonalized by a
similarity transformation, $M\to \ULT M\UR$, not a unitary transformation.
The left- and right-handed mixing angles can differ even when the
diagonalized mass matrix is proportional to the identity. Thus, in contrast
to eq.~\divcancel, the expression $\sum_{ff'\phi\phi'} \yfp
\yfpp^*[(g_\lft)^{ff'}-(g_\rht)^{ff'}]$ is not invariant under
transformations of the fields, because $\yfp$ is rotated by $\UR$ (recall
that $\yfp$ is the Yukawa coupling only for the RH $f$'s) whereas $g_\lft$
is rotated by $\UL$.

We can get some insight into eqs.~\fermoffdiag--\Sfn\ by looking at 
special values of the parameters. Let us assume there is a dominant Yukawa
coupling $y$ between the left-handed $b$ quark and a single species of
scalar and fermion, $f_1$ and $\phi_1$ in the weak basis,
\label\scaix
\eq
{\cal L}_{\rm scalar} = y \,\phi_1 \,\bar f_1 \Pl b + {\rm h.c.}
\eeq
In the mass basis the couplings will therefore be
\eq
	\yfp = y \, {\cal U}_\ssS^{1\phi}\, ({\cal U}_\rht^{1f})^*.
\eeq
Now gauge invariance only relates the $(1,1)$ elements of the
neutral-current coupling matrices in the weak basis: 
\label\scaxi\eq
 (\gs)^{11} + \gbl - (\gfr)^{11} = 0. 
\eeq
There are three limiting cases in which the results become easier to
interpret:

\topic{1} If all the scalars are degenerate with each other, and likewise
for the fermions, then the nonmixing result of eq.~\scaiv\ holds, except
one must make the replacement
\label\scaxva
\eq
	\gfl - \gfr \to (\UR\,S_m\,\ULT\>\gl\>\UL\,S_m\,\URT - \gr)^{11},
\eeq
where $S_m$ is the diagonal matrix of the signs of the fermion masses.

\topic{2} If there are only two scalars and if they are much heavier than
all of the fermions, only the term $G_{\phi\phi'}$ is significant. Let
$\phi_1$ and $\phi_2$ denote the weak-eigenstate scalars, and $\phi$ and
$\phi'$ the mass eigenstates; then
\label\scaxvi 
\eqa
\delta \gbl &= {y^2 n_c \over 16\pi^2} \,(I^{\phi_1}_3 - I^{\phi_2}_3)\, 
\cS^2\sS^2\, F_\ssS(M^2_{\phi}/M^2_{\phi'}); \eol
\label\scaxvii 
F_\ssS(r) &= {r+1\over 2(r-1)} \, \ln r -1, \eeol
\eeq
where $\cS$ and $\sS $ are the cosine and sine of the scalar mixing angle.
The function $F_\ssS(r)$ is positive except at $r=1$ where it is zero, and
so the sign of $\delta \gbl$ is completely controlled by the factor 
$(I^{\phi_1}_3 - I^{\phi_2}_3)$. We see that to increase $R_b$ it is necessary
that $I^{\phi_1}_3 <I^{\phi_2}_3$. 

\topic{3} When there are only two fermions, with weak eigenstates $f_{1}$,
$f_{2}$ and mass eigenstates $f$, $f'$ both much heavier than any of the
scalars, then
\label\scavi
\eq
\eqalign{
\delta \gbl &= {y^2 n_c \over 16\pi^2 } \Bigl\{ \gl^{11} c^2_{\ssl\ssr} + 
\gl^{22}
s^2_{\ssl\ssr} - \gr^{11} \cr
 & + (\gr^{22} - \gr^{11}) \,\cR^2\sR^2\, F_\ssr(m^2_{f}/m^2_{f'})
 - 2 (\gl^{22} - \gl^{11}) \, \cL\sL\cR\sR \, F_\ssl(m^2_{f}/m^2_{f'}) 
\Bigr\}; \cr}
\eeq
where $s_{\ssl\ssr}$ and $c_{\ssl\ssr}$ are the sine and
cosine of the difference or sum of the LH and RH mixing angles, $\theta_\ssl
- s_m \theta_\ssr$, depending on the relative sign $s_m$ of the two 
fermion mass eigenvalues, and 
\eq\label\morefns
F_\ssr(r) = F_\ssS(r) = {r+1 \over 2(r-1)} \, \ln r -1; \qquad 
\hbox{and} \qquad F_\ssl(r) = {\sqrt{r}\over r-1} \ln r - 1 .
\eeq
The function $F_\ssl$ has some of the same properties as $F_\ssS = F_\ssr$,
including invariance under $r\to 1/r$, being positive semidefinite and
vanishing at $r=1$. Plots of these functions are shown in Figure \figthree.
Note that the first line of eq.~\scavi\ is the same as \scaxva.

To get some idea of the error we have made by neglecting the mass of
the $Z$ boson one can compute the lowest order correction as in section
5.3.  The answer is more complicated than for the case of diagonal $Z$
couplings, except when the fermions are degenerate with each other and
likewise for the bosons. In that case the answer is given again by
eq.~\MZcorr\ except that $\gfr\to(\gfr)^{11}$ and $\gfl\to
(\UR\,S_m\,\ULT\>\gl\>\UL\,S_m\,\URT - \gr)^{11}$, precisely as in
eq.~\scaxva.  Thus we would still expect it to be a small correction
even when there is mixing of the particles in the loop.

These simplifying assumptions can be used to gain a semi-analytic
understanding of why certain regions of parameter space are favoured in
complicated models, which is often missing in analyses that treat the
results for the loop integrals as a black box. The observations we
make here may be useful when searching for modifications to a model
that would help to explain $R_b$. The next two sections exemplify this
by creating some new models that take advantage of our insights, and by
elucidating previous findings in an already existing model,
supersymmetry.

\subsection{Examples of Models That Work}

Besides ruling out certain classes of models, our general
considerations also suggest what {\it is} required in order to explain
$R_b$. Obviously new fermions and scalars are required, whose Yukawa
couplings allow them to circulate inside the loop. We give two
examples, one with diagonal and one with nondiagonal couplings of the new
particles to the $Z$ boson.

\table\clifftabone 

For our first example we introduce several exotic quarks $F$, $P$ and
$N$, and a new Higgs doublet $\phi$, whose quantum numbers are listed
in Table \clifftabone. The unorthodox electric charge assignments do
not ensure cancellation of electroweak anomalies, but this can be fixed
by adding additional fermions, like mirrors of those given, which do
not contribute to $R_b$.

\midinsert
$$\vbox{\tabskip=1em plus 4em minus .5em \offinterlineskip
\halign to \hsize{\strut \tabskip=1em plus 2em minus .5em \hfil#\hfil &
\hfil#\hfil &\hfil#\hfil &\hfil#\hfil &\hfil#\hfil \cr
\noalign{\hrule}\noalign{\smallskip}\noalign{\hrule}\noalign{\medskip}
Field & Spin & $SU_c(3)$ & $SU_\ssl(2)$ & $U_\ssy(1)$ \cr
\noalign{\medskip}\noalign{\hrule}
\noalign{\smallskip}
$\phi$ & 0 & {\bf 1} & {\bf 2} & $q-{1\over 6}$ \cr 
\noalign{\smallskip}
$F_\ssl$ & $\hf$ & {\bf 3} & {\bf 2} & $q+{1\over 2}$ \cr 
\noalign{\smallskip}
$P_\ssr$ & $\hf$ & {\bf 3} & {\bf 1} & $q+1$ \cr 
\noalign{\smallskip}
$N_\ssr$ & $\hf$ & {\bf 3} & {\bf 1} & $q$ \cr 
\noalign{\medskip}\noalign{\hrule}\noalign{\smallskip}\noalign{\hrule} }}
$$
\centerline{{\bf Table \clifftabone}}
\smallskip\noindent
{\eightrm Field Content and Charge Assignments: Electroweak quantum numbers
for the new fields which are added to the SM to produce the observed value
for $\ss R_b$.} 

\endinsert

The hypercharges in Table \clifftabone\ allow the following Yukawa
interactions:
\eq\label\uglyyukawas
\Scl_y = y \, {\bar N}_\rht Q_\lft^i \phi^j \epsilon_{ij} 
+ g_p \, {\bar P}_\rht F_\lft^i H^j \epsilon_{ij} 
+ g_n \, {\bar N}_\rht F_\lft^i {\tw H}^j \epsilon_{ij} + \hc,
\eeq
where $\epsilon_{ij}$ is the $2\times 2$ antisymmetric tensor, $H$ is the
usual SM Higgs doublet and $Q_\lft = \pmatrix{t_\lft \cr b_\lft\cr}$ is the
SM doublet of third generation LH quarks. When $H$ gets its VEV, $\Avg{H} =
v$, we find two fermion mass eigenstates, $p$ and $n$, whose masses are
$m_p = g_p v$ and $m_n = g_n v$ and whose electric charges are $Q_p = q+1$
and $Q_n = q$. There are also two new scalar mass eigenstates,
$\varphi_\pm$, whose electric charges are $Q_+ = q+{1\over 3}$ and $Q_- =
q-{2\over 3}$.

In the mass eigenstate basis, the Yukawa interactions with the new
scalars are
\eq\label\uglyyukmass
\Scl_y = y \, \ol{n}_\rht b_\lft \varphi_+ - y \, \ol{n}_\rht t_\lft
\varphi_- + \hc, \eeq
from which we see that the $n$ couples to the $b$-quark as in eq.~\scaiii.

The weak isospin assignments of the $n$ are $I^n_{3\ssl} = -\hf$ and
$I^n_{3\ssr} = 0$, so that $g^n_\lft - g^n_\rht = - \, \hf$. Therefore,
from eq.~\scaiv, one obtains $\delta \gbl<0$. The central value of $R_b$
can be reproduced if $\delta \gbl = -0.0067$, which is easily obtained by
taking $y\sim 1$ and $r\gg 1$, so that $\Scf(r) \simeq 1$. The Yukawa
coupling could be made smaller by putting the new scalars in a higher
colour representation like the adjoint. 

We have not explored the detailed phenomenology of this model, but it is
clearly not ruled out since we are free to make the new fermions and
scalars as heavy as we wish. And since we can always take $m_p=m_n$, there
is no contribution to the oblique parameter $T$. The contribution to $R_b$
does not vanish even as the masses become infinite, but this is consistent
with decoupling in the same way as a heavy $t$ quark, since the new
fermions get their masses through electroweak symmetry breaking. The price
we have to pay for such large masses is correspondingly large coupling
constants.

Next we build a model that uses our results for nondiagonal couplings
to the $Z$. It is a simple modification of the SM that goes in the
right direction for fixing the $R_b$ discrepancy but not quite far
enough in magnitude. Variations on the same theme can completely
explain $R_b$ at the cost of making the model somewhat more baroque.

\table\clifftabtwo

Our starting point is a two-Higgs doublet extension of the SM. We
take the two Higgs fields, $H_d = \pmatrix{H_d^0 \cr H_d^- \cr}$ and $H_u =
\pmatrix{H_u^+ \cr H_u^0 \cr}$, to transform in the usual way under the SM
gauge symmetry. It was explained earlier why this model does not by itself
produce the desired effect, but eq.~\scaxvi\ suggests how to fix this
problem by introducing a third scalar doublet, $\Delta =
\pmatrix{\Delta^{++} \cr \Delta^+}$, which mixes with the other Higgs
fields. The charge assignments of these fields, listed in Table
\clifftabtwo, ensure that the two fields $H_u^+$ and $\Delta^+$ can mix
even though they have different eigenvalues for $I_3$.

\midinsert
$$\vbox{\tabskip=1em plus 2em minus .5em \offinterlineskip
\halign to \hsize{\strut \tabskip=1em plus 2em minus .5em \hfil#\hfil &
\hfil#\hfil &\hfil#\hfil &\hfil#\hfil &\hfil#\hfil \cr
\noalign{\hrule}\noalign{\smallskip}\noalign{\hrule}\noalign{\medskip}
Field & Spin & $SU_c(3)$ & $SU_\ssl(2)$ & $U_\ssy(1)$ \cr
\noalign{\medskip}\noalign{\hrule}
\noalign{\smallskip}
$H_d$ & 0 & {\bf 1} & {\bf 2} & $- \, \hf$ \cr 
\noalign{\smallskip}
$H_u$ & 0 & {\bf 1} & {\bf 2} & $+ \hf$ \cr 
\noalign{\smallskip}
$\Delta$ & 0 & {\bf 1} & {\bf 2} & $+ \frac32$ \cr 
\noalign{\medskip}\noalign{\hrule}\noalign{\smallskip}\noalign{\hrule} }}
$$
\centerline{{\bf Table \clifftabtwo}}
\smallskip\noindent
{\eightrm Field Content and Charge Assignments: Electroweak quantum numbers
for all of the scalars --- including the SM Higgs doublet --- of the
three-doublet model.} 

\endinsert

In this model the new scalar field cannot have any Yukawa couplings to
ordinary quarks since these are forbidden by hypercharge conservation. The
only Yukawa couplings involving the LH $b$-quark are those which also
generate the mass of the $t$-quark:
\eq
\Scl_{yuk} = y_t \, \sqrt2 \; \bar t \Pl b \; H_u^+ + \hc,
\eeq
where $y_t = m_t/v_u$ is the conventionally-normalized Yukawa coupling. We
imagine $v_u$ to be of the same order as the single-Higgs SM value, and so
we expect $y_t$ to be comparable to its SM size. 

The scalar potential for such a model very naturally incorporates $H_u^+ -
\Delta^+$ mixing. Gauge invariance permits quartic scalar interactions of
the form $\lambda (H_u^\dagger \Delta) (H_u^\dagger H_d) + \hc$, which
generate the desired off-diagonal terms: $\lambda (\Delta^+ H_u^{+*} v_u^*
v_d + \Delta^+ H_d^- v_u^{2*} ) + \hc$\ 

Since the weak isospin assignments are $I_3^{\Delta^+} = - \, \hf$ and
$I_3^{H_u^+} = + \hf$, the colour factor is $n_c = 1$ and the relevant
Yukawa coupling is $y = y_t \sqrt2$, we see that eq.~\scaxvi\ predicts the
following contribution due to singly-charged Higgs loops: 
\eq
\delta \gbl = - \, {y_t^2 \over 16 \pi^2} \; 2 \cS^2 \sS^2 \; F_\ssS(r)~, 
\eeq
with $r$ being the ratio of the scalar mass eigenstates, 
$r = M^2_\phi/M^2_{\phi'}$.
Taking optimistic values for the parameters\foot\caveat{Note that the
charged-scalar mixing in this model is suppressed if one of the scalar
masses gets very large compared to the weak scale.} ($\theta_\ssS = {\pi
\over 4}$, $2 \cS^2 \sS^2 = \hf$, $y_t = 1$ and $M_\phi/M_{\phi'}= 10$), 
we find
$\delta \gbl = -0.0043$, which is two thirds of what is required: $(\delta
\gbl)_{\rm exp} = - 0.0067 \pm 0.0021$.

In addition to the contribution of the singly-charged scalar loops, one
should consider those of the other nonstandard scalar fields we introduced.
Since all of the scalars that mix have the same eigenvalue for $I_3$, their
contribution is given by eq.~\scaiv, which is small if the scalars are much
heavier than the light fermions. Then only the $t$-quark contribution is
important. In this limit there are appreciable contributions only from the
three charged scalar fields, one of which is eaten by the physical $W$
boson and so is incorporated into the SM $t$-quark calculation, and the
other two of which we have just computed.

So, for an admittedly special region of parameter space, this simple model
considerably ameliorates the $R_b$ discrepancy, reducing it to a $1\sigma$
effect. It is easy to adapt it so as to further increase $\delta \gbl$ and
also enlarge the allowed region of the model's parameter space. The
simplest way is by increasing the size of the colour factor $n_c$ or the
isospin difference $I^{\phi'}_3 -I^\phi_3$. For instance the new scalar,
$\Delta$, could be put into a {\bf 4} of $SU_\ssl(2)$ rather than a
doublet, and be given weak hypercharge $Y = + \frac52$. Then the
singly-charged state $\Delta^+$ has $I^{\Delta^+}_3 = - \frac 32$, making
$I^{\phi'}_3 - I^\phi_3 = -2$, which is twice as big as for the doublet.
More new scalars must be added to generate mixing amongst the
singly-charged scalar states.

A second variation would be let the two new Higgs doublets be colour octets
since this gives more than a five-fold enhancement of $\delta \gbl$ due to
the colour factor $n_c = {16 \over 3}$. It is still possible to write down
quartic scalar interactions which generate the desired scalar mixings.
Either of these models has much more room to relax the previously tight
requirements for optimal scalar masses and mixings.

\subsection{The Supersymmetric Case}

Let us now apply the above results to gain some insight into what would be
necessary to explain $R_b$ in supersymmetric extensions of the standard
model. There are two kinds of contributions involving the top-quark Yukawa
coupling, which one expects to give the dominant effect. These are the
couplings of the left-handed $b$ quark to the second Higgs doublet and the
top quark, or to the corresponding Higgsinos and top squarks,
\eq
y_t \bar b_\lft \htr \, \tilde t_\rht + y_t \bar b_\lft t_\rht\, \htwo.
\eeq

Of these, the second one gives a loop contribution like that of the
two-Higgs doublet models discussed above: it has the wrong sign for
explaining $R_b$. Since the mass of the charged Higgs is a free
parameter in supersymmetric models, we can imagine making it large
enough compared to $m_t$ so that, according to eq.~\scaiv, it has only
a small effect on $R_b$. We therefore concentrate on the Higgsino-squark
part. The charged Higgsino mixes with the Wino, and the right-handed
top squark mixes with its chiral counterpart, so in the notation of
\scaix, we have $f_1 = \tilde\htwo$, $f_2 = \Wt$$\!\!\!\!$,\ \ 
$\phi_1 = \tilde t_\rht$ and $\phi_2 = \tilde t_\lft$. 
The corresponding charge matrices
for the couplings to the $W_3$ are
\label\chargemats
\eq
\gs = \left(\matrix{ \frac23\sw^2 & 0\cr 0 & \frac12 + \frac23\sw^2
\cr} \right);\qquad
\gl = \gr = \left(\matrix{ \frac12-\sw^2 & 0\cr 0 & 1-\sw^2\cr}\right).
\eeq
Because there are two possible colour combinations for the internal lines
of the loops diagram, the colour factor in eqs.~\diagpart\--\scalaroffdiag\
is $n_c =2$. 

Before exploring the full expression for $\delta g^b_\lft$ we can
discover what parameter ranges are the most promising by looking at
the limiting cases described by eqs.~\scaxva--\scavi. The most
important lessons from these approximations follow from the charge
matrices \chargemats. We do not want the squarks to be much heavier
than the charginos because then eq.~\scaxvi\ would apply and give the
wrong sign for the correction due to the sign of the isospin difference
between the squarks. The other two cases, where the squarks are not
much heavier than the charginos, manifest a strong suppression of the
result unless the chargino mixing angles are such that $\sin(\theta_\lft -
s_m \theta_\rht)$ is large, where $s_m$ is the sign of the determinant of
the chargino mass matrix. If on the other hand $\sin(\theta_\lft - s_m
\theta_\rht)=0$, there is exact cancellation between $\gl$ and $\gr$ in
these equations because of the fact that $\gl=\gr$ for the charginos.
In summary, our analytic formulas indicate that the favoured regions
of parameter space for increasing $R_b$ are where
\label\scaxxiv
\eq
	\tan\theta_\rht\,\tan\theta_\lft = -s_m = -\sgn(m_f\, m_{f'}),
\eeq
and at least one of the squarks is not much heavier than the charginos.

In supersymmetric models the Yukawa coupling that controls the largest
contribution to $R_b$ is that of the top quark, and it depends on the ratio 
of the two Higgs VEV's, $\tan\beta = v_2/v_1$, by
\label\susyyuk
\eq
	\yfp = {m_t\over v\sin\beta},
\eeq
where $v=(v_1^2+v_2^2)^{1/2} = 174$ GeV. Therefore it is important to find
$\tan\beta$ in terms of the chargino masses and mixing
angles. The chargino mass matrix is given by
\label\scaxxv
\eq
\left(\matrix{ \mu & g v_2 \cr g v_1 & M_2 \cr } \right) = 
\ULT \left(\matrix{m_f&0\cr 0&m_{f'}\cr}\right) \UR = 
\left(\matrix{ \cL\cR m_f + \sL\sR m_{f'} & \sR\cL m_f - 
\cR\sL m_{f'}\cr
\cR\sL m_f - \sR\cL m_{f'}& \sL\sR m_f + \cL\cR m_{f'}\cr}\right),
\eeq
where $\mu$ is the coefficient of $H_1 H_2$ in the superpotential and
$M_2$ is the soft-SUSY-breaking mass term for the Wino. It follows that
\label\tanbeta
\eq
	\tan\beta = {m_f\tan\theta_\rht - m_{f'}\tan\theta_\lft \over
	m_f\tan\theta_\lft - m_{f'} \tan\theta_\rht}
\eeq

\ref\leplimit{L.~Rolandi, H.~Dijkstra, D.~Strickland and
G.~Wilson, representing the ALEPH, DELPHI, L3 and OPAL
collaborations, Joint Seminar on the First Results from LEP~1.5,
CERN, Dec.~12th, 1995.}

The above considerations allow us to understand why values of
$\tan\beta$ near unity are necessary for a supersymmetric solution to the
$R_b$ problem. From eq.~\tanbeta\ and the maximization condition
\scaxxiv\ we see that $\tan\beta$ is restricted to lie between
$|m_{f}/m_{f'}|$ and $|m_{f'}/m_{f}|$. Eq.~\scaxxiv\ together
with \scaxxv\ also implies
\label\susymasscond 
\eq
	\cL^2|m_{f}| +\sL^2|m_{f'}| = \sqrt{2}\MW \sin\beta;\qquad
	\cL^2|m_{f'}| +\sL^2|m_{f}| = \sqrt{2}\MW \cos\beta,
\eeq
This means that average value of the two chargino masses can be no
greater than $\MW$, so that the ratio $|m_{f}/m_{f'}|$ cannot differ
much from unity unless one of the charginos is much lighter than the
$W$ boson. Using the LEP 1.5 limit of 65 GeV for the lightest chargino
\leplimit\ this would then require that $\tan\beta < 1.5$.

\figure\figfour{The dependence of $\gbl$ on the various supersymmetric
parameters.}

\vskip 1truecm
\epsfxsize=5in\epsfbox[90 204 529 591]{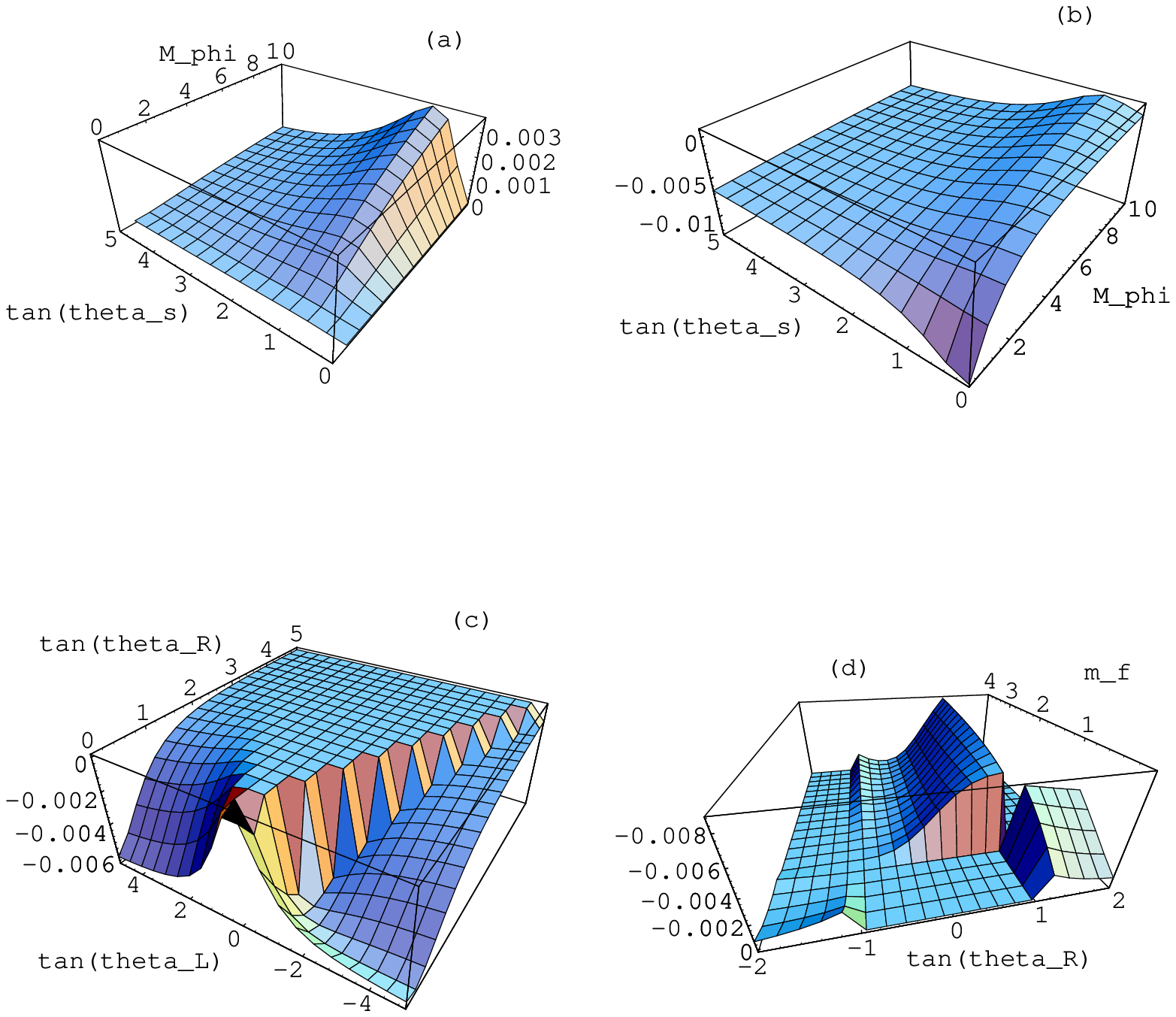}
\baselineskip = 0.8\baselineskip\noindent
{\eightrm Figure \figfour: The dependence of $\gbl$ on the various 
supersymmetric parameters.  Since $\gbl$ depends only on mass ratios
in our approximation, the units of mass are arbitrary, with the masses
of all the charginos and squarks which are not being varied 
set to unity.}
\vskip 0.5truecm
\baselineskip= 1.25\baselineskip

In the case that none of our simplifying limits apply, we have searched
the parameter space of the three independent ratios between the two
scalar masses and the two fermion masses, and the three mixing angles
$\theta_\rht$, $\theta_\lft$, $\theta_{\sss S}$ to find which regions
are favourable for increasing $R_b$. Figures \figfour a-d show the
shift in $g^b_{\sss L}$ as a function of pairs of these parameters,
using the Yukawa coupling \susyyuk\ corresponding to a top quark mass
of 174 GeV and the theoretical preference for $\tan\beta > 1$ (we
implement the latter by setting $g^b_{\sss L}=0$ for parameters that
would give $\tan\beta < 1$).  As shown in Table 1, one needs $\delta
g^b_{\sss L}=-0.0067$ in order to explain the observed value of $R_b$.
The values of the masses are taken to be $M_{\phi} \cong M_{\phi'}
\cong m_{f} \cong m_{f'}= 1$ (in arbitrary units), except for those
that are explicitly varied in each figure. In Fig.~\figfour a we look
at the situation in which $\tan\theta_\lft = \tan\theta_\rht = 1$, in
contradiction to condition \scaxxiv, and vary the scalar mixing angle
and the mass of mostly-$\tilde t_\rht$ scalar in the limit of zero
squark mixing. The sign of $g^b_{\sss L}$ has the wrong value, as
predicted by eq.~\scaxvi. Fig.~\figfour b shows the same situation
except that now $\tan\theta_\lft = - \tan\theta_\rht = 1$, in
accordance with eq.~\scaxxiv. Then the sign of $g^b_{\sss L}$ is
negative, as desired, and has the right size for substantial ranges of
$\theta_{\sss S}$ and $M_{\phi}$. In Fig.~\figfour c we keep all the
masses nearly degenerate and set $\theta_{\sss S}=0$ to show the
dependence on $\tan\theta_\lft$ and $\tan\theta_\rht$. It is easy to
see that $g^b_{\sss L}$ has the correct sign and largest magnitude
(which is also almost as large as needed) when condition \scaxxiv\ is
satisfied. Finally in Fig.~\figfour d we show the dependence on the
masses of the mostly-Wino fermion and on $\theta_\rht$ when
$\theta_{\sss S} = 0$ and $\tan\theta_\lft = -1$, showing again the
preference for mixing angles obeying \scaxxiv, as well as some
enhancement when there is a hierarchy between the two chargino masses.

\ref\feng{J.L.~Feng, N.~Polonsky and S.~Thomas, SLAC preprint SLAC-PUB-95-7050,
hep-ph/9511324.}
\ref\lizsu{E.H.~Simmons and Y.~Su, Boston University preprint, hep-ph/9602267.}
\ref\ELN{J.~Ellis, J.L.~Lopez and D.V.~Nanopoulos, CERN preprint
CERN-TH/95-314, hep-ph/9512288.}

One might therefore get the impression that it is easy to explain $R_b$
using supersymmetric contributions to the $Z b\bar b$ vertex. The
problem is that to get a large enough contribution one is driven to a
rather special region of parameter space, which comes close to
satisfying condition \scaxxiv. As mentioned above, the consequent
condition \susymasscond\ prevents one from making the chargino masses
arbitrarily heavy. This, coupled with the suppression in $R_b$ when the
squarks are heavier than the charginos, means that all the relevant
supersymmetric particles must be relatively light, except the charged
Higgs which has to be heavy to suppress the wrong-sign contribution
from $H^+$-$t$ loops. Thus in the example of Fig.~\figfour c, the
preferred values of $\cR = 1$, $\sL = \pm 1$, $\sR = \cL = 0$ imply
that $m_{f} = v\sin\beta$ and $m_{f'} = v\cos\beta$, while $\mu\cong
M_2 \cong 0$, which are precisely the circumstances of the
supersymmetric models considered in Refs.~\feng\ and \lizsu.
Fig.~\figfour d, on the other hand, has its maximum value of $R_b$ at
$\cR=\sR=\cL=-\sL =1$, implying $\tan\beta=1$ and thus from
\susymasscond\ that $|m_{f'}| + |m_{f}| = 2 \MW$. Because the lightest
chargino mass is constrained by experimental lower limits, there is
little parameter space for getting a large hierarchy between the two
chargino masses, as one would want in the present example in order to
get the full shift of $-0.0067$ in $g^b_{\sss L}$.\foot\eln{An
additional constraint is that the lightest Higgs boson mass $\ss m_{h^0}$
vanishes at tree level when $\ss \tan\beta=1$, and a very large splitting
between the top squark masses is needed for the one-loop corrections to
$\ss m_{h^0}$ to be large enough.  This is why ref.~\ELN\ finds less than
the desired shift in $\ss R_b$ in the minimal supersymmetric standard
model.  We thank J. Lopez for clarifying this point.}  Our analysis
allows one to pinpoint just where the favorable regions are for solving
the $R_b$ problem.

\ref\Wells{J.D.~Wells and G.L.~Kane, SLAC preprint SLAC-PUB-7038,
hep-ph/9510372.} 
\ref\Yamada{Y.~Yamada, K.~Hagiwara and S.~Matsumoto, KEK preprint
KEK-TH-459, hep-ph/9512227.}

We thus see that it is possible to understand many of the conclusions in
the literature \feng--\Yamada\ on supersymmetry and $R_b$
using some rather simple analytic formulas. These include the preference
for small values of $\tan\beta$ as well as light higgsinos and squarks.

\ref\glnnpb{Y. Grossmann, Z. Ligeti and E. Nardi, Weizmann Institute
preprint WIS-95/49/Oct-PH, hep-ph/9510378, \npb{465}{96}{369}.}
\ref\enplb{E. Nardi, \plb{365}{96}{327}.}
\ref\bbmrnpb{See, for example, S. Bertolini, F. Borzumati, A. Masiero and
G. Ridolfi, \npb{353}{91}{591}.} 
\ref\ua{C. Albajar {\it et al.}, UA1 Collaboration, \plb{262}{91}{63}.}
\ref\cleo{R. Balest {\it et al.}, CLEO Collaboration, Cornell preprint
CLEO-CONF-94-4.} 
\ref\cdf{C. Anway-Wiese {\it et al.}, CDF Collaboration, Fermilab preprint
Fermilab-Conf-95/201-E.} 

\section{Future Tests}

If we exclude the possibility that the experimental value of $R_b$ is
simply a 3.7$\sigma$ statistical fluctuation, we can expect that, once the
LEP collaborations have completed their analyses of all the data collected
during the five years of running at the $Z$ pole, the `$R_b$ crisis' will
become an even more serious problem for the standard model. (Of course, it
is wise to keep in mind that there may be a simple explanation, namely that
some systematic uncertainties in the analysis of the experimental data are
still not well understood or have been underestimated.) In sections 3-5 we
have discussed a variety of models of new physics which could account for
the experimental measurement of $R_b$. The next obvious step is to consider
which other measurements may be used to reveal the presence of this new
physics.

The most direct method of finding the new physics is clearly the discovery
of new particles with the correct couplings to the $Z$ and the $b$ quark.
However, failing that, there are some indirect tests. For example, many of
the new-physics mechanisms which have been analysed in this paper will
affect the rate for some rare $B$ decays in a predictable way. The rates
for the rare decays $B\to X_s \ell^+ \ell^- $ and $B\to X_s \nu \bar\nu $
are essentially controlled by the $Z\bbar s$ effective vertex
$\Gamma^\mu_{bs}\,$, since additional contributions (such as box diagrams
and $Z$--$\gamma$ interference) are largely 
subleading.\foot\bsnunu{Due to the absence of $\ss Z$--$\ss\gamma$ 
interference and of large
renormalization-group-induced QCD corrections, the process $\ss B\to X_s
\nu \bar\nu $ represents theoretically the cleanest proof of the effective
$\ss Z\bbar s$ vertex \glnnpb.} In the SM, in the approximation made
throughout this paper of neglecting the $b$-quark mass and momentum, a
simple relation holds between the dominant $m_t$ vertex effects in $R_b$
and in the effective $Z\bbar s$ vertex $\Gamma^\mu_{bs}\,$:
\eq\label\Zbssm
\Gamma^{\mu,\SM}_{bs} = {V_{tb}^*V_{ts}\over\left| V_{tb} \right|^2}\,
\delta \, \Gamma^{\mu,\SM} 
\eeq
where $\delta \, \Gamma^{\mu,\SM}$ is defined as in eq.~\Zbs\ with the SM
form factor as given in eqs.~\formf\ and \SMresult. The meaning of
eq.~\Zbssm\ is that, within the SM, the $Z\bbar s$ effective vertex
measurable in $Z$-mediated $B$ decays represents a {\it direct} measurement
of the $m_t$-dependent vertex corrections contributing to $R_b$, modulo a
ratio of the relevant CKM matrix elements. In particular, both corrections
vanish in the $m_t\to 0$ limit. The question is now: how is this relation
affected by the new physics invoked in Secs.~3-5 to explain $R_b$?

Consider first the tree-level $b$--$b'$ mixing effects analysed in Sec.~3.
It is straightforward to relate the corrections of the LH and RH $Z\bbar b$
couplings to new tree-level mixing-induced FCNC couplings
$g^{bs}_{\ssl,\ssr}$. In this case eq.~\genrotation\ reads 
\eq\label\FCgenrotation
\eqalign{
g^{bs}_{\ssl,\ssr}
= \sum_{\alpha w} (g_w^\alpha)_{\ssl,\ssr} \, \Scu_{\ssl,\ssr}^{\alpha b*}
\Scu_{\ssl,\ssr}^{\alpha s} \, \Scm^{w1} \cr}~.
\eeq
Hence $g^{bs}_{\ssl,\ssr}$ involve the same gauge couplings and mixing
matrices that determine the {\it deviation} from the SM of the
flavour-diagonal $b$ couplings. 

It is also true that, for many models of new physics, the loop corrections
to the $Z\bbar b$ vertex would change the effective $Z\bbar s$ vertex in
much the same way, therefore inducing computable modifications to the SM
electroweak penguin diagrams. In these models, for each loop diagram
involving the new states $f,\, f'$ and their coupling to the $b$-quark
$g_{ff'b}$, there will be a similar diagram contributing to
$\Gamma_{bs}^\mu$ that can be obtained by the simple replacement
$g_{ff'b}\to g_{ff's}$. For example, the general analysis of $t$-quark
mixing effects presented in Sec.~4 can be straightforwardly applied to
$Z$-mediated $B$ decays. Deviations from the SM predictions for the $B\to
X_s \ell^+ \ell^- $  and $B\to X_s \nu \bar\nu $ decay rates can be easily
evaluated by means of a few simple replacements like $|{\cal V}_{tb}|^2 \to
V^*_{tb}V_{ts}$ and  $|{\cal V}_{t'b}|^2 \to V^*_{t'b}V_{t's}$ in all our
equations.\foot\enplbt{For example, the particular case of mixing of the
top-quark with a new isosinglet $\ss T'$, and the corresponding effects
induced on the $\ss Z\bbar s$ vertex, was studied in Ref.~\enplb\ through
an analysis very similar to that of Sec.~4.} To a large extent, this is
also true for SUSY models. Indeed, the analysis  of the SUSY contributions
to the $Z\bbar s$ form factor \bbmrnpb\ can teach much about SUSY effects
in $R_b$. And once a particular region of parameter space suitable to
explain the $R_b$ problem is chosen, a definite {\it numerical} prediction
for the $B\to X_s \ell^+ \ell^- $ and $B\to X_s \nu \bar\nu $ decay rates
can be made. 

This brief discussion shows that, for a large class of new-physics models, 
the new contributions to $R_b$ and to the effective $\Gamma^{\mu}_{bs}$
vertex are computable in terms of the same set of new-physics parameters.
Therefore, for all these models, the assumption that some new physics is
responsible for the deviations of $R_b$ from the SM prediction will imply a
quantitative prediction of the corresponding deviations for $Z$-mediated
$B$ decays. 

However, this statement cannot be applied to all new-physics
possibilities.  For example, if a new $Z'$ boson is responsible for the
measured value of $R_b$, then no signal can be expected in $B$ decays,
since in this case the new physics respects the GIM mechanism. This
would also be true if $m_b$-dependent effects are responsible for the
observed deviations in $R_b$ as could happen, for example, in the very
large $\tan\beta$ region of multi-Higgs-doublet or SUSY models. More
generally, the loop contributions of the new states $f,\,f'$ can be
different, since $g_{ff's}$ is not necessarily related to $g_{ff'b}$,
and  in particular, whenever the new physics involved in $R_b$ couples
principally to the third generation, it is quite possible that no
sizeable effect will show up in $B$ decays. Still, the study of $B\to
X_s \ell^+ \ell^- $ and $B\to X_s \nu \bar\nu $ could help to
distinguish between models that do or do not significantly affect these
decays.

Unfortunately, at present only upper limits have been set on the
branching ratios for $B\to X_s \ell^+ \ell^- $ \ua--\cdf\ and $B \to
X_s \nu \bar\nu $ \glnnpb. Since these limits are a few times larger
than the SM predicitons, they cannot help to pin down the correct
solution to the $R_b$ problem. However, future measurements of these
rare decays at $B$ factories could well confirm that new physics is
affecting the rate of $b$-quark production in $Z$ decays, as well as
give some hints as to its identity. If no significant deviations from
the SM expectations are detected, this would also help to restrict the
remaining possibilities.

\section{Conclusions}

Until recently, the SM has enjoyed enormous success in explaining all
electroweak phenomena. However, a number of chinks have started to appear
in its armour. There are currently several disagreements between theory
and experiment at the $2\sigma$ level or greater. They are: $R_b \equiv
\Gamma_b/\Gamma_{\rm had}$ ($3.7\sigma$), $R_c \equiv \Gamma_c/\Gamma_{\rm
had}$ ($2.5\sigma$), the inconsistency between $A^0_e$ as measured at LEP
with that determined at SLC ($2.4\sigma$), and $\AFB{\tau}$ ($2.0\sigma$).
Taken together, the data now exclude the SM at the 98.8\% confidence level.

Of the above discrepancies, it is essentially only $R_b$ which causes
problems. If $R_b$ by itself is assumed to be accounted for by new
physics, then the fit to the data despite the other discrepancies is
reasonable ($\chisqminpdof = 15.5/11$) -- the other measurements could
thus be regarded simply as statistical fluctuations.

In this paper we have performed a systematic survey of new-physics
models in order to determine which features give corrections to $R_b$
of the right sign and magnitude. The models considered can be separated
into two broad classes: those in which new $Z{\bar b}b$ couplings
appear at tree level, by $Z$ or $b$-quark mixing with new particles,
and those which give loop corrections to the $Z{\bar b}b$ vertex. The
latter type includes $t$-quark mixing and models with new scalars and
fermions. We did not consider technicolour models or new gauge bosons
appearing in loops since these cases are much more model-dependent.

The new physics can modify  either the left-handed  or right-handed
 $Z{\bar b}b$ couplings, $\gbl$ or $\gbr$.  To increase $R_b$ to its
experimental value, $\delta\gbl$ must be negative and have a magnitude
typical of a loop correction with large Yukawa couplings.  Thus
$\delta\gbl$ could either be a small tree-level effect, or a large
one-loop effect.  On the other hand, the SM value of $\gbr$ is opposite
in sign to its LH counterpart and is about five times smaller.
Therefore one would need a large tree-level modification to $\gbr$ to
explain for $R_b$.

Here are our results:

\topic{(1) Tree-level Effects} It is straightforward to explain $R_b$
if the $Z$ or $b$ mix with new particles. With $Z$--$Z'$ mixing there
are constraints from neutral-current measurements, but these do not
exclude all models.  Using $b$--$b'$ mixing is easier since the
experimental value of $R_b$ can be accommodated by $b_\lft$--$b'_\lft$
or $b_\rht$--$b'_\rht$ mixing. If the mixing is in the LH $b$ sector,
then solutions are possible so long as $I'_{3\lft} < -1/2$.  
An additional possibility with $I'_{3\lft} > 0$ and very large LH
mixing, though perhaps unappealing, is still viable.  For RH $b$
mixing, if $I'_{3\rht} > 0$ then small mixing is permitted, while if
$I'_{3\rht} < 0$, large mixing is necessary.  Interestingly, the
required large $b$-mixing angles are still not ruled out
phenomenologically.  A number of papers in the literature have
appealed to $b$-$b'$ mixing to explain $R_b$. Our ``master formula''
\gammab\ and Table \enritableone\ include all of these models, as well
as many others.

\topic{(2) Loops: $t$--$t'$ Mixing} In the presence of $t$--$t'$ mixing,
the SM radiative correction can be reduced, depending on the weak isospin
quantum numbers of the $t'$ as well as on the LH and RH mixing angles.
However, we found that it is not possible to completely explain $R_b$ via
this method. The best we can do is to decrease the discrepancy between
theory and experiment to about $2\sigma$. Such a scenario predicts the
existence of a light ($\sim 100$ GeV) charge $2/3$ quark, decaying
primarily to $Wb$.

\topic{(3) Loops: Diagonal Couplings to the $Z$} We considered models
with exotic fermions and scalars coupling to both the $Z$ and
$b$-quark. We assumed that the couplings to the $Z$ are diagonal, \ie\
there are no flavour-changing neutral currents (FCNC's). The correction
$\delta\gbl$ can then be written in a simple form, eq.~\scaiv. The key
point is that $\delta\gbl$ is proportional to
$I^f_{3\lft}-I^f_{3\rht}$, where $I^f_{3\lft,\rht}$ is the third
component of weak isospin of the fermion field $f_{\lft,\rht}$ in the
loop. This explains at a glance why many models, such as
multi-Higgs-doublet models and Zee-type models, have difficulty
explaining $R_b$. Since the dominant contributions in these models
typically have top-type quarks ($I'_{3\lft}={1\over 2}$,
$I'_{3\rht}=0$) circulating in the loop, they give corrections of the
wrong sign to $R_b$. However, these considerations did permit us to
construct viable models of this type which do explain $R_b$. Two such
examples are given Sec.~5.4, and many others can be invented.

\topic{(4) Loops: Nondiagonal Couplings to the $Z$} We also examined
models with exotic fermions and scalars which were allowed to have
nondiagonal couplings to the $Z$. Such FCNC's can occur when particles
of different weak isospin mix. The correction $\delta\gbl$ is much more
complicated (eq.~\defofGterms) than in the previous case; even its sign
is not obvious.  However there are several interesting limiting cases
where it again becomes transparent.  The contributions to $R_b$ of
supersymmetry fall into this category, which we discussed in
some detail.

\bigskip
\centerline{\bf Acknowledgements}
\bigskip

This research was financially supported by NSERC of Canada and FCAR du
Qu\'ebec. EN wishes to acknowledge the pleasant hospitality of the Physics
Department at McGill University, during the final stage of this work. DL
would like to thank Ken Ragan for helpful conversations.

\ref\ChanPok{P.H.~Chankowski and S.~Pokorski, preprint IFT-96/6, 
hep-ph/9603310.}

\bigskip {\bf Note Added:} After completing this work we became aware
of ref.~\ChanPok, which discusses a different region of parameter space
in SUSY models than the one we focused on.  Because of our criterion of
explaining the entire $R_b$ discrepancy rather than only reducing its
statistical significance, we exclude the region in question.

\listrefs
\bye